\documentclass[letterpaper,11pt]{article}

\usepackage[utf8]{inputenc}
\usepackage[T1]{fontenc}
\usepackage{jcapmod}
\usepackage{amsmath,amsfonts,amssymb,mathtools,physics,bm,textalpha,mathrsfs,amsbsy,amstext,cancel,hyperref,graphicx}
\usepackage[normalem]{ulem}
\usepackage[letterpaper,margin=2.55cm]{geometry}

\numberwithin{equation}{section}

\def\sqrtb{\mathpalette\DHLhksqrt}
\def\DHLhksqrt#1#2{%
\setbox0=\hbox{$#1\sqrt{#2\,}$}\dimen0=\ht0
\advance\dimen0-0.2\ht0
\setbox2=\hbox{\vrule height\ht0 depth -\dimen0}%
{\box0\lower0.4pt\box2}}

\makeatletter
\DeclareFontFamily{OMX}{MnSymbolE}{}
\DeclareSymbolFont{MnLargeSymbols}{OMX}{MnSymbolE}{m}{n}
\SetSymbolFont{MnLargeSymbols}{bold}{OMX}{MnSymbolE}{b}{n}
\DeclareFontShape{OMX}{MnSymbolE}{m}{n}{
    <-6>  MnSymbolE5
   <6-7>  MnSymbolE6
   <7-8>  MnSymbolE7
   <8-9>  MnSymbolE8
   <9-10> MnSymbolE9
  <10-12> MnSymbolE10
  <12->   MnSymbolE12
}{}
\DeclareFontShape{OMX}{MnSymbolE}{b}{n}{
    <-6>  MnSymbolE-Bold5
   <6-7>  MnSymbolE-Bold6
   <7-8>  MnSymbolE-Bold7
   <8-9>  MnSymbolE-Bold8
   <9-10> MnSymbolE-Bold9
  <10-12> MnSymbolE-Bold10
  <12->   MnSymbolE-Bold12
}{}
\makeatother

\newcommand{\mpl}{M_{\scriptscriptstyle\mathrm{Pl}}}
\newcommand{\tpl}{t_{\scriptscriptstyle\mathrm{Pl}}}
\newcommand{\mplt}{\widetilde{M}_{\scriptscriptstyle\mathrm{Pl}}}
\newcommand{\epsphi}{\epsilon_\psi}
\newcommand{\mL}{M_{\scriptscriptstyle\mathrm{L}}}
\newcommand{\NL}{{\scriptscriptstyle\mathrm{NL}}}
\newcommand{\G}{{\scriptscriptstyle\mathrm{G}}}
\newcommand{\gammaE}{\gamma_{\scriptscriptstyle\mathrm{E}}}
\newcommand{\orc}{\includegraphics[height=\fontcharht\font`A]{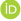}}

\begin{document}

\begin{titlepage}

\baselineskip=15.5pt \thispagestyle{empty}

\begin{center}
    {\fontsize{20.74}{24}\selectfont \bfseries
    Cuscuton Bounce Beyond the Linear Regime: \\
    \vspace*{.2cm}
    Bispectrum and Strong Coupling Constraints}
\end{center}

\vspace{0.1cm}

\begin{center}
    {\fontsize{12}{18}\selectfont
    Amir Dehghani,$^{1,2~\href{https://orcid.org/0009-0005-0395-9553}{\orc}}$ Ghazal Geshnizjani,$^{2,1~\href{https://orcid.org/0000-0002-2169-0579}{\orc}}$ and Jerome Quintin$^{1,2~\href{https://orcid.org/0000-0003-4532-7026}{\orc}}$}
\end{center}

\begin{center}
    \vskip8pt
    \textsl{$^1$ Department of Applied Mathematics and Waterloo Centre for Astrophysics,\\ University of Waterloo, Waterloo, ON N2L 3G1, Canada}\\
    \vskip4pt
    \textsl{$^2$ Perimeter Institute for Theoretical Physics, Waterloo, ON N2L 2Y5, Canada}
\end{center}

\vspace{1.2cm}

\hrule
\vspace{0.3cm}
\noindent {\bf Abstract}\\[0.1cm]
Cuscuton Gravity is characterized as a scalar field that can be added to general relativity without introducing any new dynamical degrees of freedom on a cosmological background. Yet, it modifies gravity such that spacetime singularities can be avoided. This has led to the Cuscuton bounce, a nonsingular cosmology that has been shown to be linearly stable, which is a rare feat. Upon introducing mechanisms known to generate a near-scale-invariant power spectrum of isocurvature perturbations in the prebounce contracting phase, we perform an extensive linear analysis of all scalar perturbations as they evolve through the Cuscuton bounce, both analytically and numerically. Then, after deriving the third-order perturbed action for our theory, we compare the magnitude of its terms (on shell) to those in the second-order action. We show that perturbativity is maintained in the infrared throughout the evolution, including through the bounce. In the ultraviolet, we show that a hierarchy of scales is maintained, with the strong coupling scale well above the relevant background energy scale at all times. We reconfirm these results by computing the three-point functions in various limits and demonstrate that the models do not have any strong coupling problems and furthermore that there is negligible non-Gaussianities on observable scales. Consequently, the primary potential source of observable non-Gaussianities may only arise from the conversion of isocurvature perturbations to curvature perturbations. The whole scenario is thus a robust, stable, weakly coupled nonsingular cosmological model, consistent with observations.
\vskip10pt
\hrule
\vskip10pt

\end{titlepage}

\thispagestyle{empty}
\tableofcontents
\newpage
\pagenumbering{arabic}
\setcounter{page}{1}


\section{Introduction}

Bouncing cosmology offers a compelling alternative to inflationary cosmology as the theory of the very early universe. While the inflationary paradigm provides a seemingly simple early universe scenario (resolving the puzzles of standard big bang cosmology and generating the seeds of our universe's large-scale structures; see, e.g., \cite{Guth:1980zm,Linde:1981mu,Mukhanov:1981xt,Bardeen:1983qw}), it is not free of any conceptual or theoretical issues (see, e.g., \cite{Brandenberger:2012uj,Ijjas:2013vea,Ijjas:2014nta,Agrawal:2018own,DiTucci:2019xcr,Bedroya:2019tba}). For instance, at least within its simplest (semi)classical realizations, it generally predicts an initial singularity in the past \cite{Borde:2001nh} (see \cite{Yoshida:2018ndv,Geshnizjani:2023hyd} and references therein for a detailed view of the issue). In contrast, a key aspect of bouncing cosmology is to provide a mechanism that avoids any cosmological singularity. However, this is precisely the source of one of the main challenges that bouncing cosmology faces (see, e.g., \cite{Battefeld:2014uga,Brandenberger:2016vhg} for reviews): how can the bounce --- the transition from contraction to expansion --- consistently and successfully occur?

Even considering a simple cosmological background (i.e., flat, homogeneous, and isotropic), it is already challenging to find bouncing solutions. This cannot be achieved with Einstein gravity and `normal matter' satisfying the null energy condition (NEC). Therefore, we must seek a modification or an extension to the theory, which could account for high-curvature or nonlocal effects from a complete theory of quantum gravity (e.g., string theory \cite{Gasperini:1996fu,Gasperini:2003pb,Gasperini:2004ss,Quintin:2018loc,Hohm:2019jgu,Wang:2019kez,Wang:2019dcj,Quintin:2021eup,Gasperini:2023tus}, loop quantum gravity \cite{Date:2004fj,Singh:2006im,deHaro:2012xj,Wilson-Ewing:2012lmx,Wilson-Ewing:2013bla,Cai:2014zga,Wilson-Ewing:2017vju}, etc.). Propositions for resolving big bang singularities into classically nonsingular cosmological bounces are abundant in the literature. However, while many succeed at describing the background evolution, they often suffer from instabilities or other kinds of pathologies once inhomogeneities are included. In this work, we shall focus on one particular theory, which shows great promise since it has not exhibited any of these problems thus far: the Cuscuton \cite{Afshordi:2006ad,Afshordi:2007yx}. This theory proposes to add a new scalar field to general relativity (GR) which breaks the time diffeomorphism invariance precisely such that the theory does not propagate any new local degree of freedom (DoF) compared to GR (see, e.g., \cite{Afshordi:2006ad,Afshordi:2007yx,Bhattacharyya:2016mah,Gomes:2017tzd,Boruah:2017tvg,Lin:2017oow,Iyonaga:2018vnu,Mukohyama:2019unx,Gao:2019twq,Aoki:2021zuy,DeFelice:2022uxv,Mylova:2023ddj,Lin:2017fec}).

The Cuscuton generically allows for nonsingular cosmologies that are free of instabilities at the linear level, i.e., there is no Ostrogradsky, ghost, gradient or tachyonic instability; see \cite{Lin:2017fec,Boruah:2018pvq,Quintin:2019orx,Sakakihara:2020rdy}. This is already a remarkable feature on its own since this is often very difficult to get from other types of modified gravity such as models of Horndeski theory that admit nonsingular bounces (for models that are claimed to work and others that simply do not work, see, e.g., \cite{Lin:2010pf,Easson:2011zy,Qiu:2011cy,Cai:2012va,Cai:2013kja,Easson:2013bda,Battarra:2014tga,Ijjas:2016tpn,Ijjas:2016vtq,Libanov:2016kfc,Kobayashi:2016xpl,deRham:2017aoj,Dobre:2017pnt,Ijjas:2017pei,Akama:2017jsa,Banerjee:2018svi,Mironov:2019fop,Ye:2019sth,Ye:2019frg,Creminelli:2016zwa,Cai:2016thi,Cai:2017tku,Volkova:2024mbn,Bohnenblust:2024mou,An:2025xeb}). Within Horndeski theory, one typically needs very special choices of functions (see the previous references) or introduce beyond-Horndeski terms (e.g., within degenerate higher-order scalar-tensor theories; see, e.g., \cite{Creminelli:2016zwa,Cai:2016thi,Cai:2017tku,Kolevatov:2017voe,Cai:2017dyi,Mironov:2018oec,Kolevatov:2018mhu,Volkova:2019jlj,Mironov:2019mye,Mironov:2019haz,Ilyas:2020qja,Zhu:2021whu,Mironov:2022ffa,Volkova:2024mbn}). Even in the cases that are claimed to work in (beyond-)Horndeski theory, it is often difficult to remain under perturbative control in the high-curvature regime of the bounce \cite{Koehn:2015vvy}, as corrections from higher-order terms (expanding the Lagrangian to third order and beyond) can become of the order of (or even larger than) second-order perturbations of the Lagrangian. This can be manifest, for instance, from the behavior of the sound speed, which may become very small in the bounce phase. One often finds the ratio of terms in the third-order action (computed on shell) to that of terms in the second-order action to scale as $1/c_\mathrm{s}^2$ (see, e.g., \cite{Baumann:2011dt}), and control of the perturbation theory may become problematic if this ratio becomes larger than unity. This is usually a sign of strong coupling or even nonunitarity \cite{deRham:2017aoj,Ageeva:2022byg,Cai:2022ori,Ageeva:2022asq}; at the very least, it may imply that large non-Gaussianities are generated through the bounce. (For related discussions in other alternative scenarios, see, e.g., \cite{Joyce:2011kh,Ageeva:2018lko,Ageeva:2020gti,Ageeva:2020buc,Ageeva:2021yik,Ageeva:2022fyq,Cai:2022ori,Ageeva:2022asq,Akama:2022usl,Volkova:2024mbn}.) For the Cuscuton, it appears that the sound speed may remain far from very small values \cite{Quintin:2019orx}. In fact, it might even allow for slight superluminality in the vicinity of the bounce and for high wavenumbers, without violating causality (this is not uncommon in this context; see, e.g., \cite{Dubovsky:2005xd,Mironov:2020mfo,Mironov:2020pqh}).\footnote{It is worth noting the so-called sound speed does not always correspond the waves' propagation speed. However, a proper understanding and interpretation of $c_\mathrm{s}>1$ in this context would have to be the subject of a future study on its own, as it comes with several caveats. It certainly does not appear to imply any kind of instability or pathology.} Therefore, one might be able to show that the Cuscuton theory is further well behaved, in that it remains under perturbative control through a nonsingular bounce, free of strong coupling issues, with reasonable non-Gaussianities. This shall be the main goal of this paper.

A second challenge for bouncing cosmologies, if seen as alternatives to inflation, is the generation of cosmological perturbations in the prebounce contracting phase, which can lead to predictions in agreement with cosmic microwave background (CMB) observations. Different models have been proposed such as matter domination (vanishing equation of state [EoS]) \cite{Wands:1998yp,Finelli:2001sr,Brandenberger:2012zb}, the ekpyrotic scenario (ultra-stiff EoS) \cite{Khoury:2001wf,Buchbinder:2007ad,Lehners:2008vx}, or pre-big bang cosmology (stiff EoS) \cite{Gasperini:1992em,Gasperini:2002bn,Gasperini:2007zz}. Purely adiabatic scenarios (such as matter domination) are known to face several issues, e.g., the overproduction of primordial gravitational waves or non-Gaussianities \cite{Quintin:2015rta,Li:2016xjb}, background instability with respect to anisotropies \cite{Cai:2013vm,Levy:2016xcl,Lin:2017fec,Ganguly:2021pke}, or strong coupling issues \cite{Baumann:2011dt} (see \cite{Lin:2017fec,Akama:2019qeh,Akama:2024bav,Akama:2024vtu,Akama:2025ows} for workarounds). Therefore, we will focus on `curvaton-like' scenarios, in which isocurvature modes (a.k.a.~entropy perturbations) are generated with a nearly scale-invariant power spectrum before being converted to curvature perturbations. The most recent ekpyrotic models make use of this mechanism \cite{Fertig:2013kwa,Ijjas:2014fja,Quintin:2024boj}, but it can also be implemented in a background evolving with a simple massless scalar field driving contraction with a stiff EoS \cite{Gasperini:2002bn,Kim:2020iwq}. (Ekpyrosis has the advantage of being an attractor washing out anisotropies \cite{Erickson:2003zm,Coley:2005pj,Lidsey:2005wr,Garfinkle:2008ei,Barrow:2010rx,Heinzle:2011qj,Barrow:2015wfa,Cook:2020oaj}; a massless scalar field marginally addresses this issue.) These models can be implemented such that the entropy modes remain Gaussian in the contracting phase prior to their conversion into curvature perturbations \cite{Fertig:2013kwa,Ijjas:2014fja,Fertig:2015ola,Fertig:2016czu}. This will allow us to focus on and isolate the impact of the nonsingular bounce phase, driven by the Cuscuton field, on the predicted bispectrum.

\paragraph*{Outline}
The paper is organized as follows: in Sec.~\ref{sec:linear}, we introduce and review the Cuscuton bounce model, the nonsingular background evolution, and the evolution of linear perturbations. Specifically, we derive the general equations for the background and for the cosmological perturbations in Sec.~\ref{sec:setupgen}. In Sec.~\ref{sec:modelsAnalytical}, we present different scenarios for the generation of large-scale fluctuations in the contracting phase and derive approximate analytical solutions for the evolution of the background and of the perturbations, including through the bounce phase. Some numerical examples are shown in Sec.~\ref{sec:numerical}. We move on to the issues of strong coupling and non-Gaussianities in Sec.~\ref{sec:nonlinear}. After outlining our expectations in Sec.~\ref{sec:insight}, the full third-order action is derived in Sec.~\ref{sec:3rdorder}. The cubic terms are then estimated and compared to those in the second-order action in Sec.~\ref{sec:strong-coupling} to estimate the strong coupling scale. We calculate the bispectra in various limits in Sec.~\ref{sec:bispectrum}, both analytically and numerically. This both reinforces our results regarding the validity of the perturbative regime and provides an estimate for the magnitude of the non-Gaussianities. We end with a discussion in Sec.~\ref{sec:conclusions}.

\paragraph*{Summary of the main results}
Since the paper is relatively long, we summarize the main results here for the reader's convenience.
\begin{itemize}
     \item We present two models for the contracting phase, which generate scalar perturbations that can be consistent with observations: a scalar field with an ekpyrotic potential and a massless field. For both models, we analytically and numerically show that perturbations of observational interest receive insignificant growth in the infrared (IR) as they pass through the bounce. This holds both for entropy and adiabatic perturbations, as well as if one is converted into the other right before the bounce. Adiabatic modes entering the bounce from a phase dominated by a massless scalar are the most prone to enhancement; still, the growth is at most a small fraction of the perturbations' prebounce amplitude. All other modes effectively remain constant to leading order in the IR.

     \item By comparing the terms in third- and second-order actions, we show that perturbativity is well under control in the IR, and in the UV, it appears to break down when the fluctuations' physical momenta approach the scale $\sqrtb{|H_{\mathrm{b}-}|\mpl}$, where $|H_{\mathrm{b}-}|$ is the highest Hubble rate reached in the universe (in absolute value), right by the start of the bounce phase. This scale is controlled by the Cuscuton model parameters, and it is usually well below the Planck scale. Higher interaction terms due to the Cuscuton in the bounce phase may bring the strong coupling scale down to $(H_{\mathrm{b}-}^2\mpl)^{1/3}$, but in all situations the background energy scale is shown to remain below the scale of strong coupling.

     \item We compute all nonzero 3-point correlation functions in Fourier space and find them to be highly suppressed in the IR (the bispectra are blue). Both entropy and adiabatic perturbations thus remain Gaussian and within the observational bounds to a very good approximation, at least assuming no additional enhancement during the conversion process. We show that there exists a reasonable range of Cuscuton parameters over which this remains true as the perturbations pass through the bouncing phase. Ultimately, the details of the process for converting isocurvature perturbations into curvature perturbations is the main remaining possibility responsible for producing potentially observable non-Gaussianities.
\end{itemize}

\paragraph*{Notation and conventions}
Throughout this paper, we use the mostly plus metric signature $(-,+,+,+)$ and units where the speed of light and Planck's reduced constant are set to unity, $c=\hbar=1$. The reduced Planck mass is defined in terms of Newton's gravitational constant as $\mpl\equiv(8\pi G_{\scriptscriptstyle\mathrm{N}})^{-1/2}$, and we use $\tpl\equiv 1/\mpl$ as the Planck time. Finally, Greek tensorial indices run over spacetime coordinates and are contracted with the spacetime metric $g_{\mu\nu}$, while latin tensorial indices run over spatial coordinates only.


\section{The Cuscuton bounce in the linear regime}\label{sec:linear}

\subsection{General setup, background equations, and linear perturbations}\label{sec:setupgen}

The setup shall consist of three scalar fields coupled to gravity: the Cuscuton $\varphi$ with Lagrangian
\begin{equation}
    \mathcal{L}_\varphi=-\mL^2\sqrtb{-\partial_\mu\varphi\partial^\mu\varphi}-U(\varphi)\,,\label{eq:Lcuscintro}
\end{equation}
which is responsible for modifying the gravitational dynamics, in particular allowing for nonsingular bouncing spacetimes, though without introducing any new dynamical DoFs \cite{Afshordi:2006ad,Afshordi:2007yx,Bhattacharyya:2016mah,Gomes:2017tzd,Boruah:2017tvg,Lin:2017oow,Iyonaga:2018vnu,Mukohyama:2019unx,Gao:2019twq,Aoki:2021zuy,DeFelice:2022uxv,Mylova:2023ddj}; the adiabatic scalar field $\psi$ with Lagrangian
\begin{equation}
    \mathcal{L}_\psi=X-V(\psi)\,,\qquad X\equiv -\frac{1}{2}\partial_\mu\psi\partial^\mu\psi\,,
\end{equation}
which corresponds to a dominant matter field in the theory and which is responsible for driving the background cosmological evolution; and last an entropy scalar field $\chi$ with Lagrangian
\begin{equation}
    \mathcal{L}_\chi=-\frac{1}{2}\mpl^2F(\psi,X)\partial_\mu\chi\partial^\mu\chi\,,\label{eq:defLchi}
\end{equation}
which is a spectator field (i.e., unaffecting the background), responsible for the generation of late-time scalar perturbations.
The total action is thus
\begin{equation}
    S=\int\dd^4x\,\sqrtb{-g}\left(\frac{\mpl^2}{2}R+\mathcal{L}_\varphi+\mathcal{L}_\psi+\mathcal{L}_\chi\right)\,,\label{eq:actionFullGen}
\end{equation}
where $g$ is the determinant of the metric and $R$ is the Ricci scalar.
A few comments are in order: we choose a negative sign for the Cuscuton's `kinetic term' since this is the one that allows for nonsingular bouncing cosmology \cite{Boruah:2018pvq,Quintin:2019orx,Sakakihara:2020rdy} --- both signs are equally valid \cite{Afshordi:2006ad,Afshordi:2007yx} and the Cuscuton is nondynamical (there is no ghost). The Cuscuton field has dimensions of mass, and we use $\mL$ to denote the new mass scale associated with the Cuscuton `kinetic term' --- the subscript {\small `L'} is meant to indicate that this is a \underline{L}imiting curvature scale, in the spirit of a limiting curvature theory (e.g., \cite{Mukhanov:1991zn,Brandenberger:1993ef,Yoshida:2017swb,Sakakihara:2020rdy} and references therein). It is called $\mu$ in \cite{Afshordi:2006ad,Afshordi:2007yx,Boruah:2017tvg,Boruah:2018pvq}, but we shall reserve this variable for the dimensionless ratio of the Cuscuton's mass scale to the Planck scale, i.e., $\mu\equiv \mL/\mpl$. We note that the scalar field $\psi$ is simply a proxy for the dominant matter DoF, in this case taken to be a canonical scalar field with some arbitrary potential $V(\psi)$ at this point. Finally, we note that the scalar field $\chi$ is kinetically coupled to the matter field via the dimensionless function $F(\psi,X)$ (in the spirit of, e.g., \cite{Ijjas:2014fja,Levy:2015awa,Kim:2020iwq}) --- the possible form of this function will be specified shortly, but the idea is that it should be a function of either $\psi$ or its derivative. It was previously mentioned that $\chi$ acts as a spectator field; it shall soon become clear that this results in $\chi$ being relevant only perturbatively. For this reason, we define $\chi$ as a dimensionless scalar by explicit insertion of $\mpl^2$ in \eqref{eq:defLchi}.

Let us write the metric in the Arnowitt-Deser-Misner formalism as $g_{\mu\nu}\dd x^\mu\dd x^\nu=-N^2\dd t^2+h_{ij}(N^i\dd t+\dd x^i)(N^j\dd t+\dd x^j)$, where the lapse, shift, and spatial metric are perturbed about a Friedmann-Lema\^itre-Robertson-Walker (FLRW) background as
\begin{equation}
    N(t,\bm{x})=1+\alpha(t,\bm{x})\,,\qquad N_i(t,\bm{x})=\partial_i\beta(t,\bm{x})\,,\qquad h_{ij}(t,\bm{x})=a(t)^2e^{2\zeta(t,\bm{x})}\delta_{ij}\,.\label{eq:ADMpert}
\end{equation}
These are only the scalar perturbations, which are going to be the perturbations of interest throughout this work --- we will briefly comment on tensor perturbations at the end. We further perturb the scalar fields as
\begin{equation}
    \varphi(t,\bm{x})=\bar\varphi(t)+\delta\varphi(t,\bm{x})\,,\qquad\psi(t,\bm{x})=\bar\psi(t)\,,\qquad\chi(t,\bm{x})=\bar\chi(t)+\delta\chi(t,\bm{x})\,.\label{eq:fieldspert}
\end{equation}
Here we are explicitly choosing the comoving gauge with respect to the matter field $\psi$ by setting $\delta\psi(t,\bm{x})\equiv\psi(t,\bm{x})-\bar\psi(t)=0$. In what follows, we will abuse notation and simply express the matter field as $\psi(t)$, and its kinetic energy will be $X=\dot\psi^2/2$. Note that we are free to perform this choice of gauge (under some reasonable assumptions, e.g., such that close to FLRW, constant-$\psi$ surfaces form spacelike hypersurfaces), and any other gauge would result in the same second-order action (see, however, \cite{Quintin:2019orx} for a slight subtlety when using the spatially flat gauge).
In the above, $a(t)$ is the background scale factor and $\zeta(t,\bm{x})$ is the curvature perturbation.

The background equation of motion (EoM) for $\chi$ is given by $\ddot{\bar\chi}+\big[3H+\dot F(\psi,X)\big]\dot{\bar\chi}=0$, where $H\equiv\dot a/a$ is the Hubble parameter. We note that $\dot{\bar\chi}=0$ is always a solution, in which case $\bar\chi$ never contributes to the background evolution --- in this regime $\chi$ is only a spectator field. Therefore, going forward, we shall assume $\bar\chi=0$ --- the value of the constant is irrelevant, and this solution has been shown to be stable \cite{Ijjas:2014fja,Levy:2015awa,Kim:2020iwq} for the models of interest that will be explained later. The field $\chi$ thus only enters at the perturbation level as $\delta\chi$, so we will simply write $\delta\chi(t,\bm{x})=\chi(t,\bm{x})$ and directly use the latter variable to describe the perturbation.

Given the spectator field $\chi$ does not contribute to the backgound, we may write the remaining background equations as
\begin{subequations}
{\allowdisplaybreaks
\begin{align}
    &3\mpl^2H^2=\frac{1}{2}\dot{\psi}^2+V(\psi)+U(\bar\varphi)\,,\label{eq:Friedconstraint}\\
    &-2\dot H=\mpl^{-2}\dot{\psi}^2-\mu^2|\dot{\bar\varphi}|\,,\label{eq:FriedEvoEoM}\\
    &\ddot{\psi}+3H\dot{\psi}+V_{,\psi}(\psi)=0\,,\label{eq:phibackEoM}\\
    &3\mu^2H=\mpl^{-2}U_{,\varphi}(\bar\varphi)\,,\label{eq:backCuscutonConstraint}
\end{align}
}%
\end{subequations}
where a comma denotes a partial derivative.
Those are the Friedmann constraint equation, the background evolution equation, the matter EoM, and the Cuscuton constraint equation, respectively.
Note that from here on, without loss of generality, we will work under the assumption that $\dot{\bar\varphi}>0$ in order to avoid the use of absolute values. We are free to make this choice since the Cuscuton on a cosmological background is defined for timelike $\partial_\mu\varphi$, so the sign of $\dot{\bar\varphi}$ will never change.\footnote{In fact, it is known that the theories with $\partial_\mu\varphi$ timelike vs spacelike are fundamentally different, in particular with regard to the number of propagating DoFs \cite{Afshordi:2006ad,Gomes:2017tzd,Iyonaga:2018vnu}. In this work, we always consider the theory that has $\partial_\mu\varphi$ timelike.}
Defining
\begin{equation}
    \epsilon\equiv-\frac{\dot H}{H^2}\,,\qquad\epsphi\equiv\frac{\dot{\psi}^2}{2\mpl^2H^2}\,,\qquad\sigma\equiv\epsphi-\epsilon\,,\label{eq:defepsepsphi}
\end{equation}
the Friedmann constraint and evolution equations above can also be rewritten as $U(\bar\varphi)=\mpl^2(3-\epsphi)H^2-V(\psi)$ and $\mu^2\dot{\bar\varphi}=2\sigma H^2$. Note that since $\dot{\bar\varphi}>0$, it follows that $\sigma>0$, and in the GR limit $\mu^2\to 0$ implies $\sigma\to 0$ for $H\neq 0$.
We also notice from the above that the Cuscuton does not follow any dynamical equation but rather only a constraint equation --- it is the nature of the Cuscuton not to introduce any new DoF but to modify the gravitational dynamics.
Taking a derivative of the Cuscuton constraint \eqref{eq:backCuscutonConstraint}, we note that 
\begin{equation}
    U_{,\varphi\varphi}(\bar\varphi)=\frac{3\mu^2\mpl^2\dot H}{\dot{\bar\varphi}}\,.\label{eq:cuscConstrDeriv}
\end{equation}
We can thus see from \eqref{eq:backCuscutonConstraint} and \eqref{eq:cuscConstrDeriv} how an appropriate choice of Cuscuton potential allows for nonsingular solutions. In particular, if $U_{,\varphi}$ is bounded, so will be $H$; additionally, where it crosses zero corresponds to a bouncing point. If this occurs while $U_{,\varphi\varphi}$ admits an interval where it is positive, then according to \eqref{eq:backCuscutonConstraint} this will also allow for a bouncing phase (a transition from contraction [$H\propto U_{,\varphi}<0$] to expansion [$H\propto U_{,\varphi}>0$]) where $\dot H>0$, implying that the effective NEC is violated. Conversely, away from the bounce phase, the effective NEC is restored as long as $U_{,\varphi\varphi}$ becomes negative. Examples of such solutions are given in \cite{Boruah:2018pvq,Quintin:2019orx}, and we will show this explicitly in the next subsection.

We will soon consider the case where $\psi$ is massless as an example. For such a case, it is actually straightforward to see in what sense the Cuscuton is nondynamical. Furthermore, if $V(\psi)$ is set to zero, the constraint equations allow for a simple change of variable from $t$ to $\bar\varphi$ and significant simplifications of the differential equations. So let us explore this example. First, we note that once the potential $U(\varphi)$ is prescribed, the Hubble parameter is set by the constraint equation \eqref{eq:backCuscutonConstraint} as a function of the $\bar\varphi$ profile.
Then, if $\psi$ has vanishing potential, its energy density is $\rho_\psi=\dot{\psi}^2/2$, and we can write [using \eqref{eq:Friedconstraint} and \eqref{eq:backCuscutonConstraint}]
\begin{equation}
    \rho_\psi=3\mpl^2H^2-U(\bar\varphi)=\frac{U_{,\varphi}(\bar\varphi)^2}{3\mu^4\mpl^2}-U(\bar\varphi)\,,\label{eq:rhophiofvarphi}
\end{equation}
hence once the Cuscuton potential is chosen, one immediately obtains $\rho_\psi(\bar\varphi)$. The second time derivative of the scalar field as a function of $\bar\varphi$, if ever needed, is then obtained from the $\psi$ EoM \eqref{eq:phibackEoM}, $\ddot{\psi}=-3H\dot{\psi}=\mp(\mu\mpl)^{-2}U_{,\varphi}(\bar\varphi)\sqrtb{2\rho_\psi(\bar\varphi)}$, where the sign depends on whether $\dot{\psi}$ is positive or negative, which depends on the chosen initial conditions. Finally, one can isolate $\dot H$ (and then find $\epsilon$) by combining \eqref{eq:FriedEvoEoM}, \eqref{eq:cuscConstrDeriv}, and \eqref{eq:rhophiofvarphi}:
\begin{equation}
    \dot H(\bar\varphi)=\frac{2\rho_\psi(\bar\varphi)U_{,\varphi\varphi}(\bar\varphi)}{\mpl^2\big(3\mu^4\mpl^2-2U_{,\varphi\varphi}(\bar\varphi)\big)}
    \quad\Rightarrow\quad\epsilon(\bar\varphi)=3\left(1-\frac{3\mu^4\mpl^2U(\bar\varphi)}{U_{,\varphi}(\bar\varphi)^2}\right)\left(1-\frac{3\mu^4\mpl^2}{2U_{,\varphi\varphi}(\bar\varphi)}\right)^{-1}\,.\label{eq:Hepsofvarphi}
\end{equation}
From those equations, one can see that every background quantity can be expressed as a function of $\bar\varphi$. To get temporal evolution, a strategy is to find the solution for $\bar\varphi(t)$, after which every function of $\bar\varphi$ found above becomes a function of time, e.g., $\dot H(t)=\dot H(\bar\varphi(t))$. The solution for $\bar\varphi(t)$ can be found from \eqref{eq:cuscConstrDeriv} and \eqref{eq:Hepsofvarphi}:
\begin{equation}
\label{eq:varphi_for_solv}
    \dot{\bar\varphi}=\frac{3\mu^2\mpl^2\dot H(\bar\varphi)}{U_{,\varphi\varphi}(\bar\varphi)}=\frac{2}{\mu^2\mpl^2}\frac{U_{,\varphi}(\bar\varphi)^2-3\mu^4\mpl^2U(\bar\varphi)}{3\mu^4\mpl^2-2U_{,\varphi\varphi}(\bar\varphi)}\,.
\end{equation}
Given a potential $U(\varphi)$, this is a first-order ordinary differential equation that can be solved given some initial condition. The initial condition is just an arbitrary choice of reference time; often we shall set $\bar\varphi(t=0)=0$ to be the bounce point as in the numerical examples of Sec.~\ref{sec:numerical}.

Let us move on to the general description of the linear perturbations. (What follows reproduces calculations performed in \cite{Boruah:2017tvg,Boruah:2018pvq,Lin:2017fec,Quintin:2019orx,Kim:2020iwq}; the reader familiar with this work may skip ahead to Sec.~\ref{sec:modelsAnalytical}.) Upon perturbing the action to second order in $\alpha$, $\beta$, $\delta\varphi$, $\zeta$, and $\chi$ [defined in \eqref{eq:ADMpert}--\eqref{eq:fieldspert}], one finds
\begin{align}
    S^{(2)}=\mpl^2\int\dd^3\bm{x}\,\dd t\,a^3\bigg(&H^2(\epsphi-3)\alpha^2-3\mu^2H\alpha\,\delta\varphi+6H\alpha\dot\zeta+3\mu^2\dot\zeta\,\delta\varphi-3\dot\zeta^2+\frac{(\partial_i\zeta)^2}{a^2}\nonumber\\
    &-2\alpha\frac{\partial^2\zeta}{a^2}-\left(2H\alpha-2\dot\zeta+\mu^2\delta\varphi\right)\frac{\partial^2\beta}{a^2}+\frac{\mu^4}{4\sigma}\Big(3\epsilon\delta\varphi^2+\frac{(\partial_i\delta\varphi)^2}{a^2H^2}\Big)\nonumber\\
    &+\frac{1}{2}F(\psi,X)\Big(\dot{\chi}^2-\frac{(\partial_i\chi)^2}{a^2}\Big)\bigg)\,,\label{eq:S21}
\end{align}
where we used integration by parts and the background EoMs to simplify the expression.
Note that $\partial^2\equiv\delta^{ij}\partial_i\partial_j$ denotes the spatial Laplacian.
We notice that the perturbations of the lapse and shift ($\alpha$ and $\partial_i\beta$) are nondynamical, hence they just represent constraints (the Hamiltonian and momentum constraints), and they can be removed from the action.
Variation with respect to $\alpha$ and $\partial_i\beta$, respectively, yields the constraints
\begin{equation}
    2H^2(\epsphi-3)\alpha+6H\dot\zeta-\frac{2}{a^2}\left(\partial^2\zeta+H\partial^2\beta\right)=3\mu^2H\,\delta\varphi\,,\qquad
    2(H\alpha-\dot\zeta)=-\mu^2\,\delta\varphi\,,
\end{equation}
which can be solved for $\alpha$ and $\beta$ as
\begin{equation}
    \alpha=\frac{1}{H}\left(\dot\zeta-\frac{\mu^2}{2}\delta\varphi\right)\,,\qquad\beta=-\frac{\zeta}{H}+a^2\epsphi Q\,,\label{eq:solalphabeta}
\end{equation}
where $Q$ is defined according to
\begin{equation}
    \partial^2Q\equiv\dot\zeta-\frac{\mu^2}{2}\delta\varphi\,.\label{eq:psi}
\end{equation}
It is understood that $\partial^{-2}(\partial^2Q)=Q$, so $\partial^{-2}$ is the inverse spatial Laplace operator.
Similar to the lapse and shift, $\delta\varphi$ also appears nondynamically in the action (as expected since the Cuscuton is nondynamical).
Upon substituting the solutions for $\alpha$ and $\beta$ in \eqref{eq:S21}, $\delta\varphi$ is found to satisfy a constraint equation, whose solution is
\begin{equation}
    \delta\varphi=-\frac{2\sigma H}{\mu^2}\mathcal{D}^{-2}\left(\epsilon_\psi H\dot\zeta-\frac{\partial^2\zeta}{a^2}\right)\,,\qquad\mathcal{D}^2\equiv\frac{\partial^2}{a^2}-(\sigma+3)\epsilon_\psi H^2\,;\label{eq:deltacusc}
\end{equation}
here $\mathcal{D}^{-2}$ is a transformed inverse Laplace operator.

Note that the solutions \eqref{eq:solalphabeta} may appear invalid at a bounce point when $H=0$, but it turns out that no divergences actually arise \cite{Quintin:2019orx}. Indeed, one can appreciate this by expanding first the solution to the Cuscuton constraint \eqref{eq:deltacusc} about $H=0$, which yields
\begin{equation}
    \frac{\mu^2}{2}\delta\varphi\stackrel{H\approx 0}{\simeq}\dot\zeta-\frac{1}{\epsphi H}\frac{\partial^2\zeta}{a^2}-\frac{3}{\sigma}\dot\zeta+\frac{1}{\sigma\epsphi H^2}\frac{\partial^2\dot\zeta}{a^2}+\mathcal{O}(H^3)\,,\label{eq:deltavarphiH0}
\end{equation}
recalling $\epsilon$, $\epsphi$, and $\sigma$ --- defined in \eqref{eq:defepsepsphi} --- are $\mathcal{O}(H^{-2})$ about $H=0$. This means from \eqref{eq:solalphabeta},
\begin{subequations}\label{eq:alphabetaH0}
\begin{align}
    \alpha&\stackrel{H\approx 0}{\simeq}\frac{1}{\epsphi H^2}\frac{\partial^2\zeta}{a^2}+\frac{3}{\sigma H}\dot\zeta-\frac{1}{\sigma\epsphi H^3}\frac{\partial^2\dot\zeta}{a^2}+\mathcal{O}(H^2)\,,\\
    \beta&\stackrel{H\approx 0}{\simeq}-\frac{1}{\sigma H^2}\dot\zeta+\frac{3\epsphi}{\sigma}a^2\partial^{-2}\dot\zeta+\mathcal{O}(H)\,.
\end{align}
\end{subequations}
This will be useful later.

Next, inserting the solution \eqref{eq:deltacusc} back into the second-order action, performing more integration by parts, and using the background equations, one finally finds an action that separates as $S^{(2)}=S^{(2)}_{\zeta}+S^{(2)}_{\chi}$, with
\begin{equation}
    S^{(2)}_\zeta=\mpl^2\int\dd^3\bm{x}\,\dd t\,a^3\left(\mathcal{G}_1\dot\zeta^2+\mathcal{G}_2\frac{(\partial_i\dot\zeta)^2}{a^2}-\mathcal{F}_1\frac{(\partial_i\zeta)^2}{a^2}-\mathcal{F}_2\frac{(\partial_i\partial_j\zeta)^2}{a^4}-\mathcal{F}_3\frac{(\partial_i\partial_j\partial_k\zeta)^2}{a^6}\right)\label{eq:S23}
\end{equation}
and
\begin{equation}
    S^{(2)}_{\chi}=\frac{\mpl^2}{2}\int\dd^3\bm{x}\,\dd t\,a^3F(\psi,X)\left(\dot{\chi}^2-\frac{1}{a^2}(\partial_i\chi)^2\right)\,.\label{eq:S2deltachi}
\end{equation}
In the above, the coefficients are $\mathcal{G}_1=\epsphi\mathcal{D}^{-4}\left(a^{-4}\partial^4-3\epsphi H^2\left(a^{-2}\partial^2+\mathcal{D}^2\right)\right)$, $\mathcal{G}_2=\sigma\epsphi^2H^2\mathcal{D}^{-4}$,
{\allowdisplaybreaks
\begin{align}
    \mathcal{F}_1=&~\epsilon\epsphi H^2\mathcal{D}^{-6}\bigg(\frac{1}{\epsphi H^2}\frac{\partial^6}{a^6}+\left[5\epsilon+2\frac{\epsphi}{\epsilon}(3+2\eta_\psi)-(15+5\epsphi+2\eta+2\eta_\psi)\right]\frac{\partial^4}{a^4}\nonumber\\
    &-\Big[4\epsilon-5\epsphi-3(4+\eta)+\frac{\epsphi}{\epsilon}(3+\epsphi+3\eta_\psi)\Big]\Big[3\epsphi H^2\frac{\partial^2}{a^2}-(\sigma+3)\epsphi^2H^4\Big]\bigg)\,,\nonumber\\
    \mathcal{F}_2=&~\mathcal{D}^{-6}\bigg(-2\sigma\frac{\partial^4}{a^4}+\epsphi H^2\Big[4\epsilon^2-\epsilon(12+7\epsphi+\eta+\eta_\psi)+\epsphi\big(3\epsphi+2(6+\eta_\psi)\big)\Big]\frac{\partial^2}{a^2}\nonumber\\
    &-\epsphi^2H^4\Big[2\epsilon^3+\epsphi\big(6+\epsphi^2+\epsphi(5-2\eta_\psi)\big)-\epsilon^2(4+3\epsphi+\eta+\eta_\psi)\nonumber\\
    &+\epsilon\big(3(\eta_\psi-\eta-2)+\epsphi(-1+\eta+3\eta_\psi)\big)\Big]\bigg)\,,
\end{align}
}%
and $\mathcal{F}_3=-\sigma\mathcal{D}^{-4}$, where we further defined
\begin{equation}
    \eta\equiv\frac{\dot\epsilon}{H\epsilon}\,,\qquad\eta_\psi\equiv\frac{\dot\epsilon_\psi}{H\epsphi}\,.
\end{equation}
The above actions are somewhat easier to read when expressed in Fourier space\footnote{Our convention is \[\zeta(t,\bm{x})=\int_{\bm{k}}\zeta_{\bm{k}}(t)\exp(i\bm{k}\cdot\bm{x})\,,\qquad\int_{\bm{k}}\equiv\int\frac{\dd^3\bm{k}}{(2\pi)^3}\,,\] and we denote $k\equiv\lVert\bm{k}\rVert\equiv\sqrtb{\bm{k}\cdot\bm{k}}\equiv\sqrtb{\delta_{j\ell}k^jk^\ell}$ (the standard Euclidean norm and dot product).} and in conformal time (defined as $\dd\tau\equiv a^{-1}\dd t$):
\begin{equation}
    S^{(2)}_{\zeta}=\frac{\mpl^2}{2}\int_{\bm{k}}\int\dd \tau\,z^2\left(|\zeta_{\bm{k}}^{\prime}|^2-c_\mathrm{s}^2k^2|\zeta_{\bm{k}}|^2\right)\,;\qquad S^{(2)}_{\chi}=\frac{\mpl^2}{2}\int_{\bm{k}}\int\dd \tau\,z_\chi^2\left(|\chi_{\bm{k}}^{\prime}|^2-k^2|\chi_{\bm{k}}|^2\right)\,.\label{eq:S2zetaS2chi}
\end{equation}
Note that a prime denotes a derivative with respect to $\tau$, and defining $\mathcal{H}\equiv a'/a=aH$, the coefficients in the actions are now
\begin{equation}
    z^2=2a^2\epsphi\frac{k^2+3\epsphi\mathcal{H}^2}{k^2+(\sigma+3)\epsphi\mathcal{H}^2}\,,\qquad c_\mathrm{s}^2=\frac{k^4+\mathcal{H}^2\mathcal{B}_1k^2+\mathcal{H}^4\mathcal{B}_2}{k^4+\mathcal{H}^2\mathcal{A}_1k^2+\mathcal{H}^4\mathcal{A}_2}\,,\qquad z_\chi^2=a^2F(\psi,X)\,,\label{eq:z-cs2-zs}
\end{equation}
with
\begin{align}
    \mathcal{A}_1&=\epsphi(\sigma+6)\,,&\mathcal{B}_1&=\epsphi(6+3\sigma+\eta-\eta_\psi)-\sigma(6+2\sigma+\eta+\eta_\psi)\,,\nonumber\\
    \mathcal{A}_2&=3\epsphi^2(\sigma+3)\,,&\mathcal{B}_2&=\epsphi\big(3\epsphi(3+\sigma+\eta-\eta_\psi)-4\sigma(3+\sigma)-3\sigma\eta\big)\,.\label{eq:A1A2B1B2coeffs}
\end{align}
The resulting EoMs for the mode functions $\zeta_k$ and $\chi_k$ are
\begin{equation}
    \zeta_k''+2\frac{z'}{z}\zeta_k'+c_\mathrm{s}^2k^2\zeta_k=0\,,\qquad\chi_k''+2\frac{z_\chi'}{z_\chi}\chi_k'+k^2\chi_k=0\,,\label{eq:zeta-chi-eoms-k}
\end{equation}
whose specific solutions depend on the models of interest. Note that $z^2$ and $c_\mathrm{s}^2$ are both time and wavenumber dependent. It is straightforward to see that $z^2$ is always positive \cite{Boruah:2017tvg}, so there is no ghost, and $c_\mathrm{s}^2$ can also remain positive (and close to unity) given a reasonable Cuscuton potential (see \cite{Boruah:2018pvq,Quintin:2019orx}; this will be discussed later as well).

For future reference, it is useful to look at the action for $\zeta$ in different regimes of interest: in the IR to leading order in the gradient expansion, we find
\begin{equation}
    S_\zeta^{(2,\mathrm{IR})}\simeq\mpl^2\int\dd^3\bm{x}\,\dd t\,a^3\frac{\epsphi}{1+\sigma/3}\left(\dot\zeta^2-\frac{3\epsphi(3+\sigma+\eta-\eta_\psi)-4\sigma(3+\sigma)-3\sigma\eta}{3\epsphi(\sigma+3)}\frac{(\partial_i\zeta)^2}{a^2}\right)\,;\label{eq:S2zetaIR}
\end{equation}
oppositely, in the UV where spatial gradients dominate, we have
\begin{equation}
    S^{(2,\mathrm{UV})}_\zeta\simeq\mpl^2\int\dd^3\bm{x}\,\dd t\,a^3\epsphi\left(\dot\zeta^2-\frac{(\partial_i\zeta)^2}{a^2}\right)\,.\label{eq:S2zetaUV}
\end{equation}
This makes it explicit in what sense the Cuscuton affects IR perturbations, while leaving the UV sector of the theory untouched. The story in the UV is a little more complicated at the bounce point where $H=0$, though (see \cite{Quintin:2019orx}). Indeed, since $\epsphi$ is $\mathcal{O}(H^{-2})$ about $H=0$, it would appear that \eqref{eq:S2zetaUV} diverges at the point where $H=0$. However, this is solely an issue about how to take the UV limit since first expanding about $H\approx 0$ and then taking a UV limit yields a finite expression,
\begin{equation}
    S_\zeta^{(2,\textrm{UV})}\stackrel{H\approx 0}{\simeq}\mpl^2\int\dd^3\bm{x}\,\dd t\,\frac{a^3}{\sigma H^2}\left(\frac{(\partial_i\dot\zeta)^2}{a^2}-\frac{\sigma}{\epsphi}\frac{(\partial^2\zeta)^2}{a^4}\right)\,,\label{eq:S2zetaUVH0}
\end{equation}
since $\sigma H^2$ and $\sigma/\epsphi$ are $\mathcal{O}(H^{0})$.
Note that everywhere else ($H\neq 0$), \eqref{eq:S2zetaUV} is a perfectly good approximation, and this subtlety does not affect the evolution of the perturbations in the UV \cite{Quintin:2019orx}.

\subsection{Explicit models and analytical approximations}\label{sec:modelsAnalytical}

\subsubsection{Prebounce contraction}

We shall consider two main models in what follows for the prebounce contracting phase: the case where $\psi$ is an ekpyrotic field; and the case where $\psi$ is a massless field. In the former situation, this means the background scalar field has a steep and negative potential of the form
\begin{equation}
    V(\psi)=-V_0e^{-\sqrtb{2\bar\epsilon}\psi/\mpl}\label{eq:ekPotential}
\end{equation}
for some real constants $V_0>0$ and $\bar\epsilon>3$, while in the latter situation this simply means $V(\psi)$ vanishes identically. The motivation for these particular models is as follows: in \cite{Boruah:2018pvq}, a massless scalar field was considered in the matter sector as the simplest proxy for matter --- the focus was on the physics of the bouncing phase and on the stability of the solution. Then, in order to explore the evolution of cosmological perturbations through the bounce, especially scale-invariant perturbations that could account for the observations, an appropriate coupling to the spectator field $\chi$ was considered in \cite{Kim:2020iwq}. In the spirit of testing the Cuscuton bounce, we will revisit this model here again. The other model where one has an early ekpyrotic phase has the advantage of being a proper attractor \cite{Erickson:2003zm,Lidsey:2005wr}, thus properly smoothing the universe (just like inflation) before the bounce \cite{Garfinkle:2008ei,Cook:2020oaj,Ijjas:2020dws,Ijjas:2021gkf,Ijjas:2021wml,Ijjas:2021zyf,Ijjas:2022qsv,Kist:2022mew,Ijjas:2023dnb,Ijjas:2024oqn}. Even when $\psi$ is massless, anisotropies are under control as long as the evolution starts close enough to FLRW. Let us note that in concrete realizations of ekpyrotic cosmology, it is often the case that the ekpyrotic phase of contraction is followed by a phase of kination, i.e., the potential goes to $0$, so the scalar field becomes massless and kinetic driven.\footnote{This could be modeled with a potential of the form $V(\psi)=-V_0/(e^{\sqrtb{2\bar\epsilon}\psi/\mpl}+e^{-b_V\sqrtb{2\bar\epsilon}\psi/\mpl})$, for some constant $b_V>0$, as in \cite{Cai:2012va,Cai:2013kja}. At the level of the discussion for this paper, we will treat a single-exponential phase and a massless one distinctly.\label{foot:twoExp}} Therefore, the two models we will consider could actually be considered as being two separate phases of the whole prebounce contraction.

\paragraph{Ekpyrotic potential}
Let us review these models in more detail. In the far past, well before the bounce phase, we will ensure the Cuscuton field has a negligible contribution, hence we neglect it in the discussion for now. If the scalar field $\psi$ has the ekpyrotic potential \eqref{eq:ekPotential}, then this will lead, early in the contracting phase, to the well-known scaling solution (see, e.g., \cite{Lehners:2008vx,Lehners:2011kr})
\begin{align}
	a(\tau)&=\left(\frac{\tau}{\tau_{\mathrm{b}-}}\right)^{\frac{1}{\bar\epsilon-1}}\,,\nonumber\\
    \psi(\tau)&=\mpl\left(\frac{\sqrtb{2\bar\epsilon}}{\bar\epsilon-1}\ln\left(-\tau\mpl\right)+\frac{1}{\sqrtb{2\bar\epsilon}}\ln\left(\frac{(\bar\epsilon-1)^2}{\bar\epsilon-3}\frac{V_0}{\mpl^4}\frac{1}{(-\tau_{\mathrm{b}-}\mpl)^\frac{2}{\bar\epsilon-1}}\right)\right)\,,\label{eq:scalingBackEk}
\end{align}
for $\tau<\tau_{\mathrm{b}-}<0$. Note that, for practical reasons, we choose to normalize the scale factor to unity at the time $\tau_{\mathrm{b}-}$, where the subscript `$\mathrm{b}-$' indicates the end of the standard contracting phase (equivalently the onset of the bounce phase where the Hubble parameter is maximal in absolute value), but physical quantities do not depend on this choice of scale factor normalization. From the above, we note
\begin{equation}
	\mathcal{H}(\tau)=\frac{1}{(\bar\epsilon-1)\tau}\,,\qquad\epsilon=\epsphi=\bar\epsilon\,,
\end{equation}
and the energy density in the ekpyrotic field scales as $\rho_\psi\propto a^{-2\bar\epsilon}$, hence $\psi$ dominates over other fields as $a\to 0^+$, including massless fields and shear anisotropies whose energy densities scale as $a^{-6}$ (recall $\bar\epsilon>3$). In fact, an ekpyrotic field often has $\bar\epsilon\gg 3$, hence its isotropizing power.

Let us then discuss the evolution of adiabatic perturbations in a phase of ekpyrotic contraction. Defining the Mukhanov-Sasaki variable $v_k\equiv\mpl z\zeta_k$, the EoM for $\zeta_k$ in \eqref{eq:zeta-chi-eoms-k} may be rewritten as
\begin{equation}
	v_k''+\left(c_\mathrm{s}^2k^2-\frac{z''}{z}\right)v_k=0\,.\label{eq:MSeq}
\end{equation}
It is straightforward to see from \eqref{eq:z-cs2-zs} that when $\epsilon=\epsphi$, we have $z^2=2a^2\epsphi$, and here $\epsphi=\bar\epsilon$ is a constant, so
\begin{equation}
	\frac{z''}{z}=\frac{a''}{a}=\frac{\nu^2-1/4}{\tau^2}\,,\qquad\nu=\frac{\bar\epsilon-3}{2(\bar\epsilon-1)}\,,
\end{equation}
and furthermore $c_\mathrm{s}^2=1$.
The general solution is thus
\begin{equation}
	v_k(\tau)=\sqrtb{-\tau}\left(C_1\mathrm{H}_\nu^{(1)}(-k\tau)+C_2\mathrm{H}_\nu^{(2)}(-k\tau)\right)\,,\label{eq:genvksolek}
\end{equation}
where $C_{1,2}$ are integration constants and $\mathrm{H}_\nu^{(1,2)}$ are the Hankel functions of the first and second kind, respectively.
Imposing an adiabatic vacuum in the early time subhorizon\footnote{In this work, the subhorizon limit corresponds to the regime where the Mukhanov-Sasaki variable oscillates in time. Oppositely, the superhorizon limit corresponds to the regime where the Mukhanov-Sasaki variable stops oscillating. It could still be time dependent, as there may be a growing mode, but in the present work superhorizon perturbations are found to freeze, meaning that either the constant mode dominates over a decaying mode or that the growing mode is marginal (e.g., at most logarithmic). Deep in the contraction, $1/k\sim -\tau$ gives a good approximation of the corresponding `horizon scale', which is also proportional to the comoving Hubble radius $1/(a|H|)$ --- in fact, it is very well approximated by the Hubble radius when $\epsilon\sim 3$. More precisely, by inspection of \eqref{eq:MSeq} the horizon is controlled by $\sqrtb{|z''|/z}/c_\mathrm{s}$, and through the bounce phase, this may significantly differ from the Hubble radius.\label{foot:horizonDef}} limit,
\begin{equation}
	v_k(\tau)\stackrel{-k\tau\gg 1}{\simeq}\frac{1}{\sqrtb{2k}}e^{-ik\tau}\,,\label{eq:vBD}
\end{equation}
determines the integration constants, and up to an irrelevant phase, the solution is
\begin{equation}
	v_k(\tau)=-\frac{\sqrt{-\pi\tau}}{2}\mathrm{H}_\nu^{(1)}(-k\tau)\,.\label{eq:vkfullek}
\end{equation}
In the late-time superhorizon limit, the solution is well approximated by
\begin{equation}
	v_k(\tau)\stackrel{-k\tau\ll 1}{\simeq}i\frac{\Gamma(\nu)}{2}\sqrtb{\frac{-\tau}{\pi}}\left(\frac{2}{-k\tau}\right)^{\nu}\quad\Rightarrow\quad\zeta_k(\tau)\simeq -i\frac{\Gamma(\nu)}{\sqrtb{\pi\bar\epsilon}\mpl\tau_{\mathrm{b}-}k^{3/2}}\left(\frac{-k\tau_{\mathrm{b}-}}{2}\right)^{\frac{\bar\epsilon}{\bar\epsilon-1}}\,,\label{eq:vk-zetak-ek-superHubble}
\end{equation}
where $\Gamma(\nu)$ is the Gamma function, and where we ignored the decaying term as $k\tau\to 0^-$.
For subsequent use, it is also useful to keep the next-to-leading-order term to compute the time derivative,
\begin{equation}
	\zeta_k'(\tau)\stackrel{-k\tau\ll 1}{\simeq}\left(\frac{\pi}{\Gamma\Big(\frac{1}{2}-\frac{1}{\bar\epsilon-1}\Big)}-i\Gamma\Big(\frac{1}{2}+\frac{1}{\bar\epsilon-1}\Big)\sin\Big(\frac{\pi\bar\epsilon}{\bar\epsilon-1}\Big)\right)\frac{(-\tau_{\mathrm{b}-})^{\frac{1}{\bar\epsilon-1}}}{\sqrtb{2\pi\bar\epsilon}\mpl}\left(\frac{k}{2}\right)^{\frac{1}{2}-\frac{1}{\bar\epsilon-1}}(-\tau)^{-\frac{2}{\bar\epsilon-1}}\,.\label{eq:zetakpek}
\end{equation}
To zeroth order in $1/\bar\epsilon\ll 1$, this tells us that $\zeta_k$ on superhorizon scales will enter the bounce phase with an amplitude and a time derivative of the order of
\begin{equation}
    \zeta_k(\tau_{\mathrm{b}-})\simeq\frac{i}{2\mpl\sqrtb{\bar\epsilon k}}\,,\qquad\zeta_k'(\tau_{\mathrm{b}-})\simeq -ik\zeta_k(\tau_{\mathrm{b}-})\simeq\frac{1}{2\mpl}\sqrtb{\frac{k}{\bar\epsilon}}\,,\label{eq:zetakpek0}
\end{equation}
respectively. In fact, $|\zeta_k'|^2\simeq k^2|\zeta_k|^2$ holds at all times in the limit $1/\bar\epsilon\to 0^+$ (the Minkowski limit), so the kinetic and gradient terms in the second-order on-shell action for $\zeta$ contribute equally. In other words, $z''/z\to 0^-$ when $\bar\epsilon\to\infty$, so the solution for $v_k$ is approximately that of a simple harmonic oscillator. Then from \eqref{eq:vk-zetak-ek-superHubble}, the dimensionless power spectrum on superhorizon scales can be expressed as
\begin{equation}
	\mathcal{P}_\zeta(k)\equiv\frac{k^3}{2\pi^2}|\zeta_k|^2\simeq\frac{1}{2\pi^3\bar\epsilon}\Gamma\Big(\frac{\bar\epsilon-3}{2(\bar\epsilon-1)}\Big)^2\frac{1}{(-\mpl\tau_{\mathrm{b}-})^2}\left(\frac{-k\tau_{\mathrm{b}-}}{2}\right)^{\frac{2\bar\epsilon}{\bar\epsilon-1}}\,.\label{eq:Pzetaek}
\end{equation}
The scalar power spectrum is usually parametrized as $\mathcal{P}_s(k)=A_s(k/k_\star)^{n_s-1}$, with an amplitude $A_s$, pivot scale $k_\star$, and spectral index $n_s$; in particular, $n_s-1=\frac{\dd\ln\mathcal{P}_s}{\dd\ln k}$. We thus note that in the large-$\bar\epsilon$ limit the power spectrum \eqref{eq:Pzetaek} is blue with spectral index
\begin{equation}
	n_s-1=\frac{2\bar\epsilon}{\bar\epsilon-1}\simeq 2\label{eq:nsek},
\end{equation}
while $A_s\simeq 1/(8\pi^2\bar\epsilon)$ at $k_\star=\mpl$.
This is a very blue spectrum, which is clearly ruled out by observations \cite{Planck:2018vyg}. It also means the adiabatic perturbations have essentially remained in their vacuum (as naively expected from the Minkowski limit $1/\bar\epsilon\to 0$). Thus, they should be unenhanced and unobservable on cosmological scales, unless something significant occurs in the bounce phase, but we will shortly see that, to a good approximation, perturbations remain constant through the bounce.

To see why they are not observable, let us assume that \eqref{eq:Pzetaek} describes the post bounce power spectrum and estimate the real-space `amplitude' of the curvature perturbation, denoted $\bar\zeta$, on a relevant cosmological scale $\mathscr{R}$. One can find the variance of $\zeta$  coarse-grained over a region of comoving size $\mathscr{R}$ by smoothing\footnote{For more details on smoothing the power spectrum, see, e.g., \cite{Liddle:2000cg}.} the power spectrum with a Window function,
\begin{equation}
	\bar\zeta^2(\mathscr{R})\equiv\int_0^\infty\frac{\dd k}{k}\,\mathcal{\widetilde{W}}^2(k\mathscr{R})\mathcal{P}_\zeta(k)=\mathcal{P}_\zeta(1/\mathscr{R})\simeq\frac{1}{8\pi^2\bar\epsilon}\left(\frac{1}{\mpl \mathscr{R}}\right)^2\,,\label{eq:zetaReal}
\end{equation}
where the result of the integral follows by taking a delta function for the squared Fourier-space Window function,
\begin{equation}
    \mathcal{\widetilde{W}}^2(k\mathscr{R}) = \delta(k\mathscr{R}-1)\,,\label{eq:Windowdelta}
\end{equation}
and the last equality follows from \eqref{eq:Pzetaek} in the large-$\bar\epsilon$ limit. Considering the comoving cosmological scales $\mathscr{R}$ over which we observe correlations in $\zeta$ are many orders of magnitude larger than the Planck length $\mpl^{-1}$, we conclude that $|\bar\zeta(\mathscr{R})|$ is minute, as expected for `vacuum fluctuations'. For example, if we take $\mathscr{R}=\mathrm{Mpc}/0.05$, then we find $\mpl \mathscr{R}\sim 10^{58}$, hence $|\bar\zeta(\mathscr{R})|$ would be of the order of $10^{-59}/\sqrtb{\bar\epsilon}$, which is further suppressed by the largeness of $\bar\epsilon$.

Let us move on to the discussion of the entropy perturbation $\chi$. First, one has to specify the coupling function $F(\psi,X)$ between the ekpyrotic field $\psi$ and the entropy field $\chi$. As in \cite{Ijjas:2014fja,Levy:2015awa}, we choose
\begin{equation}
	F(\psi)=F_0e^{-\sqrtb{2\bar\epsilon(1+\delta)}\psi/\mpl}\,,\label{eq:Fofphi-ek}
\end{equation}
for some $F_0>0$ and some small parameter $0<\delta\ll 1$. We do not assume any $X$ dependence here. The EoM for $\chi_k$ in \eqref{eq:zeta-chi-eoms-k} can be expressed in terms of its Mukhanov-Sasaki variable $u_k\equiv\mpl z_\chi\chi_k$ as
\begin{equation}\label{EOMuk}
	u_k''+\left(k^2-\frac{z_\chi''}{z_\chi}\right)u_k=0\,,
\end{equation}
where we recall $z_\chi^2=a^2F$, so here we find
\begin{equation}
	\frac{z_\chi''}{z_\chi}=\frac{\nu_s^2-1/4}{\tau^2}\,,\qquad\nu_s\simeq\frac{3}{2}+\frac{\bar\epsilon\delta}{2(\bar\epsilon-1)}\,,\label{eq:zsppozs-ek}
\end{equation}
where the approximation is to leading order in small $\delta$. The general solution for $u_k(\tau)$ will be of the same form as \eqref{eq:genvksolek}, with $\nu$ replaced by $\nu_s$, and so imposing an adiabatic vacuum for $u_k$ as in \eqref{eq:vBD} yields the same expressions as \eqref{eq:vkfullek} and \eqref{eq:vk-zetak-ek-superHubble}, though again with $\nu$ replaced by $\nu_s$:
\begin{subequations}
\begin{align}
	&u_k(\tau)=-\frac{\sqrtb{-\pi\tau}}{2}\mathrm{H}_{\nu_s}^{(1)}(-k\tau)\stackrel{-k\tau\ll 1}{\simeq}i\frac{\Gamma(\nu_s)}{2}\sqrtb{\frac{-\tau}{\pi}}\left(\frac{2}{-k\tau}\right)^{\nu_s}\label{eq:uk-chik-full-super-ek1}\\
	\Rightarrow\quad &
    \begin{cases}
    \chi_k(\tau)\simeq i\Gamma(\nu_s)\left(\frac{(\bar\epsilon-1)^2}{\bar\epsilon-3}\right)^{\frac{2+\delta}{4}}\left(\frac{-\tau_{\mathrm{b}-}\sqrtb{V_0}}{\mpl}\right)^{\delta/2}\sqrtb{\frac{2V_0}{\pi F_0\mpl^{4}k^3}}\left(-\frac{k\tau_{\mathrm{b}-}}{2}\right)^{-\frac{\delta}{2}(1+\frac{1}{\bar\epsilon})}\\
    \chi_k'(\tau)\simeq i\sqrtb{\frac{\bar\epsilon V_0k}{2F_0}}\frac{\tau}{\mpl^2}
    \end{cases}\,.
    \label{eq:uk-chik-full-super-ek}
\end{align}
\end{subequations}
Note that small-$\delta$ and large-$\bar\epsilon$ approximations are made to obtain the last expressions, and the next-to-leading-order term in the expansion of the Hankel function is kept to find the time derivative.
Correspondingly, the dimensionless power spectrum is
\begin{align}
	\mathcal{P}_\chi(k)&\equiv\frac{k^3}{2\pi^2}|\chi_k|^2\nonumber\\
    &\simeq\frac{1}{\pi^3}\left(\frac{(\bar\epsilon-1)^2}{\bar\epsilon-3}\right)^{1+\frac{\delta}{2}}\Gamma\Big(\frac{3}{2}+\frac{\delta}{2}\Big(1+\frac{1}{\bar\epsilon}\Big)\Big)^2\left(\frac{-\tau_{\mathrm{b}-}\sqrtb{V_0}}{\mpl}\right)^{\delta}\frac{V_0}{F_0\mpl^{4}}\left(-\frac{k\tau_{\mathrm{b}-}}{2}\right)^{-\delta(1+\frac{1}{\bar\epsilon})}\,. \label{eq:Pchikek}
\end{align}
We can read the spectral index in this case to be
\begin{equation}
	n_s-1\simeq-\delta\label{eq:ns-chi-ek}
\end{equation}
in the limit of large $\bar\epsilon$. By taking the small-$\delta$ limit, we can also estimate the amplitude as
\begin{equation}
    A_s\simeq\frac{\bar\epsilon V_0}{4\pi^2F_0\mpl^{4}}\,.\label{eq:Asek}
\end{equation}
Since $\delta$ is a small number, the spectral index \eqref{eq:ns-chi-ek} can be in accordance with the observations, which indicate $n_s\sim 0.97$ (e.g., \cite{Planck:2018vyg}), provided these isocurvature perturbations convert to curvature perturbations nearly one-to-one (this will be rediscussed later). Similarly, the amplitude can be compatible with the current observational estimates of $A_s\sim 10^{-9}$ (e.g., \cite{Planck:2018vyg}), given some parameter values for $\bar\epsilon$, $V_0$, and $F_0$ (e.g., $\bar\epsilon\sim 100$, $V_0\sim 10^{-9}\,\mpl^4$, $F_0\sim 1$). This also tells us that the amplitude of the real-space perturbation $\chi$ is of the order of $10^{-5}$. Finally, let us note from \eqref{eq:uk-chik-full-super-ek} and \eqref{eq:Asek} that
\begin{equation}
    \chi_k(\tau_{\mathrm{b}-})\simeq i\pi\sqrtb{\frac{2A_s}{k^3}}\,,\qquad\chi_k'(\tau_{\mathrm{b}-})\simeq i\pi\sqrtb{2A_sk}\tau_{\mathrm{b}-}\,,\label{eq:chichiptaubmek}
\end{equation}
by the time superhorizon perturbations enter the bounce phase.

\paragraph{Massless case}
Let us move on to the discussion of the second model, i.e., the one where the scalar field $\psi$ is massless. As we said, this can either be viewed as the kination phase that often follows ekpyrotic contraction or it can be viewed as a toy model of its own. In such a case, the background scaling solution early in the contracting phase generally reads
\begin{equation}
	a(\tau)=\left(\frac{\tau}{\tau_{\mathrm{b}-}}\right)^{1/2}\,,\qquad\psi(\tau)=\psi_0\pm\sqrtb{\frac{3}{2}}\mpl\ln\left(\frac{\tau}{\tau_0}\right)\,,\qquad\tau<\tau_{\mathrm{b}-}<0\,,\label{eq:scalingBackMassless}
\end{equation}
where $\psi_0=\psi(\tau_0)$ is an integration constant at a time $\tau_0<\tau_{\mathrm{b}-}<0$, and $\psi'(\tau_0)=\pm\sqrtb{3/2}\mpl/\tau_0$ is taken in accordance with the Friedmann constraint equation, which reads $3\mpl^2(a'/a)^2\simeq\psi'^2/2$ when $\psi$ dominates [which implies $\tau_0\ll\tau_{\mathrm{b}-}$ with our choice of calibration for $a(\tau)$]. If such a phase is to follow a phase of ekpyrosis [cf.~\eqref{eq:scalingBackEk}], then we pick the plus sign in the expression for $\psi'(\tau_0)$ [so that $\psi'(\tau_0)\propto 1/\tau_0<0$] and $\psi_0$ is fixed at the transition\footnote{The goal is not to model such a transition, so we omit details. A refined treatment would consider a more elaborate potential function that asymptotes zero --- recall footnote \ref{foot:twoExp}.} point; otherwise, if there is no ekpyrosis, the sign of $\psi'(\tau_0)$ and the value of $\psi_0$ are completely arbitrary. From \eqref{eq:scalingBackMassless}, we note
\begin{equation}
	\mathcal{H}=\frac{1}{2\tau}\,,\qquad\epsilon=3\,,\label{eq:scalingBackMassless_H}
\end{equation}
so the energy density of the scalar field scales as $\rho_\psi\propto a^{-6}$, and the EoS in this case is referred to as stiff. This is marginally stable against the growth of anisotropies, which also scale as $a^{-6}$, and if anisotropies are diluted in an earlier phase of ekpyrotic contraction, then they should not be an issue closer to the bounce either.

The calculation of the perturbations is similar to the previous one, where now $z(\tau)=\sqrtb{6}a(\tau)$, so $z''/z=a''/a=(\nu^2-1/4)/\tau^2$ with $\nu=0$,
so the solution for $v_k=\mpl z\zeta_k$ is
\begin{align}
	&v_k(\tau)=-\frac{\sqrtb{-\pi\tau}}{2}\mathrm{H}_0^{(1)}(-k\tau)\stackrel{-k\tau\ll 1}{\simeq}i\sqrtb{\frac{-\tau}{\pi}}\left(\ln(2)-\gammaE +\frac{i\pi}{2}-\ln(-k\tau)\right)\nonumber\\
	\Rightarrow\quad & \zeta_k(\tau)\simeq \frac{i}{\mpl}\sqrtb{\frac{-\tau_{\mathrm{b}-}}{6\pi}}\left(\ln(2)-\gammaE +\frac{i\pi}{2}-\ln(-k\tau)\right)\quad\Rightarrow\quad\zeta_k'(\tau)\simeq-\frac{i}{\mpl}\sqrtb{\frac{-\tau_{\mathrm{b}-}}{6\pi}}\frac{1}{\tau}\,,
 \label{eq:zeta_super}
\end{align}
where $\gammaE $ is the Euler-Mascheroni constant. Note that as $\tau\to 0^-$, the log term represents a (slowly) growing mode, as is evident from the time derivative. The power spectrum is
\begin{align}
	\mathcal{P}_\zeta(k,\tau)&\simeq\frac{-\tau_{\mathrm{b}-}k^3}{12\pi^3\mpl^2}\left|\ln(2)-\gammaE +\frac{i\pi}{2}-\ln(-k\tau)\right|^2\nonumber\\
    \Rightarrow\ \mathcal{P}_\zeta(k_\mathrm{p},t)&\simeq\frac{k_\mathrm{p}(t)^3}{24\pi^3\mpl^2|H(t)|}\left|\ln(2)-\gammaE +\frac{i\pi}{2}-\ln\left(\frac{k_\mathrm{p}(t)}{2|H(t)|}\right)\right|^2\,,
\end{align}
where $k_\mathrm{p}(t)\equiv k/a(t)$ represents the physical wavenumber on the second line.
Since $k$ is a superhorizon mode above, the deep-IR limit implies that the log term should dominate over the constant coefficients inside the absolute values, so we may approximately write the power spectrum right at the end of the standard contracting phase as
\begin{equation}
	\mathcal{P}_\zeta(k_\mathrm{p},t_{\mathrm{b}-})\simeq\frac{k_\mathrm{p}(t_{\mathrm{b}-})^3}{24\pi^3\mpl^2|H_{\mathrm{b}-}|}\left(\ln\left[\frac{k_\mathrm{p}(t_{\mathrm{b}-})}{2|H_{\mathrm{b}-}|}\right]\right)^2\,,\label{eq:Pzetamassless}
\end{equation}
where $|H_{\mathrm{b}-}|$ represents the maximal Hubble scale that is cosmologically reached (in absolute value). Note that the spectral index of the above power spectrum runs, i.e., it is not simply a constant as before, but the log term is only a slight correction to a deeply blue $n_s-1=3$ spectrum. Indeed, one finds
\begin{equation}
    n_s-1=3+\frac{2}{\ln(\frac{k}{2a|H|})}\,,\qquad\alpha_s\equiv\frac{\dd n_s}{\dd\ln k}=-\frac{2}{\left(\ln\left[\frac{k}{2a|H|}\right]\right)^2}\,.\label{eq:nsmassless}
\end{equation}
The fact that the spectrum is deeply blue implies that it is not cosmologically observable, as was the case for the $\zeta$ spectrum in ekpyrosis. Indeed, even if one takes a very small bouncing energy scale, say $|H_{\mathrm{b}-}|\sim 10^{-15}\,\mpl\sim\mathrm{TeV}$, then for $k\sim 10^{-58}\,\mpl\sim 0.04\,\mathrm{Mpc}^{-1}$, \eqref{eq:Pzetamassless} yields $\mathcal{P}_\zeta\sim 10^{-158}$, hence in real space we expect $\bar\zeta\sim 10^{-79}$.

Turning to the entropy perturbation, we once again have to specify what is the coupling $F(\psi,X)$ between the background (massless) scalar $\psi$ and the perturbation $\chi$. If we think of the regime where $\psi$ is massless as following ekpyrosis, then it is sensible to take the same coupling function as \eqref{eq:Fofphi-ek}, except\footnote{Again, a proper model that transitions from ekpyrosis to kination would have a smooth potential that asymptotes $0$ and correspondingly $\epsilon$ would smoothly transition from a large constant to $3$. The point here is that, in principle, even if we treat $V(\psi)$ as being exactly vanishing, we can use the same couplings and solutions as in ekpyrosis and just replace $\bar\epsilon$ by $3$.} with $\bar\epsilon=3$:
\begin{equation}
	F(\psi)=F_0e^{-\sqrtb{6(1+\delta)}\psi/\mpl}\,.\label{eq:Fofphi-massless}
\end{equation}
From $\psi(\tau)=\psi_0+\sqrtb{3/2}\mpl\ln(\tau/\tau_0)$, this implies $F\propto(-\tau)^{-3(1+\delta/2)}$ in the small-$\delta$ limit. Note, though, that we can just as well choose something like
\begin{equation}
\label{eq:Fofphi-massless2}
	F(X)=\left(\frac{X}{\Lambda^2\mpl^2}\right)^{1+\delta/2}
\end{equation}
for some new energy scale $\Lambda$, which also yields $F\propto(-\tau)^{-3(1+\delta/2)}$ and thus achieves the same power spectrum for $\chi_k$. The latter was the approach of \cite{Kim:2020iwq} --- they essentially had a term of the form $(\partial_\mu\psi\partial^\mu\psi)(\partial_\nu\chi\partial^\nu\chi)$ in their Lagrangian leading to $F\sim (H/\Lambda)^2$ in the prebounce phase --- and it is better motivated if there is no ekpyrosis, i.e., if $\psi$ is fundamentally massless, since the coupling function \eqref{eq:Fofphi-massless2} is shift symmetric. It is, however, a noncanonical higher-dimensional operator if one wishes to obtain the desired red tilt in the scalar power spectrum (i.e., when $\delta\gneqq 0$). A term of the form $e^{\lambda\psi}(\partial_\mu\chi)^2$ in the Lagrangian, for some real constant $\lambda$, might find motivation in string theory (where $\psi$ and $\chi$ would be akin to the dilaton and the axion\footnote{Note that this is essentially the setup of pre-big bang cosmology \cite{Gasperini:1992em}, a string cosmology scenario where the dilaton acts as a massless scalar driving an Einstein-frame contracting phase with stiff EoS and where the axion is responsible for generating a nearly scale-invariant scalar power spectrum of isocurvature perturbations \cite{Gasperini:2002bn,Gasperini:2007zz}.}, respectively), though it is beyond the scope of this paper to derive \eqref{eq:Fofphi-massless} from a more fundamental theory (see, e.g., \cite{Lehners:2018vgi,Bernardo:2021wnv,Shiu:2023yzt} for comments about this) --- we have in mind either \eqref{eq:Fofphi-massless} or \eqref{eq:Fofphi-massless2} from a phenomenological point of view.

The EoM for $u_k=\mpl z_\chi\chi_k$ is the same as before [recall \eqref{EOMuk}], except this time $z_\chi''/z_\chi=(\nu_s^2-1/4)/\tau^2$ with
\begin{equation}
	\nu_s\simeq\frac{3}{2}+\frac{3\delta}{4}\label{eq:nus-massless}
\end{equation}
in the small-$\delta$ limit, though this is really the same as \eqref{eq:zsppozs-ek} with $\bar\epsilon=3$. The solution for $u_k(\tau)$ is then as in \eqref{eq:uk-chik-full-super-ek1}, with the expression for $\nu_s$ replaced by \eqref{eq:nus-massless}, so in the end, we have [taking the coupling \eqref{eq:Fofphi-massless2} for concreteness, but similar expressions can be derived from \eqref{eq:Fofphi-massless} as the coupling]
\begin{align}
\label{eq:chi_anysol}
	&\chi_k(\tau)\stackrel{-k\tau\ll 1}{\simeq}i\frac{2^{\frac{3+\delta}{2}}}{3^{\frac{2+\delta}{4}}\sqrtb{\pi}}\Gamma\Big[\frac{3}{2}\Big(1+\frac{\delta}{2}\Big)\Big]\frac{\Lambda}{\mpl}(-\tau_{\mathrm{b}-}\Lambda)^{\delta/2}\left(\frac{2}{-k\tau_{\mathrm{b}-}}\right)^{3\delta/4}k^{-3/2}\stackrel{\delta\ll 1}{\simeq}i\sqrtb{\frac{2}{3}}\frac{\Lambda}{k^{3/2}\mpl}\,,\nonumber\\
    &\chi_k'(\tau)\stackrel{-k\tau\ll 1}{\simeq}i\frac{2^{\frac{1+\delta}{2}}}{3^{\frac{2+\delta}{4}}\sqrtb{\pi}}\Gamma\Big[\frac{1}{2}\Big(1+\frac{3\delta}{2}\Big)\Big]\frac{\Lambda}{\mpl}(-\tau_{\mathrm{b}-}\Lambda)^{\delta/2}\left(\frac{2}{-k\tau_{\mathrm{b}-}}\right)^{3\delta/4}\sqrtb{k}\tau\stackrel{\delta\ll 1}{\simeq}i\sqrtb{\frac{2}{3}}\frac{\Lambda\sqrtb{k}\tau}{\mpl}\,.
\end{align}
Then, the power spectrum is
\begin{equation}
    \mathcal{P}_\chi(k)\simeq\frac{2^{2+\delta}\Gamma\left[\frac{3}{2}\left(1+\frac{\delta}{2}\right)\right]^2}{3^{1+\delta/2}\pi^3}\left(\frac{\Lambda}{\mpl}\right)^2(-\tau_{\mathrm{b}-}\Lambda)^\delta\left(\frac{2}{-k\tau_{\mathrm{b}-}}\right)^{3\delta/2}\,,\label{eq:Pchimassless2}
\end{equation}
and the spectral index reads
\begin{equation}
	n_s-1\simeq-\frac{3}{2}\delta\,.\label{eq:ns-chi-massless}
\end{equation}
This is not exactly the same as \eqref{eq:ns-chi-ek} --- it simply means that if the same model is used through subsequent phases of ekpyrosis and kination, the spectral index will simply change by a factor of $3/2$ from one phase to the other. If the model consists purely of a massless field, then certainly \eqref{eq:ns-chi-massless} can be in agreement with $n_s\sim 0.97$ (e.g., $\delta\sim 0.02$). The amplitude of \eqref{eq:Pchimassless2} in the small-$\delta$ limit can be estimated as
\begin{equation}
    A_s\simeq\frac{\Lambda^2}{3\pi^2\mpl^2}\,,\label{eq:AsLambda}
\end{equation}
hence the amplitude is controlled by $\Lambda/\mpl$ (e.g., $\Lambda\approx 2.67\times 10^{-4}\,\mpl$ achieves $A_s\approx 2.4\times 10^{-9}$). From this and \eqref{eq:chi_anysol}, we obtain
\begin{equation}
    \chi_k(\tau_{\mathrm{b}-})\simeq i\pi\sqrtb{\frac{2A_s}{k^3}}\,,\qquad\chi_k'(\tau_{\mathrm{b}-})\simeq i\pi\sqrtb{2A_sk}\tau_{\mathrm{b}-}\,,\label{eq:chichipmasslesstbm}
\end{equation}
by the time the superhorizon perturbations enter the bounce phase. Unsurprisingly, these expressions are the same as \eqref{eq:chichiptaubmek}.

\subsubsection{Bounce phase, conversion, and reheating}\label{eq:bounceanalytical}

So far, we have reviewed two models of contracting cosmology in which the matter field $\psi$ dominates: when $\psi$ has an ekpyrotic potential and when $\psi$ is massless. In both cases, adiabatic perturbations ($\zeta$) have a blue spectrum [$n_s\sim 3$ \eqref{eq:nsek} and $n_s\sim 4$ \eqref{eq:nsmassless}, respectively], with $\zeta_k$ being approximately constant on superhorizon scales [recall \eqref{eq:vk-zetak-ek-superHubble} and \eqref{eq:zeta_super}, respectively; there is a slight logarithmic growth in the latter case]. In both cases, though, an appropriate kinetic coupling to a spectator field $\chi$ generates a (nearly) scale-invariant power spectrum of entropy perturbations, where $\chi_k$ is approximately constant on superhorizon scales [recall \eqref{eq:uk-chik-full-super-ek} and \eqref{eq:chi_anysol}].

Three things still need to happen at this point: the background cosmology must bounce, i.e., there must be a transition from contraction to expansion; the isocurvature perturbations must be converted into curvature perturbations, i.e., $\chi$ must source $\zeta$, so that before the radiation era, it is the $\zeta$ power spectrum that becomes nearly scale invariant; and finally the primordial matter field $\psi$ must also decay into standard model fields, i.e., the universe must reheat, so that radiation-dominated expansion (standard hot big bang cosmology) may begin. Our focus will be on modeling and studying the bounce carefully; the conversion of isocurvature perturbations into curvature perturbations and reheating have been the subject of other studies (e.g., \cite{Takamizu:2004rq,Lehners:2007wc,Battefeld:2007st,Lehners:2008my,Lehners:2009qu,Lehners:2009ja,Lehners:2010fy,Quintin:2014oea,Ijjas:2014fja,Fertig:2015ola,Fertig:2016czu,Hipolito-Ricaldi:2016kqq,Ijjas:2020cyh,Ijjas:2021ewd}), whose results we will simply quote when needed. However, it is still worth noting that some of the details of the conversion process have important implications for perturbations through the bounce. First, we need to distinguish two possibilities (e.g., as studied in \cite{Fertig:2016czu}): if the conversion occurs before the bounce or after. If the latter prevails, then we need to understand how both $\zeta$ (which has a blue spectrum) and $\chi$ (which has a scale-invariant spectrum) evolve through the bounce, matching them to their prebounce properties. In the former possibility, i.e., if $\chi$ gets converted into $\zeta$ perturbations before the bounce, then we simply need to study $\zeta$ (which has acquired the prebounce properties of the $\chi$ field, so it has a scale-invariant power spectrum) through the bounce.\footnote{We shall always assume that the conversion process is fully efficient, in other words that negligible isocurvature perturbations are left after sourcing the curvature perturbations. This is a good approximation according to \cite{Lehners:2007wc,Lehners:2008my,Lehners:2009qu,Lehners:2009ja,Lehners:2010fy,Ijjas:2014fja,Fertig:2015ola,Fertig:2016czu,Ijjas:2020cyh,Ijjas:2021ewd}.}

With that in mind, we now focus on the nonsingular transition from contraction to expansion, i.e., the bounce phase, which is driven by the Cuscuton field. We shall present specific examples of the resulting cosmological dynamics using numerical computations in a subsequent subsection, but before that, it is also useful to provide some analytical approximations and some intuitive picture regarding the general dynamics of the background and of the perturbations through the bounce phase.

First, the Cuscuton bounce phase is a regime in which the geometric null convergence condition\footnote{The geometric null convergence condition states that the contraction of the Ricci tensor with any null vector twice should be nonnegative.}
(which  in FLRW says that $\dot H\leq 0$) is violated, so $\dot H>0$ for some time interval. This ensures the bounce is nonsingular. In other words, $H(t)$ never blows up; instead, it crosses $0$, i.e., we have a smooth transition from contraction ($H<0$) to expansion ($H>0$). The bounce phase begins when $\dot H$ crosses zero for the first time, which we denote by the subscript `{\small $\mathrm{b}-$}', and it ends the second time $\dot H$ crosses zero, which we denote by the subscript `{\small $\mathrm{b}+$}'. As this phase is usually short, the scale factor is roughly constant to leading order ($a\approx 1$ in our normalization), hence conformal time and physical time are approximately equal in the bounce phase. The following is often a good approximate parameterization of a bounce phase,
\begin{equation}
    a(\tau)\stackrel{\tau\approx 0}{=}1+\mathcal{O}(\tau^2)\,,\qquad\mathcal{H}(\tau)\stackrel{\tau\approx 0}{=}\Upsilon\tau+\mathcal{O}(\tau^3)\,,\qquad\mathcal{H}'(\tau)\stackrel{\tau\approx 0}{=}\Upsilon+\mathcal{O}(\tau^2)\,,\label{eq:aHHpb}
\end{equation}
for some dimension-2 constant $\Upsilon>0$, where we choose the origin of the time axis, $\tau=0$, to correspond to the bounce point.\footnote{Naturally, since the series expansion is performed about the bounce where $\tau=0$, the approximation becomes less accurate the farther away from the bounce point at $\tau=0$. The same is true if the bounce duration (scale) is too long (high) --- it holds best as long as $\Upsilon\tau^2\lesssim\mathcal{O}(1)$. Nevertheless, it shall capture the leading-order behavior in what follows and hence remain valid as an order-of-magnitude approximation.} In fact, \eqref{eq:aHHpb} has been derived in \cite{Quintin:2019orx} as a series solution for a generic Cuscuton bounce. Thus, \eqref{eq:aHHpb} applies to the interval $\tau_{\mathrm{b}-}<\tau<\tau_{\mathrm{b}+}$. When a variable is evaluated at $\tau=0$, we denote it by a subscript `b'. Note that the parameter $\Upsilon$ and the bounce duration $\tau_{\mathrm{b}+}-\tau_{\mathrm{b}-}$ control the energy scale of the bounce phase.

With such a parameterization, we can find approximate solutions to the mode functions \eqref{eq:zeta-chi-eoms-k}. Focusing on deep IR modes ($k\to 0$), which is the relevant regime of observational CMB modes as they cross a nonsingular bounce, and using \eqref{eq:z-cs2-zs}, the equations (which include the contributions from the Cuscuton) can be expressed as
\begin{equation}
    \zeta_k''+\frac{(z^2)'}{z^2}\zeta_k'\stackrel{k\to 0}{\simeq} 0\,,\qquad z^2\stackrel{k\to 0}{\simeq}\frac{2\epsphi a^2}{1+\sigma/3}\,,\label{eq:zetaIR}
\end{equation}
and
\begin{equation}
    \chi_k''+2\frac{(z_\chi^2)'}{z_\chi^2}\chi_k'\stackrel{k\to 0}{\simeq}0\,.\label{eq:chiIR}
\end{equation}
To be more quantitative, the regime of validity of the IR approximation for the $\zeta_k$ and $\chi_k$ EoMs above are, respectively,
\begin{equation}
    kc_\mathrm{s}\ll\sqrtb{\frac{\left|z''\right|}{z}}\,,\qquad k\ll\sqrtb{\frac{\left|z_\chi''\right|}{z_\chi}}\,,\label{eq:kIRregimes}
\end{equation}
where we recall from \eqref{eq:z-cs2-zs} that $c_\mathrm{s}$ and $z$ are complicated background- and $k$-dependent quantities, while $z_\chi=a\sqrtb{F}$ is simply a power-law in $\tau$. Therefore, the first expression of \eqref{eq:kIRregimes} is a somewhat involved time-dependent nonlinear inequality in $k$. This may be simplified, in principle, but for the sake of simplicity here, we assume that the approximation is respected. For a range of IR modes of cosmological relevance, the approximation has been confirmed numerically.
The time-dependent solutions to \eqref{eq:zetaIR} and \eqref{eq:chiIR} are
\begin{equation}
    \zeta_k'(\tau)\stackrel{k\to 0}{\simeq}\zeta_k'(\tau_{\mathrm{b}-})\frac{a(\tau_{\mathrm{b}-})^2}{a(\tau)^2}\frac{1-\frac{\epsilon(\tau)}{\epsphi(\tau)}+\frac{3}{\epsphi(\tau)}}{1-\frac{\epsilon(\tau_{\mathrm{b}-})}{\epsphi(\tau_{\mathrm{b}-})}+\frac{3}{\epsphi(\tau_{\mathrm{b}-})}}\,,\qquad\chi_k'(\tau)\stackrel{k\to 0}{\simeq}\chi'_k(\tau_{\mathrm{b}-})\frac{a(\tau_{\mathrm{b}-})^2F(\tau_{\mathrm{b}-})}{a(\tau)^2F(\tau)}\,,\label{eq:zetaprimeandchiprimebounceIR}
\end{equation}
where the `initial' time derivatives $\zeta_k'(\tau_{\mathrm{b}-})$ and $\chi'_k(\tau_{\mathrm{b}-})$ can be estimated by matching the solutions to the end of the prebounce contracting phase. The equations also admit the usual superhorizon constant solutions, $\zeta_k'\simeq 0$ and $\chi_k'\simeq 0$, but by knowing the approximate value of the perturbations and their first derivatives at the onset of the bounce, the above captures the correct time evolution of the perturbations through the bounce.

Assuming for now that the conversion of isocurvature perturbations into curvature perturbations occurs only after the bounce, let us solve the $\chi_k$ equation, which is controlled by the time dependence of $F$, itself a function of $\psi$.
Let us assume that the matter field $\psi$ is massless in the bounce phase in any situation; recall the earlier discussion according to which even if $\psi$ is an ekpyrotic field, its potential should go to zero by the time of the bounce.
Then, its background EoM is $\psi''=-2\mathcal{H}\psi'$, hence from \eqref{eq:aHHpb} one finds
\begin{equation}
    \psi(\tau)\stackrel{\tau\approx 0}{=}\psi_\mathrm{b}+\psi'_{\mathrm{b}}\tau+\mathcal{O}(\tau^3)\,.\label{eq:phib}
\end{equation}
Recalling the Friedmann constraint \eqref{eq:Friedconstraint} in conformal time and substituting $V(\psi)=0$ for the duration of the bounce, we have $3\mpl^2\mathcal{H}^2=\psi'^2/2+a^2U(\bar\varphi)$, so at the bounce point
\begin{equation}
    \psi_{\mathrm{b}}^{\prime 2}=-2U_\mathrm{b}\,,\label{eq:phipbasUb}
\end{equation}
where the Cuscuton potential at the bounce point must be negative (a well-known fact; see \cite{Boruah:2018pvq,Quintin:2019orx}), i.e., $U_\mathrm{b}\equiv U(\bar\varphi=0)<0$. Considering the case where the coupling function is given by \eqref{eq:Fofphi-massless2} when $\psi$ is fundamentally massless, and in the $\delta\ll 1$ limit, we find $a(\tau)^2F(\tau)\approx\psi'(\tau)^2/(2\Lambda^2\mpl^2)$, which near $\tau = 0$ can be estimated as $a(\tau)^2F(\tau)\approx\mathrm{constant}+\mathcal{O}(\tau^2)$. Hence, substituting this estimate into \eqref{eq:zetaprimeandchiprimebounceIR} for the near-bounce regime implies $\chi_k'(\tau)\approx\chi_k'(\tau_{\mathrm{b}-})$ and
\begin{equation}
\label{eq:chi_solbounce}
    \chi_k(\tau)\approx\chi_k(\tau_{\mathrm{b}-})+\chi'_k(\tau_{\mathrm{b}-})(\tau-\tau_{\mathrm{b}-})\,.
\end{equation}
In the case where $\psi$ is an ekpyrotic field, we may be tempted to use \eqref{eq:Fofphi-ek} or \eqref{eq:Fofphi-massless} for $F$, but as argued in, e.g., \cite{Fertig:2015ola,Fertig:2016czu}, these functions should really be modified to reach unity by the time of the bounce, i.e., one assumes that the Lagrangian for $\chi$ approaches that of a massless canonical scalar (which has decoupled from $\psi$). In that case, we are left with $\chi_k''+2\mathcal{H}\chi_k'\simeq 0$ for the IR modes, and hence we recover \eqref{eq:chi_solbounce} again, provided $a\approx\mathrm{constant}$ through the bounce.

If we use the expected value of $\chi_k$ and $\chi_k'$ at the onset of the bounce phase [recall \eqref{eq:chichipmasslesstbm}] and assume that the background dynamics of the bounce phase is symmetric, so that $\tau_{\mathrm{b}+}=-\tau_{\mathrm{b}-}$, then we may infer from \eqref{eq:chi_solbounce} the value of the perturbation after the bounce phase as
\begin{equation}
    \chi_k(\tau_{\mathrm{b}+})\simeq\chi_k(\tau_{\mathrm{b}-})\left(1-2k^2\tau_{\mathrm{b}+}^2\right)\,.\label{eq:chiktaubp}
\end{equation}
For an IR mode, $k^2\tau_{\mathrm{b}+}^2\ll 1$ is an extremely small correction. Therefore, it is well justified to treat $\chi_k$ as constant through the bounce phase.

Moving on to the description of $\zeta_k$ through the bounce, we see from \eqref{eq:zetaprimeandchiprimebounceIR} that we need to know the behavior of $\epsilon$ and $\epsilon_\psi$. First, recall that prebounce, when the Cuscuton is negligible, we have $\epsilon\approx\epsilon_\psi\approx 3$. It is exactly 3 for a massless scalar, and we expect the ekpyrotic potential to go to zero by the time of the bounce, so we assume that this approximation holds by the time the bounce starts at $\tau_{\mathrm{b}-}$. We are thus left with evaluating (treating the scale factor as approximately constant again)
\begin{equation}
    \zeta_k'(\tau)\approx\zeta_k'(\tau_{\mathrm{b}-})\left(1-\frac{\epsilon(\tau)}{\epsphi(\tau)}+\frac{3}{\epsphi(\tau)}\right)\,.\label{eq:zetakprimeapproxbounce0}
\end{equation}
Gathering \eqref{eq:defepsepsphi}, \eqref{eq:aHHpb}, and \eqref{eq:phib}, we note
\begin{equation}
    \frac{1}{\epsphi}=\frac{2\mpl^2\mathcal{H}^2}{\psi'^2}\stackrel{\tau\approx 0}{=}\frac{2\mpl^2\Upsilon^2\tau^2}{\psi_\mathrm{b}^{\prime 2}}+\mathcal{O}(\tau^4)\,,\label{eq:1oepspsib}
\end{equation}
while
\begin{equation}
    \frac{\epsilon}{\epsphi}=-\frac{2\mpl^2\dot H}{\dot\psi^2}\stackrel{\tau\approx 0}{=}-\frac{2\mpl^2\Upsilon}{\psi_\mathrm{b}^{\prime 2}}+\mathcal{O}(\tau^2)\,,\label{eq:epsoepsphib}
\end{equation}
so to leading order in small $\tau$ we only have
\begin{equation}
    \zeta_k'(\tau)\approx\zeta_k'(\tau_{\mathrm{b}-})\left(1-\frac{\epsilon(\tau)}{\epsphi(\tau)}\right)\,.\label{eq:zetakprimeapproxbounce}
\end{equation}
At this point, we note that by combining equations \eqref{eq:FriedEvoEoM} and \eqref{eq:cuscConstrDeriv}, we may generally write
\begin{equation}
    \mpl^2\left(\frac{3\mu^4\mpl^2}{U_{,\varphi\varphi}(\bar\varphi)}-2\right)\dot H=\dot\psi^2\,,\label{eq:dotHphidotbackrel}
\end{equation}
so that near the bounce we have the relation
\begin{equation}
    \Upsilon=\frac{\psi_\mathrm{b}^{\prime 2}}{\mpl^2(3\mu^4\mpl^2/m_\mathrm{b}^2-2)}=-\frac{2U_\mathrm{b}}{\mpl^2(3\mu^4\mpl^2/m_\mathrm{b}^2-2)}\,,\label{eq:Upsilonpotentialparams}
\end{equation}
where we defined $m_\mathrm{b}^2\equiv U_{,\varphi\varphi}(\bar\varphi=0)$ and used \eqref{eq:phipbasUb}. This relation tells us the Hubble rate of change approximately throughout the bounce phase (or exactly at the bounce point) solely as a function of the potential parameters. It is also useful since --- combining \eqref{eq:epsoepsphib}, \eqref{eq:dotHphidotbackrel}, and \eqref{eq:Upsilonpotentialparams} --- we find
\begin{equation}
\label{eq:eps_app}
    \frac{\epsphi}{\epsilon}=1-\frac{3\mu^4\mpl^2}{2U_{,\varphi\varphi}(\bar\varphi)}\qquad\Rightarrow\qquad\frac{\epsilon}{\epsphi}\stackrel{\tau\approx 0}{\approx}\left(1-\frac{3\mu^4\mpl^2}{2m_\mathrm{b}^2}\right)^{-1}\,.
\end{equation}
Following \cite{Quintin:2019orx}, we may thus define
\begin{equation}
    \mplt^2\equiv\mpl^2\left(1-\frac{3\mu^4\mpl^2}{2m_\mathrm{b}^2}\right)\,,\label{eq:mplt}
\end{equation}
hence we are left with
\begin{equation}
\label{eq:zeta_b_approx}
    \zeta_k'(\tau)\approx\zeta_k'(\tau_{\mathrm{b}-})\left(1-\frac{\mpl^2}{\mplt^2}\right)\qquad\Rightarrow\qquad\zeta_k(\tau)\approx\zeta_k(\tau_{\mathrm{b}-})+\zeta_k'(\tau_{\mathrm{b}-})\left(1-\frac{\mpl^2}{\mplt^2}\right)(\tau-\tau_{\mathrm{b}-})\,.
\end{equation}
According to \cite{Quintin:2019orx}, if the sound speed is to remain close to unity at all times and on all scales\footnote{It is worth noting that in the IR limit, the term `sound speed' is a bit misleading since the Wentzel-Kramers-Brillouin (WKB) approximation breaks down in that regime; thus, this quantity no longer reflects the waves' propagation speed.}, then we should set $m_\mathrm{b}\approx\mu^2\mpl$, which would result in $\mplt^2\approx-\mpl^2/2$. As explained in \cite{Quintin:2019orx}, the negativity of the effective background gravitational coupling about the bounce point, $\mplt^2$, can be understood as the fact that the effective NEC is violated there.\footnote{One can generally think of the Cuscuton as a modified gravity theory, which yields a set of modified Friedmann equations in FLRW. However, those equations can be recast as the standard Friedmann equations of GR, except with a modified Planck mass \cite{Afshordi:2007yx}. This is what is meant by the effective background gravitational coupling here.} Finally, this would mean $1-\mpl^2/\mplt^2\approx 3$, and thus $\zeta_k(\tau)\approx\zeta_k(\tau_{\mathrm{b}-})+3\zeta_k'(\tau_{\mathrm{b}-})(\tau-\tau_{\mathrm{b}-})$.

We note that, because we are doing an expansion about $\tau=0$ and only keeping leading-order terms, the solution when extrapolated to $\tau_{\mathrm{b}\pm}$ may be quite approximate. In particular, we note that we lost continuity of $\zeta_k'(\tau)$ at $\tau_{\mathrm{b}-}$ in going from \eqref{eq:zetakprimeapproxbounce0} to \eqref{eq:zetakprimeapproxbounce}. While this may appear bad, it is actually very close to the reality. This is because the behavior of $\zeta_k'(\tau)$ is controlled by the ratios $3/\epsilon_\psi$ and $\epsilon/\epsilon_\psi$ according to \eqref{eq:zetakprimeapproxbounce0}, with $3/\epsilon_\psi\approx\epsilon/\epsilon_\psi\approx 1$ before the bounce and $3/\epsilon_\psi\approx0$ and $\epsilon/\epsilon_\psi\approx\mpl^2/\mplt^2\approx -2$ during the bounce, and it turns out that these ratios change abruptly at $\tau_{\mathrm{b}-}$ and $\tau_{\mathrm{b}+}$. In fact, in all fully numerical solutions we computed, we confirm that $\zeta_k'(\tau)$ abruptly jumps up and then back down at the beginning and end of the bounce phase, respectively, with a near constant value in between. In other words, $\zeta_k'(\tau)$ is well approximated by a step function in that interval. This agrees with our approximate solution \eqref{eq:zeta_b_approx}, which suggests a discontinuous jump in $\zeta_k'(\tau)$ at $\tau_{\mathrm{b}-}$. We will show numerical examples confirming this in the next subsection (cf.~Fig.~\ref{fig:zeta}).

If we consider adiabatic perturbations from an ekpyrotic phase entering the bounce, recalling \eqref{eq:zetakpek0}, then we expect from \eqref{eq:zeta_b_approx}
\begin{equation}
    \zeta_k(\tau_{\mathrm{b}+})\simeq\zeta_k(\tau_{\mathrm{b}-})\bigg(1+2i\Big(1-\frac{\mpl^2}{\mplt^2}\Big)k\tau_{\mathrm{b}+}\bigg)\,,
\end{equation}
assuming again $\tau_{\mathrm{b}+}=-\tau_{\mathrm{b}-}$. Given $k\tau_{\mathrm{b}+}\ll 1$ for IR modes, this implies $\zeta_k$ receives marginal growth through the bounce phase considering $1-\mpl^2/\mplt^2\sim\mathcal{O}(1)$. If we consider adiabatic perturbations from a massless phase entering the bounce, recalling \eqref{eq:zeta_super}, then we expect
\begin{equation}
    \zeta_k(\tau_{\mathrm{b}+})\simeq\zeta_k(\tau_{\mathrm{b}-})\bigg(1+\frac{2}{-\ln(-k\tau_{\mathrm{b}-})}\Big(1-\frac{\mpl^2}{\mplt^2}\Big)\bigg)\,,\label{eq:zetakbouncemassless}
\end{equation}
where we further assumed $-k\tau_{\mathrm{b}-}$ is sufficiently small so that $-\ln(-k\tau_{\mathrm{b}-})\gg|\ln(2)-\gammaE +i\pi/2|\approx 1.575$. This implies that $\zeta_k$ may grow through the bounce, but only by a limited amount, small relative to its size when it entered the bounce. Finally, in the situation isocurvature perturbations from the prebounce phase source curvature perturbations before the bounce starts, we may expect [e.g., from \eqref{eq:chichipmasslesstbm}] that $\zeta_k(\tau_{\mathrm{b}-})\simeq i\pi\sqrtb{2A_s/k^3}$ and $\zeta_k'(\tau_{\mathrm{b}-})\simeq i\pi\sqrtb{2A_sk}\tau_{\mathrm{b}-}$. Thus, \eqref{eq:zeta_b_approx} implies that
\begin{equation}
    \zeta_k(\tau_{\mathrm{b}+})\simeq\zeta_k(\tau_{\mathrm{b}-})\bigg(1-2\Big(1-\frac{\mpl^2}{\mplt^2}\Big)k^2\tau_{\mathrm{b}+}^2\bigg)\,.\label{eq:zetakscaleinvariantthroughbounce}
\end{equation}
From this, we conclude again that $\zeta_k$ remains more or less constant throughout the bounce phase. This matches the expectation that $\zeta_k$ can only receive limited growth through the bounce, as derived in \cite{Quintin:2019orx}.

\subsection{Numerical examples}\label{sec:numerical}

In the previous subsection, we provided general details for different contracting models that then undergo a bounce due to the Cuscuton, and we presented how the different perturbation modes evolve across the different phases of evolution. Various approximations were made along the way to isolate the different regimes of interest and to derive analytical results. However, everything can be computed numerically to verify the validity of these approximations. We present such examples in this subsection for both the background and linear perturbations.

Since our focus is on testing the physics of the Cuscuton bouncing phase and since treating $\psi$ as an ekpyrotic field or a fundamentally massless one does not affect the physics of the bounce phase significantly, we only present here numerical results for the case where $\psi$ is massless. Provided a Cuscuton potential $U(\varphi)$, this means that we can numerically solve \eqref{eq:varphi_for_solv} for $\bar\varphi(t)$, and then any other background quantities can be derived from the solution for $\bar\varphi(t)$. For instance, recalling \eqref{eq:backCuscutonConstraint}, we have $H(t)=U_{,\varphi}(\bar\varphi(t))/(3\mu^2\mpl^2)$. In the spirit of following up on \cite{Boruah:2018pvq,Kim:2020iwq}, we choose the following Cuscuton potential in our numerical tests,
\begin{equation}
\label{eq:U_phi}
    U(\varphi)=\frac{1}{2}m^2(\varphi^2-\varphi^2_\infty)-\frac{m^4}{4}\left(e^{2(\varphi^2-\varphi^2_\infty)/m^2}-1\right)\,,
\end{equation}
for some constants $m,\varphi_\infty>0$.
For more details on why this particular form of the potential works, we refer the reader to \cite{Boruah:2018pvq,Kim:2020iwq}, but in short, provided $\varphi_\infty^2\gg m^2$, one has $U_\mathrm{b}=U(0)\simeq-m^2\varphi_\infty^2/2<0$ and $m_\mathrm{b}^2=U_{,\varphi\varphi}(0)\simeq m^2>0$ as desired. This choice of potential is in no way unique, but the numerical results presented here with this choice of potential are qualitatively quite generic.

\begin{figure}[t]
\centering 
\includegraphics[width=0.45\columnwidth]{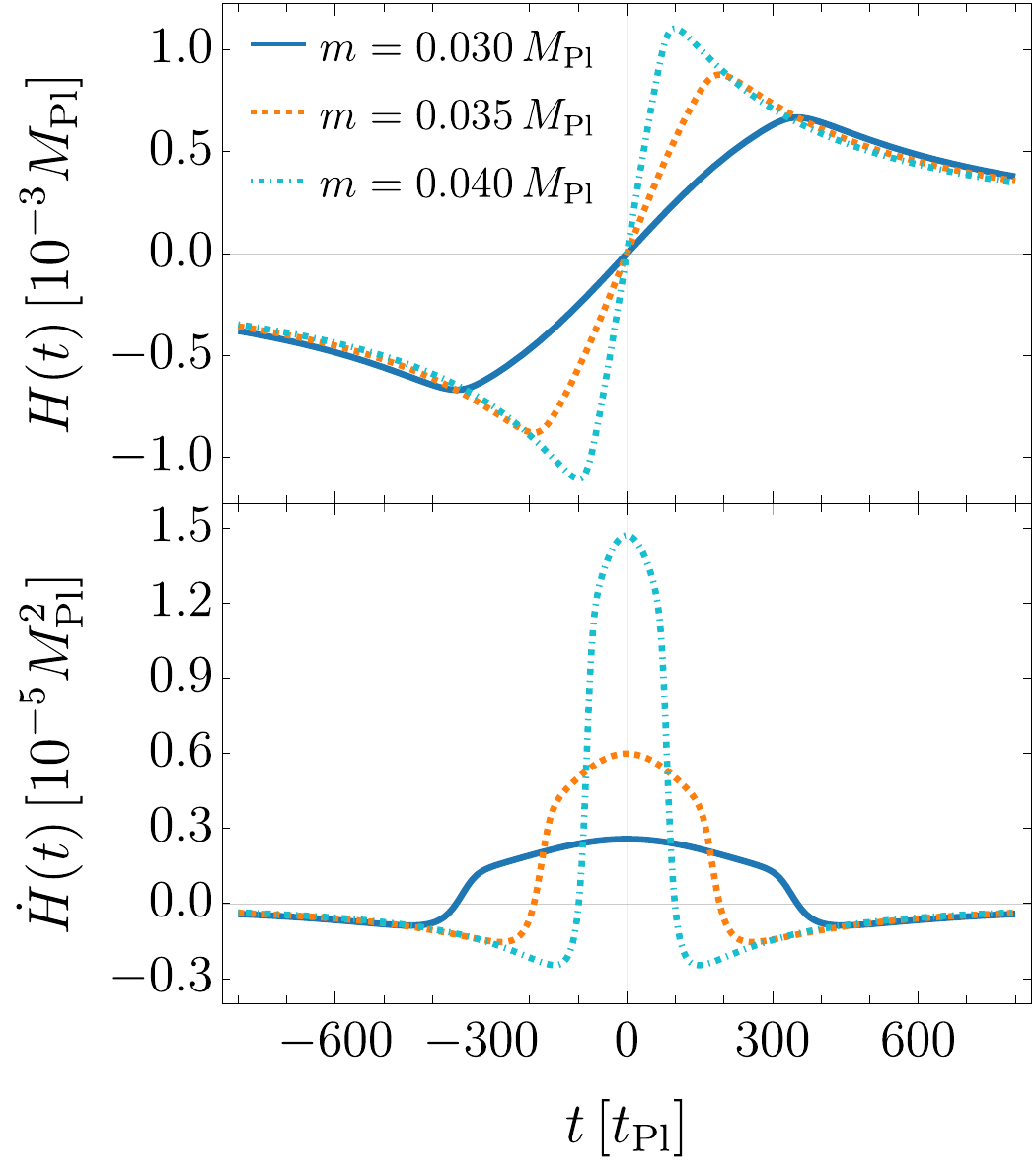}
\hspace{1cm}
\includegraphics[width=0.45\columnwidth]{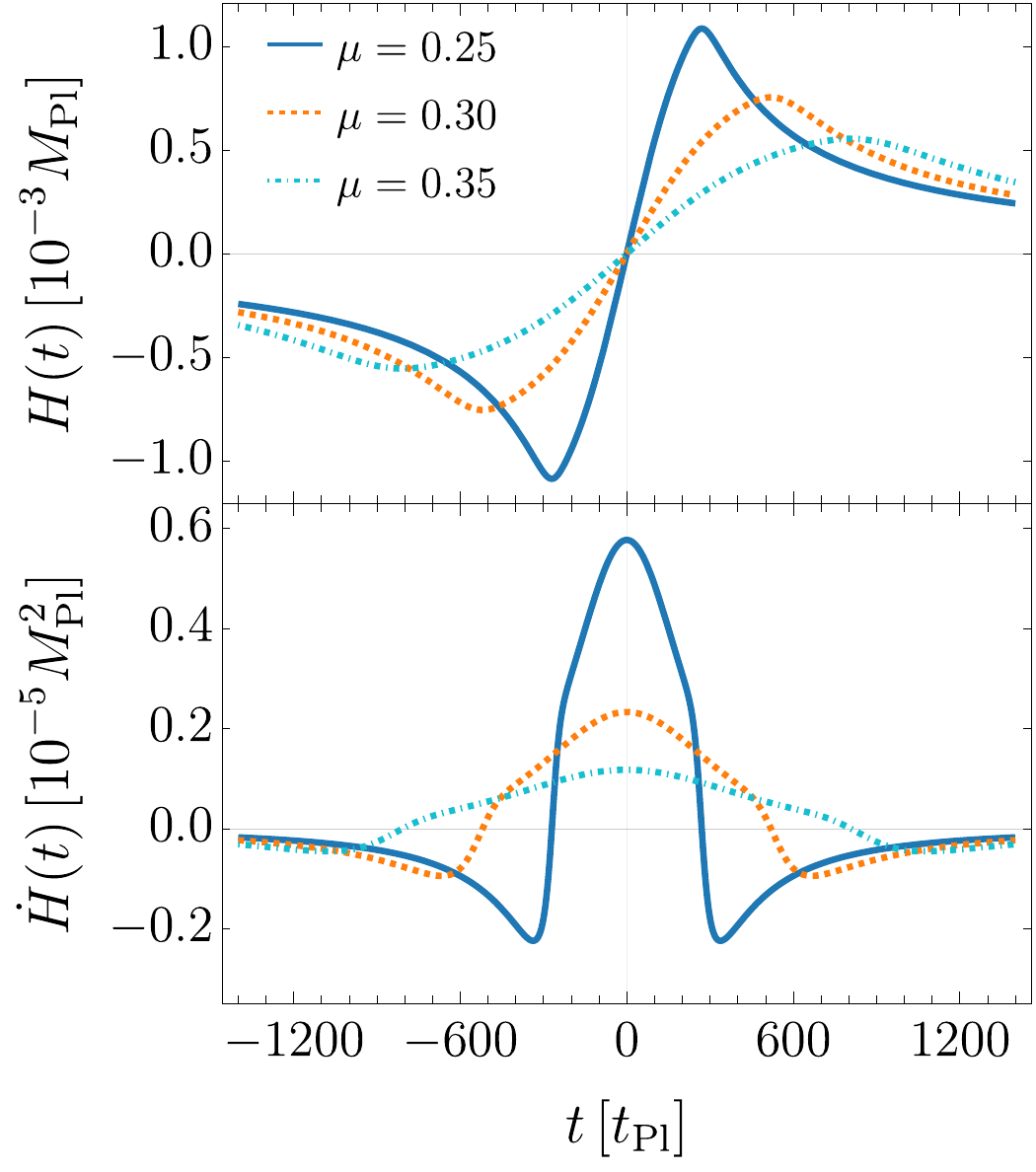}
\caption{{\footnotesize Numerical background solutions for the Hubble parameter $H$ and its time derivative $\dot H$ as functions of the physical time $t$. The Cuscuton parameters $m$ and $\mu$ are varied in the left- and right-hand plots, respectively. In the left panel, we fix $\mu=0.2$ and $\varphi_\infty=0.1\,\mpl$. In the right panel, we fix $m=0.04\,\mpl$ and $\varphi_\infty=0.2\,\mpl$.}}
\label{fig:bounceH-Hdot}
\end{figure}

A set of background solutions for different Cuscuton potential parameters is shown in Fig.~\ref{fig:bounceH-Hdot}. The time range is chosen as to put the emphasis mostly on the bouncing phase since this is the nontrivial part of the background evolution. For the prebounce contracting phase, the numerical solutions approach an $\epsilon=3$ constant EoS, with $H(t)\simeq 1/(3t)$ as $-t$ becomes large (recall that we assume $\psi$ is massless here). Note that the same scaling is found at large $t$ (i.e., after the bounce), because we have not implemented reheating in this example --- $\psi$ is the only matter field ever present in this toy model. All examples shown exhibit a bounce, that is, $H$ crosses 0 and $\dot H$ has a regime where it is positive; hence, the effective NEC is violated.
We can quantitatively verify from the solutions that $\dot H\Delta t^2\lesssim\mathcal{O}(1)$, where $\Delta t$ indicates the duration of the bounce phase [defined as the time interval where $\dot H(t)>0$].
This supports the approximation made in \eqref{eq:aHHpb} and throughout Sec.~\ref{eq:bounceanalytical} that the scale factor is roughly constant to leading order through the bounce phase, so that in this regime $t\approx \tau$, $H\approx\mathcal{H}$, $\dot H\approx\mathcal{H}'$, etc.
Moreover, the linear behavior of $H$ as a function of time is manifest about the bounce point $t=0$, supporting the approximations in \eqref{eq:aHHpb}, which were made to solve the perturbation equations analytically about the bounce point. Of course, $\dot H$ is not exactly constant in the bounce phase in the examples of Fig.~\ref{fig:bounceH-Hdot} --- it is more like a flattened bell curve --- but nevertheless, this shows $\dot H$ is dominated by a more or less constant value once $\dot H$ has sharply transitioned from negative to positive values.

Given numerical solutions for the background, we may numerically solve the equations for the perturbations without any approximations, i.e., the equations of \eqref{eq:zeta-chi-eoms-k} with all the coefficients described in \eqref{eq:z-cs2-zs} and \eqref{eq:A1A2B1B2coeffs}. From this point on, for all the numerical solutions shown as examples, we choose the following model parameters: $m=0.04\,\mpl$, $\varphi_\infty=0.1\,\mpl$, and $\mu=0.2$; the corresponding background is the cyan dot-dashed curve in the left panel of Fig.~\ref{fig:bounceH-Hdot}. The initial conditions for the perturbations are set such that we have an adiabatic vacuum on subhorizon scales in the far past (at a prebounce time $\tau\ll-1/k$ for a given $k$ mode). For modes of observational interest, the transition from sub- to superhorizon scales occurs far away from the bounce phase, in a regime where the contributions from the Cuscuton are completely negligible (a fact that we checked numerically). Correspondingly, we confirm that the numerical solutions are nearly identical to the analytical solutions expressed in terms of the Hankel function of the first kind shown in the previous subsection. As such, we may use the analytical solutions (in terms of the Hankel function) to set the initial conditions on superhorizon scales, as long as it is sufficiently far away from the bounce, and numerically evolve from there until after the bounce phase. Since the computational cost associated with numerically solving perturbations from the subhorizon regime to the superhorizon regime is high, it is advantageous to initiate the analysis using the Hankel function on superhorizon scales. This approach is particularly beneficial given that the primary interest lies in the physics of superhorizon modes throughout the bounce phase.

\begin{figure}[t]
\centering 
\includegraphics[width=\columnwidth]{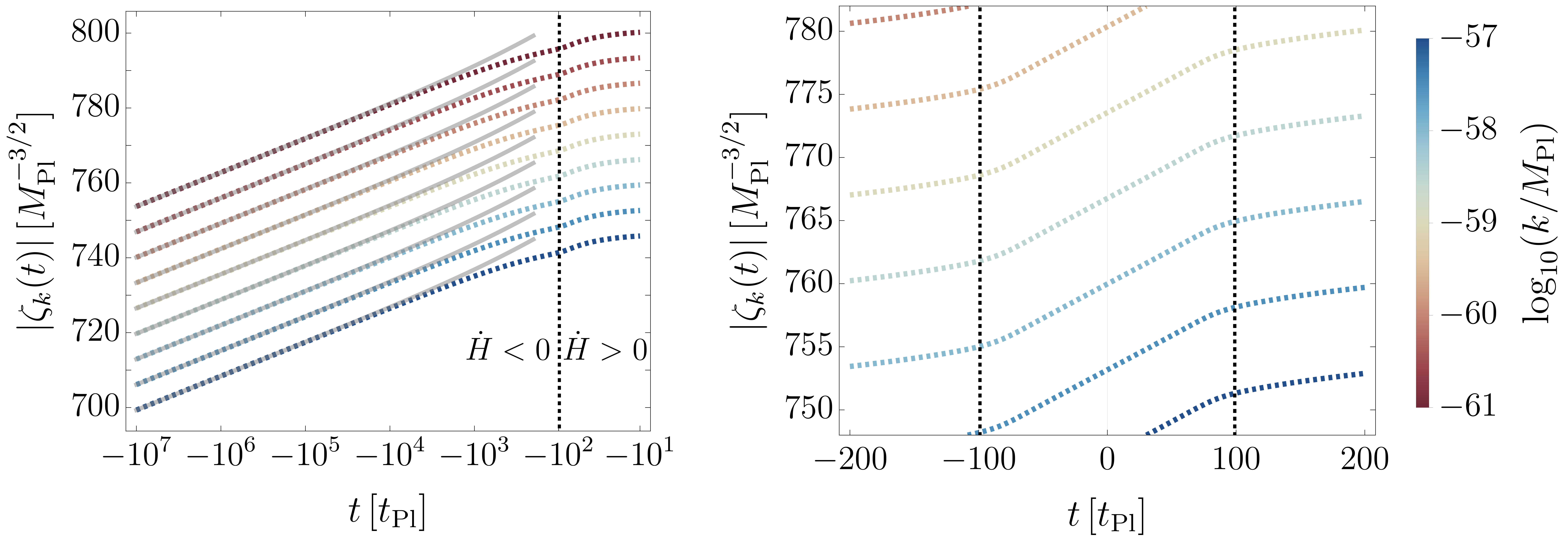}
\caption{{\footnotesize Numerical solutions for the absolute value of the curvature perturbations $|\zeta_k(t)|$ on superhorizon scales. The left panel shows the contracting phase alongside the beginning of the bounce phase on a logarithmic time scale. Dashed colored curves show the numerical solutions, and the grey solid lines show the corresponding analytic approximation. The color represents the wavenumber $k$ according to the color bar on the far right. The vertical black dotted lines indicate when $\dot H=0$, representing the start and end of the bounce phase. The right panel shows a zoomed-in version of the whole bouncing phase on a linear time scale.}}
\label{fig:zeta}
\end{figure}

The numerical solutions for the absolute value of the curvature perturbations, $|\zeta_k(t)|$, are shown in Fig.~\ref{fig:zeta} in the case where $\psi$ is massless. The colored dashed curves show the numerical solutions, where the color coding represents the wavenumber $k$ of the perturbations (we choose a range that roughly corresponds to the observable scales in the CMB). The black vertical dotted lines indicate the moments $\dot H$ crosses zero, so in the left plot the focus of the logarithmic time scale is on the contracting phase and the beginning of the bounce phase, while in the right plot the focus of the linear time scale is on the whole bounce phase. In the left plot, the faint gray solid lines show the analytical approximation \eqref{eq:zeta_super}, which should hold in the contracting phase well before the bounce phase, for the same set of $k$ modes as the numerical solutions. As we can see, the numerical and analytical curves match almost perfectly for $t/\tpl\lesssim -10^4$. Even by $t/\tpl\sim -10^{2.2}$ (where we cut the gray lines for visual purposes, though the bounce phase really starts at $t/\tpl\approx -99$), the relative difference between the analytical and numerical curves remains very small ($\sim 1\%$).
This deviation is due to the Cuscuton becoming important in the approach to the bounce --- indeed, the bounce is triggered because of the Cuscuton, so its effect should become manifest in that regime.

The plot on the left of Fig.~\ref{fig:zeta} shows that superhorizon adiabatic perturbations grow logarithmically in a contracting phase dominated by a massless scalar (note that the horizontal axis is logarithmic, while the vertical axis is linear). Similarly, we see that the $k$ dependence is very weak, again logarithmic in fact: the amplitude of $|\zeta_k|$ at any fixed time changes by less than $10\%$ across $4$ orders of magnitude in $k$. While the plots show that the dimensionful quantity $|\zeta_k|$ is nearly independent of $k$, we can infer that the dimensionless power spectrum $\sim k^3|\zeta_k|^2$ is strongly blue, as anticipated analytically.

The right plot of Fig.~\ref{fig:zeta} shows that the numerical solutions match the analytical expectation for $\zeta_k(t)$ also through the bounce, which is that it grows linearly as a function of time, with a rate that is an $\mathcal{O}(1)$ factor times its rate when it entered the bounce phase; recall \eqref{eq:zeta_b_approx}.
Indeed, we see quite neatly the change of slope in $\dot\zeta_k(t)$ (roughly equivalent to $\zeta'_k(\tau)$ in the bounce phase) going from the contracting phase to the bouncing phase. Moreover, the amplification $|\zeta_k|$ receives passing through the bounce is relatively small compared to its amplitude, again as anticipated analytically in \eqref{eq:zetakbouncemassless}.

Moving on to entropy perturbations, we take \eqref{eq:Fofphi-massless2} as the choice of coupling function, with $\Lambda=10^{-3.5}\,\mpl$ and $\delta=0$. This will yield an exactly scale-invariant power spectrum, which is sufficient for the sake of presenting the general behavior of $\chi_k$, but the computation is straightforward to repeat with a slight $\delta>0$.
We could also take \eqref{eq:Fofphi-massless} instead of \eqref{eq:Fofphi-massless2} for the coupling function. However, they have the same time dependence when the Cuscuton is negligible. Therefore, they produce identical results for $\chi_k$ during the contracting phase and expectedly through the bounce, as discussed in the previous subsection.
We have confirmed this numerically.

\begin{figure}[t]
\centering 
\includegraphics[width=\columnwidth]{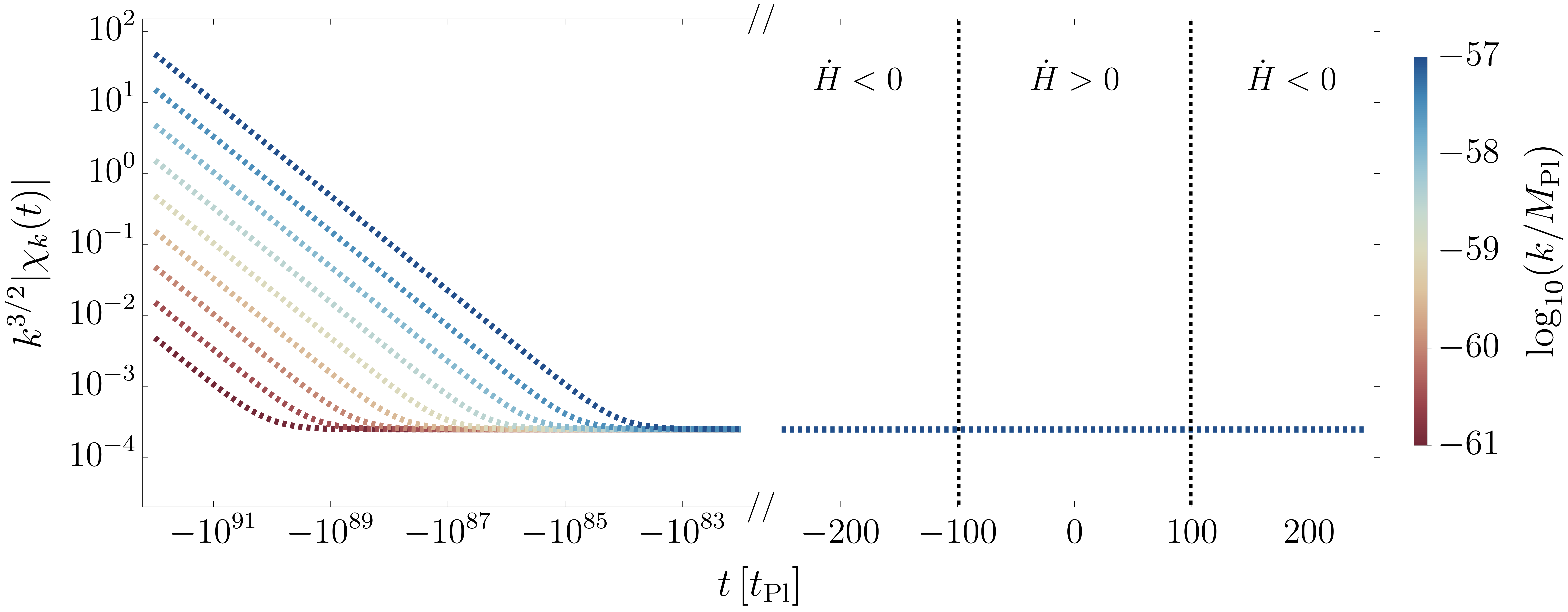}
\caption{{\footnotesize Behavior of $k^{3/2}|\chi_k(t)|$ from the numerical solutions for the entropy perturbations, $\chi_k(t)$. The plot shows part of the subhorizon evolution, horizon crossing, and part of the superhorizon evolution (on a logarithmic time scale to the left of the break), and it shows the evolution through the bounce phase (on a linear time scale to the right of the break). The colors indicate the wavenumber $k$. The vertical black dotted lines show when $\dot H=0$, indicating the beginning and end of the bounce phase.}}
\label{fig:chi}
\end{figure}

The numerical solutions for the entropy perturbation $\chi_k(t)$ are shown in Fig.~\ref{fig:chi}, where we plot the dimensionless quantity $k^{3/2}|\chi_k(t)|$, essentially the square root of the dimensionless power spectrum as a function of time. Curves of different colors represent different $k$ modes as in the previous figure. The time axis is broken as to put the emphasis on two processes happening in very different time regimes: when the modes exit the horizon; and when they pass through the bounce phase (in between the two black vertical dotted lines). The decrease in $|\chi_k(t)|$ when the modes are subhorizon is due to the fact that\footnote{A word on notation: $\sim$ should be read as `approximately scales as' here (similar to $\simeq$, but up to proportionality constants, ignoring numerical factors). Elsewhere, it may mean `of the order of'. Its interpretation should be clear from context.} $z_\chi=a\sqrt{F}\sim 1/(-\tau)$, so then $|\chi_k(\tau)|\propto |u_k|/z_\chi(\tau)\sim(-\tau)$ if $u_k$ is in its adiabatic vacuum state ($|u_k|\sim 1/\sqrtb{2k}$). Once the modes cross the horizon, they freeze and $\chi_k$ becomes nearly constant; in fact, the numerical results match the analytical approximation \eqref{eq:chi_anysol} according to which $\chi'_k(\tau)/\chi_k(\tau)\simeq k^2\tau$, which means that the growth of $|\chi_k(\tau)|$ in the supper-Hubble regime as $k\tau\to 0^-$ is very much suppressed. Figure \ref{fig:chi} also shows that as the modes of different $k$ freeze on superhorizon scales, they all approach the same value for $k^{3/2}|\chi_k|\sim 2\times 10^{-4}$ --- this is the behavior to be expected for an exact scale-invariant dimensionless power spectrum (recall that we took $\delta=0$) of amplitude $\sim 10^{-9}$. The plots further show that the numerical solutions remain essentially constant as they pass through the bounce phase, which matches the analytical approximation given in \eqref{eq:chiktaubp} --- the possible amplification through the bounce phase is IR-suppressed as $(k\tau_{\mathrm{b}-})^2\ll 1$. The lesson here is that the bounce phase does not affect the scale-invariant spectrum. Note that if the $\chi_k$ isocurvature perturbations are converted into curvature perturbations $\zeta_k$ right before the bounce phase, then the same conclusion would be reached according to the discussion around \eqref{eq:zetakscaleinvariantthroughbounce}.


\section{Beyond the linear regime}\label{sec:nonlinear}

Since this section covers multiple aspects, we provide a brief guideline on how to read it. The main objective is to compute the third-order action, determine the strong coupling scale, and analyze the bispectrum to quantify non-Gaussianities. We carry out this analysis both analytically and numerically across different phases and scales. We begin in Sec.~\ref{sec:insight} with a brief literature review and theoretical expectations. In Sec.~\ref{sec:3rdorder}, we derive the third-order action. Section \ref{sec:strong-coupling} focuses on the strong coupling scale, considering linear perturbations and interactions, analyzed separately for the prebounce and bounce phases across IR and UV scales. Section \ref{sec:bispectrum} then presents the bispectrum and non-Gaussianity analysis, addressing the same cases as before. The strong coupling scale is revisited using bispectrum calculations and compared to previous results. Finally, we end the section with numerical results that support the analytical estimates.

\subsection{Theoretical insights and expectations}\label{sec:insight}

So far, we have seen how the Cuscuton may yield a nonsingular bouncing cosmology at the background level, and we have seen examples of cosmological perturbations that may evolve in such a background. Specifically, a large-scale nearly scale-invariant 2-point function of entropy perturbations may be generated either during an ekpyrotic or stiff contracting phase, together with unamplified adiabatic perturbations. Provided that the isocurvature perturbations are efficiently converted into curvature perturbations, the models could match the current observational constraints from the CMB.

However, these findings are based on the study of linear perturbations and their corresponding 2-point statistics, but a detailed comparison with CMB constraints requires advancing the analysis beyond the linear regime. In particular, nonlinearities should not compromise the results. For example, the existence of strong coupling issues or the production of (very large) large non-Gaussianities would represent severe drawbacks to the theory.

Naturally, when the Cuscuton contribution is negligible (i.e., deep in the contracting phase before the bounce) and before the conversion of isocurvature to curvature perturbations, we should recover the results of the literature, according to which the perturbations should remain nearly perfectly Gaussian (see, e.g., \cite{Fertig:2013kwa,Ijjas:2014fja,Fertig:2015ola} and references therein). This makes intuitive sense: during that phase, $\zeta$ is effectively unexcited and remains very close to its Gaussian vacuum state, and $\chi$ has no cubic interaction term in the action (since, in particular, it has no potential), hence, it has a vanishing 3-point correlation function. To better understand the rationale behind this intuition, we note that in this regime, the theory is very close to GR with canonical scalar fields, so in particular the scalars' sound speed is very close to unity, and initial conditions are taken according to a Lorentzian adiabatic vacuum; hence non-Gaussianities of equilateral and orthogonal shape should not get generated significantly (see, e.g., \cite{Chen:2006nt,Langlois:2008qf}).\footnote{The intuition really comes from inflation, where $\epsilon<1$. Since $\epsilon\gtrsim\mathcal{O}(1)$ for contracting models, the non-Gaussianity results from inflation are not directly applicable. Still, they may provide insight.} Furthermore, local non-Gaussianities are similarly not generated, as can be assessed through the EoMs for $\zeta$ and $\chi$. That is, if we expand the perturbations to second order at the level of their EoMs, we find no source for the entropy mode at quadratic order since there is no potential for $\chi$ and since there is no turning of the trajectory in $\{\psi,\chi\}$-space \cite{Fertig:2013kwa,Ijjas:2014fja,Fertig:2015ola}. In other words, $\zeta$ and $\chi$ are decoupled at the perturbative level.

Therefore, no intrinsic local non-Gaussianities are generated for the entropy perturbations. Similarly, if we expand the curvature perturbation to second order at the level of its EoM on large scales, one finds that it is sourced to second order according to (in the superhorizon limit; see again, e.g., \cite{Fertig:2013kwa,Ijjas:2014fja,Fertig:2015ola})
\begin{equation}
	\dot\zeta^{(2)}\simeq\frac{HV_{,\psi}}{\dot{\psi}^2}s\left(\frac{1}{2}\frac{F_{,\psi}}{F}s-\frac{\dot{s}}{\dot{\psi}}\right)=-\frac{HV_{,\psi}F}{\dot\psi^3}\chi\dot\chi\,,
\end{equation}
where $s\equiv\sqrtb{F}\chi$. However, we know that the superhorizon solution for $\chi$ is constant in time to leading order [i.e., $\dot\chi\simeq 0$, recall \eqref{eq:uk-chik-full-super-ek} or \eqref{eq:chi_anysol}], hence $s\sim\sqrtb{F}$ and $\dot\zeta^{(2)}\simeq 0$. Therefore, the curvature perturbation is not sourced by entropy modes at this point, indicating once more no deviations from Gaussianity.\footnote{Subleading effects could still generate small local non-Gaussianities at this point, such as the decaying contribution to $\chi$. Also, more refined models (especially of ekpyrosis) would have a slightly time-dependent background EoS, which would also have the effect of sourcing $\zeta$ to second order. Such effects remain tiny, though \cite{Fertig:2013kwa}.} We note that these statements remain true even to third order \cite{Fertig:2015ola}, hence we expect $f^\textrm{(local)}_\NL\approx 0$ and $g^\textrm{(local)}_\NL\approx 0$. These quantities are the coefficients of the expansion of the full real-space perturbation $\zeta(\bm{x})$ in powers of a Gaussian variable $\zeta_\G(\bm{x})$, schematically $\zeta(\bm{x})=\zeta_\G(\bm{x})+f_\NL^\textrm{(local)}\zeta_\G^2(\bm{x})+g_\NL^\textrm{(local)}\zeta_\G^3(\bm{x})+\cdots$. We will provide a more formal definition when we discuss the computation of the bispectrum in a subsequent subsection.

Two things can still generate non-Gaussianities at this point: the conversion of isocurvature perturbations into curvature perturbations; and the Cuscuton's contribution. The former has already been considered quite extensively in the literature \cite{Lehners:2007wc,Lehners:2008my,Lehners:2009qu,Lehners:2009ja,Lehners:2010fy,Ijjas:2014fja,Fertig:2015ola,Fertig:2016czu}. The conversion can be modeled in different ways, but the general idea is that there must exist a potential barrier that induces a rotation in $\{\psi,\chi\}$-space. If the rotation rate is $\dot\theta=V_{,s}/\dot\psi$, where we recall $s(\psi,\chi)=\sqrtb{F(\psi)}\chi$, then curvature perturbations are sourced on large scales following $\dot\zeta\simeq 2H\dot\theta s/\dot\psi$, after which $\zeta$ roughly acquires the amplitude and spectral dependence of $\chi$. This conversion generates local non-Gaussianities of the order of $f^\textrm{(local)}_\NL=\mathcal{O}(1-10)$, depending on the precise modeling of the conversion process \cite{Lehners:2007wc,Lehners:2008my,Lehners:2009qu,Lehners:2009ja,Lehners:2010fy,Ijjas:2014fja,Fertig:2015ola,Fertig:2016czu}.
This is very close to current observational bounds, e.g., $f^\textrm{(local)}_\NL=-0.9\pm 5.1$ (65\% CL) from \textit{Planck} \cite{Planck:2019kim}.
This tells us that if there is any Cuscuton contribution to $f^\textrm{(local)}_\NL$, it should not exceed that constraint either, that is, not much larger than $\mathcal{O}(1)$.
Future observational constraints on $f_\NL$ will certainly constrain this even more (see, e.g., \cite{Camera:2014bwa,Karagiannis:2018jdt,SimonsObservatory:2018koc,Meerburg:2019qqi,Karagiannis:2019jjx,Biagetti:2022qjl} and references therein).

This naturally leads us to compute the bispectrum with the Cuscuton included to determine both that the theoretical analysis of the model is sound and that the model is not excluded by observations. In particular, the behavior of the bispectrum going through the bounce will be crucial since it is precisely the presence of the Cuscuton that enables the nonsingular transition. Two things could go wrong: the first one, as already pointed out, would be that large non-Gaussianities could be generated through the bounce such that observational bounds rule out the models; and the second one would be that the theory becomes strongly coupled, meaning in this context that the perturbative expansion would essentially break down, indicating that the linear perturbations would be untrustworthy. Many nonsingular bounces (using different physics from Cuscuton) are known to suffer from either issue, if not both (e.g., \cite{Gao:2014hea,Gao:2014eaa,Quintin:2015rta,deRham:2017aoj}). We will show that the Cuscuton bounce is robust in that respect: no sizable non-Gaussianities are produced through the bounce, whatsoever, and the theory remains well under perturbative control throughout. We estimate the strong coupling scale along the way and show that it is well above the scale of the background.

\subsection{Third-order action}\label{sec:3rdorder}

The strategy we shall employ to investigate strong coupling and compute bispectra is to use the action expanded to third order in perturbations. We show the details of how this is obtained in this subsection.

From this point on, we only consider the case where $\psi$ is massless at the level of the interactions. This would not be applicable deep in the contracting phase for ekpyrosis, but we already know that the perturbations remain Gaussian in that regime. Thus, to investigate the effects of the Cuscuton and the nonsingular bounce, it is fair to set the potential for $\psi$ to zero from the start. We then expand the action given in \eqref{eq:actionFullGen} to third order in $\alpha$, $\beta$, $\zeta$, $\delta\varphi$, and $\chi$ defined in \eqref{eq:ADMpert} and \eqref{eq:fieldspert}, always performing some integration by parts and using the background EoMs for simplifications. For better illustration, we classify the third-order action in three parts:
\begin{equation}
    S^{(3)}=\mpl^2\int\dd^3\bm{x}\,\dd t\,a^3\left(\mathcal{L}^{(3)}_{(\chi)}+\mathcal{L}^{(3)}_{(\delta\varphi)}+\mathcal{L}^{(3)}_{(\mathrm{rest})}\right)\,.\label{eq:S33p}
\end{equation}
Note that $\mathcal{L}^{(3)}_{(\chi)}$ is not meant to represent the third-order expansion of $\mathcal{L}_{\chi}$ as defined in \eqref{eq:defLchi} and likewise for the other terms. Instead, we have in mind expanding the full action to third order and gathering all the cubic terms involving two $\chi$s first in $\mathcal{L}^{(3)}_{(\chi)}$, then all the remaining cubic terms involving at least one $\delta\varphi$ in $\mathcal{L}^{(3)}_{(\delta\varphi)}$, and finally all the remaining cubic terms in $\mathcal{L}^{(3)}_{(\mathrm{rest})}$ (before the substitution of any constraint equations for $\alpha$, $\beta$, and $\delta\varphi$).\footnote{Note also that the `Lagrangian density' as defined by $S^{(3)}=\mpl^2\int\dd^3\bm{x}\,\dd t\,a^3\mathcal{L}^{(3)}$ does not have the appropriate dimensions given the factorization of $\mpl^2$, i.e., henceforth $\mathcal{L}$ has dimensions of mass squared. Still, we abuse notation given the practicality of this factorization.} This yields
\begin{subequations}\label{eq:L30}
{\allowdisplaybreaks
\begin{align}
    \mathcal{L}^{(3)}_{(\chi)}=&-\frac{1}{2}F(\psi,X)\left((\alpha-3\zeta)\dot\chi^2+(\alpha+\zeta)\frac{(\partial_i\chi)^2}{a^2}+2\dot\chi\frac{\partial^i\beta\partial_i\chi}{a^2}\right)-XF_{,X}\alpha\left(\dot\chi^2-\frac{(\partial_i\chi)^2}{a^2}\right)\,,\\
    \mathcal{L}^{(3)}_{(\delta\varphi)}=&~9\mu^2(\dot\zeta-H\alpha)\zeta\,\delta\varphi+\frac{3\mu^4\epsilon(\alpha+3\zeta)}{4\sigma}\delta\varphi^2-\frac{\mu^6\epsphi\epsilon(2\epsilon-\eta-6)}{8\sigma^3H}\delta\varphi^3+\mu^2\zeta\frac{\partial^i\beta\partial_i\delta\varphi}{a^2}\nonumber\\
    &+\frac{\mu^4}{4\sigma H^2}\left(2\alpha+\zeta-\frac{\mu^2\dot{\delta\varphi}}{2\sigma H^2}\right)\frac{(\partial_i\delta\varphi)^2}{a^2}\,,\\
    \mathcal{L}^{(3)}_{(\mathrm{rest})}=&~(3-\epsphi)H^2\alpha^3+3(\epsphi-3)H^2\alpha^2\zeta-6H\alpha^2\dot\zeta+18H\alpha\zeta\dot\zeta+3\alpha\dot\zeta^2-9\zeta\dot\zeta^2+2H\zeta\frac{\partial^i\alpha\partial_i\beta}{a^2}\nonumber\\
    &+(\zeta-\alpha)\frac{(\partial_i\zeta)^2}{a^2}-2\zeta\frac{\partial^i\beta\partial_i\dot\zeta+\alpha\partial^2\zeta}{a^2}+2(H\alpha-\dot\zeta)\alpha\frac{\partial^2\beta}{a^2}+\alpha\frac{(\partial^2\beta)^2-(\partial_i\partial_j\beta)^2}{2a^4}\nonumber
    \\
    &+\frac{\partial^2\beta\partial^i\beta\partial_i\zeta}{a^4}+\frac{3\partial^i\beta\partial^j\beta\partial_i\partial_j\zeta}{2a^4}\,.
\end{align}
}%
\end{subequations}
Note that partial derivatives are always meant to act on immediate variables only.
Upon substituting the solutions for the perturbed lapse and shift \eqref{eq:solalphabeta} and recalling the definition of $Q$ in \eqref{eq:psi}, the action can be recast in the form of \eqref{eq:S33p}, except now with different $\mathcal{L}$s. First, the terms that couple to $\chi$ are
\begin{align}
    \mathcal{L}^{(3)}_{(\chi)}=&-\frac{F(\psi,X)}{2H}\Big(\dot\zeta-3H\zeta-\frac{\mu^2}{2}\delta\varphi\Big)\dot\chi^2
    -\frac{F(\psi,X)}{2H}\Big(\dot\zeta+H\zeta-\frac{\mu^2}{2}\delta\varphi\Big)\frac{(\partial_i\chi)^2}{a^2}\nonumber\\
    &+F(\psi,X)\left(\frac{\partial_i\zeta}{aH}-a\epsphi\Big(\partial_i\partial^{-2}\dot\zeta-\frac{\mu^2}{2}\partial_i\partial^{-2}\delta\varphi\Big)\right)\frac{\partial^i\chi}{a}\dot\chi\nonumber\\
    &-\frac{XF_{,X}}{H}\left(\dot\zeta-\frac{\mu^2}{2}\delta\varphi\right)\left(\dot\chi^2-\frac{(\partial_i\chi)^2}{a^2}\right)\,.\label{eq:L3chir}
\end{align}
Then, terms that still include the Cuscuton contribution can be divided as $\mathcal{L}^{(3)}_{(\delta\varphi)}=\mathcal{L}^{(3)}_{(\delta\varphi)^3}+\mathcal{L}^{(3)}_{(\delta\varphi)^2}+\mathcal{L}^{(3)}_{(\delta\varphi)^1}$ with
\begin{subequations}\label{eq:L3deltavarphiallr}
{\small
{\allowdisplaybreaks
\begin{align}
    \mathcal{L}^{(3)}_{(\delta\varphi)^3}=&-\frac{\mu^6}{8\sigma^2a^2H^4}\left(2\sigma H\,\delta\varphi+\dot{\delta\varphi}\right)(\partial_i\delta\varphi)^2
    -\frac{\mu^6\epsphi C_3}{16\sigma^3H}\delta\varphi^3
    +\frac{\mu^6\epsphi^2}{16H}(\partial_i\partial_j\partial^{-2}\delta\varphi)^2\,\delta\varphi\,,\label{eq:L3deltavarphi23}\\
    \mathcal{L}^{(3)}_{(\delta\varphi)^2}=&~\frac{\mu^4}{4\sigma a^2H^3}(2\dot\zeta+H\zeta)(\partial_i\delta\varphi)^2
    +\frac{\mu^4}{a^2H^2}\delta\varphi\,\partial^i\zeta\partial_i\delta\varphi
    -\frac{\mu^4\epsphi}{4a^2H^2}\delta\varphi^2\,\partial^2\zeta\nonumber\\
    &+\frac{\mu^4\epsphi}{8\sigma H}\left(6(\sigma+3)H\zeta+\big(\sigma(2+3\epsphi)+6\big)\dot\zeta\right)\delta\varphi^2
    +\frac{\mu^4\epsphi}{4a^2H^2}\delta\varphi\,\partial^i\partial^j\zeta\partial_i\partial_j\partial^{-2}\delta\varphi\nonumber\\
    &-\frac{\mu^4\epsphi^2}{8H}\left(2\,\delta\varphi\,\partial^i\partial^j\partial^{-2}\delta\varphi\,\partial_i\partial_j\partial^{-2}\dot\zeta+\dot\zeta(\partial_i\partial_j\partial^{-2}\delta\varphi)^2-H\partial^i\partial^{-2}\delta\varphi\left(2\,\delta\varphi\,\partial_i\zeta+3\partial^j\partial^{-2}\delta\varphi\,\partial_i\partial_j\zeta\right)\right)\,,\\
    \mathcal{L}^{(3)}_{(\delta\varphi)^1}=&~\frac{\mu^2\epsphi}{4H}\left((2-3\epsphi)\dot\zeta-12H\zeta\right)\dot\zeta\,\delta\varphi
    +\frac{\mu^2}{2a^2H^2}\left(H(1+\epsphi)(\partial_i\zeta)^2+2\big(H\zeta+(1+\epsphi)\dot\zeta\big)\partial^2\zeta\right)\delta\varphi\nonumber\\
    &+\frac{\mu^2}{4a^4H^3}\left((\partial_i\partial_j\zeta)^2-(\partial^2\zeta)^2\right)\delta\varphi
    +\frac{\mu^2\epsphi^2}{4H}\partial^i\partial^j\partial^{-2}\dot\zeta\left(\delta\varphi\,\partial_i\partial_j\partial^{-2}\dot\zeta+2\dot\zeta\partial_i\partial_j\partial^{-2}\delta\varphi\right)\nonumber\\
    &-\frac{\mu^2\epsphi}{2a^2H^2}\left(\partial^i\partial^j\zeta\big(\delta\varphi\,\partial_i\partial_j\partial^{-2}\dot\zeta-2H\partial_i\partial^{-2}\delta\varphi\partial_j\zeta\big)+\big(H\partial^i\zeta\partial^j\zeta+\dot\zeta\partial^i\partial^j\zeta\big)\partial_i\partial_j\partial^{-2}\delta\varphi\right)\nonumber\\
    &-\frac{\mu^2\epsphi^2}{2}\left(\delta\varphi\,\partial^i\zeta\partial_i\partial^{-2}\dot\zeta+\partial^i\partial^{-2}\delta\varphi\big(\dot\zeta\partial_i\zeta+3\partial^j\partial^{-2}\dot\zeta\partial_i\partial_j\zeta\big)\right)\,.\label{eq:L3deltavarphi2}
\end{align}
}}%
\end{subequations}
In \eqref{eq:L3deltavarphi23} to lighten the (already very long) expression we defined
\begin{equation}
    C_3=-\epsilon^3(2+\epsphi)+\epsphi^2(6+\epsphi(2+\epsphi))+\epsilon^2(10+3\epsphi(2+\epsphi))-3\epsilon(4+\epsphi(4+\epsphi(2+\epsphi)))-2\epsilon\eta\,.
\end{equation}
Finally, the remaining third-order terms containing $\zeta$ are
\begin{align}
    \mathcal{L}^{(3)}_\mathrm{(rest)}=&~\frac{\epsphi(\epsphi-2)}{2H}\dot\zeta^3+3\epsphi\zeta\dot\zeta^2-\frac{\epsilon\zeta(\partial_i\zeta)^2}{a^2}-\frac{\epsphi\dot\zeta^2\partial^2\zeta}{a^2H^2}+\frac{3\epsilon\partial^2\zeta(\partial_i\zeta)^2}{4a^4H^2}\nonumber\\
    &+\frac{\epsphi^2}{2H}\left(H\partial^i\partial^{-2}\dot\zeta\big(2\dot\zeta\partial_i\zeta+3\partial^j\partial^{-2}\dot\zeta\partial_i\partial_j\zeta\big)-\dot\zeta(\partial_i\partial_j\partial^{-2}\dot\zeta)^2\right)\nonumber\\
    &+\frac{\epsphi}{a^2H^2}\partial^i\partial^j\partial^{-2}\dot\zeta\left(H\partial_i\zeta\partial_j\zeta+\dot\zeta\partial_i\partial_j\zeta\right)\,.\label{eq:L3restr}
\end{align}
In principle at this point, the solution to the constraint $\delta\varphi$, \eqref{eq:deltacusc}, could be substituted to further obtain an action that depends solely on the two dynamical degrees of freedom $\chi$ and $\zeta$. However, the resulting expression is tremendously long and is not particularly instructive. One could hope that integration by parts further simplifies the action as is the case in GR \cite{Collins:2011mz}, but we have not found this to be the case here (more comments about this below). Still, in what follows, we will find that the third-order action as expressed above is already useful.

An important observation here is about the GR limit. If one sets the Cuscuton and the entropy field to zero ($\mu$, $U$, and $F$ all equivalently $0$), then the theory reduces to GR with a free (noninteracting, massless) scalar. In the third-order action above, this means only $\mathcal{L}^{(3)}_\mathrm{(rest)}$ [eq.~\eqref{eq:L3restr}] survives with $\epsphi=\epsilon=3$. In such a case, one can verify that the action is indeed equivalent to what one expects in GR with a free scalar (see, e.g., \cite{Collins:2011mz,Wang:2013zva}), namely, after integrating by parts and using the GR EoM for $\zeta$,
\begin{equation}
    \mathcal{L}^{(3)}_\textrm{(GR)}=9\zeta\dot\zeta^2+9\zeta\frac{(\partial_i\zeta)^2}{a^2}-\frac{9}{2}\dot\zeta\partial_i\zeta\partial^i\partial^{-2}\dot\zeta+\frac{27}{4}\partial^2\zeta(\partial_i\partial^{-2}\dot\zeta)^2\,.\label{eq:L3GR}
\end{equation}
When we have the Cuscuton, the `GR part' of our \emph{full} third-order action, which we called $\mathcal{L}^{(3)}_\mathrm{(rest)}$, is generally different from $\mathcal{L}^{(3)}_\textrm{(GR)}$. Indeed, one could contemplate extracting the GR terms by performing integration by parts and using the full linear EoM for $\zeta$ as in the standard calculation (e.g., \cite{Collins:2011mz,Wang:2013zva}). However, the second-order action for $\zeta$, \eqref{eq:S23}, does not straightforwardly factorize into a `GR part' and a `Cuscuton part', and the nonlocality of the corresponding EoM for $\zeta$ and of the constraint equation for $\delta\varphi$ makes it practically impossible to further simplify \eqref{eq:L3restr} [and \eqref{eq:L3deltavarphiallr} for that matter]. One implication of this is that, when the Cuscuton is important, \eqref{eq:L3restr} has higher-order operators of the form $\nabla^4\zeta^3/H^2$, while in the GR limit we only expect operators of the form $\nabla^2\zeta^3$ because higher-order operators like $\nabla^4\zeta^3/H^2$ amount to boundary terms. [Henceforth, we use $\nabla$ to schematically denote the power counting of derivatives (spatial or temporal).] Yet, formally in the UV limit $k\to\infty$, the Cuscuton reduces to GR \cite{Afshordi:2007yx}, so one could contemplate the possibility that terms of the form $\nabla^4\zeta^3/H^2$ may be removed from the third-order Cuscuton action by integration by parts perturbatively in $1/k$. We checked and confirm that, before integration by parts, the third-order action of the Cuscuton and of GR match to leading order in $1/k$, and they are of the form $\nabla^4\zeta^3/H^2$. More comments on this will follow later.

\subsection{Toward estimating the strong coupling scale}
\label{sec:strong-coupling}

\subsubsection{Strategy}\label{sec:strongCouplingStrategy}

The first thing we wish to do with the knowledge of the third-order action is to compare the `size' of its terms to those of the second-order action. If we express these actions as, say, $S^{(3)}=\mpl^2\int\dd^3\bm{x}\,\dd t\,a^3\mathcal{L}^{(3)}$ and $S^{(2)}=\mpl^2\int\dd^3\bm{x}\,\dd t\,a^3\mathcal{L}^{(2)}$, respectively, then the ratio $\mathcal{L}^{(3)}(t,\bm{x})/\mathcal{L}^{(2)}(t,\bm{x})$ is informative of the perturbative expansion's regime of validity and of the theory's strong coupling scale (e.g., \cite{Leblond:2008gg,Baumann:2011dt,Joyce:2011kh,Ageeva:2020gti,Ageeva:2020buc}). Indeed, in order for $\zeta$ and $\chi$ to be meaningful perturbation variables and for the theory to be controllable, one should have
\begin{equation}
    \mathcal{L}=\mathcal{L}^{(0)}+\mathcal{L}^{(2)}+\mathcal{L}^{(3)}+\ldots
\end{equation}
with every term smaller than the previous one. (The expansion generally includes $\mathcal{L}^{(1)}$, but it is often omitted since up to boundary terms it vanishes on shell, i.e., upon applying the background EoMs.) As well explained in \cite{Baumann:2011dt}, two things could go wrong: for subhorizon fluctuations, a ratio $|\mathcal{L}^{(3)}/\mathcal{L}^{(2)}|\gtrsim 1$ is indicative of a quantum strong coupling problem, with unsuppressed loop corrections and potential violation of unitarity; while for superhorizon fluctuations, a ratio $|\mathcal{L}^{(3)}/\mathcal{L}^{(2)}|\gtrsim 1$ is indicative of large classical nonlinearities, which may spoil the analysis of the linear perturbations and perhaps even of the background. In the same spirit, $|\mathcal{L}^{(2)}/\mathcal{L}^{(0)}|\gtrsim 1$ would signify a breakdown of perturbativity. While this would have nothing to do with interactions being strongly coupled, we nevertheless abuse terminology and refer to scales where $|\mathcal{L}^{(2)}/\mathcal{L}^{(0)}|\sim 1$ as strong coupling scales. A proper analysis of the quantum strong coupling scale in cosmological contexts involves the computation of loops or nonlinear corrections to various observables to determine at what point they become substantial. As a sensible alternative, one might instead check whether these corrections are significant within the $S$-matrix (e.g., \cite{deRham:2017aoj,Ageeva:2022byg,Cai:2022ori,Ageeva:2022asq}; see \cite{DuasoPueyo:2024rsa} for yet another alternative involving entanglement), but a rough estimate can already be found by estimating $\mathcal{L}^{(2)}/\mathcal{L}^{(0)}$ and especially $\mathcal{L}^{(3)}/\mathcal{L}^{(2)}$ (e.g., \cite{Joyce:2011kh,Koehn:2015vvy,Ageeva:2018lko,Ageeva:2020gti,Ageeva:2020buc,Ageeva:2021yik,Ageeva:2022fyq,Ageeva:2022asq,Akama:2022usl}). Moreover, since our primary goal is to test the physics of the Cuscuton in the bounce phase, we will be mostly concerned with the classical strong coupling problem for superhorizon perturbations (which are of observational interest), but we will comment on the subhorizon regime too. Generally speaking, the goal is to find the regime of validity of the model at hand, which from an effective field theory perspective may come with UV (and potentially IR) cutoffs.

A comparison of perturbed Lagrangians as described above can only make sense when performed with real-space perturbations, because cosmological perturbations (through the perturbative expansion of the metric and the matter fields) are defined in real space; cf.~\eqref{eq:ADMpert} and \eqref{eq:fieldspert}. Let us denote either $\zeta$ or $\chi$ by the variable $\xi$. Note that since these variables are technically quantum operators, $\hat{\xi}$, they are not deterministic functions of $\bm{x}$ in real space; instead their expectation values manifest themselves as stochastic fields. However, defining $\bar{\xi}(t;\mathscr{R})$ as the standard deviation of the coarse-grained $\hat{\xi}$ with some Window function $\mathcal{W}*\hat{\xi}(t,\bm{x})$ over some radius $\mathscr{R}$ enables us to quantify the real-space `size' (i.e., amplitude) of the coarse-grained $\hat{\xi}$ (and likewise for the derivatives thereof). Recalling \eqref{eq:zetaReal}, we can evaluate $\bar{\xi}(t;\mathscr{R})$ in terms of the Fourier-space mode function $\xi_k$ and Fourier transform of $\mathcal{W}$ as\footnote{Since acting on a function that depends only on the magnitude of $\bm{k}$, note
\[
    \int_{\bm{k}}\mathcal{\widetilde{W}}^2(k\mathscr{R})|\xi_k(t)|^2=\int_0^\infty\frac{\dd k~4\pi k^2}{(2\pi)^3}\,\mathcal{\widetilde{W}}^2(k\mathscr{R})|\xi_k(t)|^2=\int_0^\infty\frac{\dd k}{k}~\mathcal{\widetilde{W}}^2(k\mathscr{R})\mathcal{P}_\xi(k,t)\,,
\]
where we recall $\mathcal{P}_\xi(k,t)\equiv k^3|\xi_k(t)|^2/(2\pi^2)$.}
\begin{equation}
    \bar{\xi}(t;\mathscr{R})\equiv\sqrtb{\int_{\bm{k}}\mathcal{\widetilde{W}}^2(k\mathscr{R})|\xi_k(t)|^2}\,,\qquad\bar{\dot\xi}(t;\mathscr{R})\equiv\sqrtb{\int_{\bm{k}}\mathcal{\widetilde{W}}^2(k\mathscr{R})|\dot\xi_k(t)|^2}\,.
\end{equation}
Ignoring $\mathcal{O}(1)$ numerical factors, estimating the square of the Window function in Fourier space using a delta function as in \eqref{eq:Windowdelta}, and denoting the smoothing comoving scale at which we evaluate the `size' of the perturbation $\xi$ as $\mathscr{R}=1/k_\mathscr{R}$, we have
\begin{equation}
    \xi(t;1/k_\mathscr{R})\sim k_\mathscr{R}^{3/2}|\xi_{k_\mathscr{R}}(t)|\,,\qquad\dot\xi(t;1/k_\mathscr{R})\sim k_\mathscr{R}^{3/2}|\dot\xi_{k_\mathscr{R}}(t)|\,,\label{eq:xiW}
\end{equation}
where to simplify the notation the bar indicating the coarse-grained estimates is omitted from this point onward.
From this, we can similarly estimate the size of spatial derivatives as
\begin{equation}
    \partial\xi(t;1/k_\mathscr{R})\sim k_\mathscr{R}\xi(t;1/k_\mathscr{R})\,,\label{eq:pxiW}
\end{equation}
and so on. For instance, if $\xi_k$ is related to its Mukhanov-Sasaki variable $\upsilon_k$ through a dimensionless function $f_\xi$ as $\upsilon_k=\mpl af_\xi\xi_k$, then for $k_\mathscr{R}$ in the far UV we expect for an adiabatic vacuum\footnote{Note that the time variation of $a$ and $f_\xi$ should be such that the WKB approximation holds in the UV limit, so that a well-defined adiabatic vacuum for $\upsilon_k$ in that regime exists. In most scenarios in the literature, this is ensured by construction using the generalized `slow-roll conditions'; otherwise one may seek more sophisticated transformations to check if a proper canonical variable exists to set the adiabatic vacuum conditions \cite{Geshnizjani:2013lza}. Also, \eqref{eq:UV_est_xi} really holds only if $f_\xi\neq 0$ and as long as it has little $k$ dependence (subleading in the UV).} with Lorentzian dispersion relation $\omega_k\sim k$ ($\upsilon_k\sim k^{-1/2}$ and $\dot\upsilon_k\sim \omega_k\upsilon_k\sim k^{1/2}$)
\begin{equation}
    \xi\sim\frac{k_\mathscr{R}/a}{\mpl f_\xi}\,,\qquad\dot\xi^2\sim\frac{(\partial_i\xi)^2}{a^2}\sim\frac{(k_\mathscr{R}/a)^4}{\mpl^2f_\xi^2}\,.\qquad\textrm{(UV)}\label{eq:UV_est_xi}
\end{equation}
As another example, when $\xi$ represents $\chi$ which has a scale-invariant power spectrum on superhorizon scales in the prebounce phase [as in \eqref{eq:Pchikek} and \eqref{eq:Pchimassless2} in the limit $\delta=0$], then for $k_\mathscr{R}$ in the far IR we find
\begin{equation}
    \xi\sim\sqrtb{A_s}\,,\qquad\dot\xi\sim\frac{(k_\mathscr{R}/a)^2}{|H|}\sqrtb{A_s}\,,\qquad\frac{\partial\xi}{a}\sim\frac{k_\mathscr{R}}{a}\sqrtb{A_s}\,.\qquad\textrm{(IR)}\label{eq:IR_est_s-i}
    \end{equation}
Such estimates will be used repeatedly in what follows.

\subsubsection{Background}

Let us start by estimating the relevant energy scales of the background cosmology (see, e.g., \cite{deRham:2017aoj}),
\begin{equation}
    E_\mathrm{back}(t)\sim\max\bigg\{\left|H\right|,\big|\dot H\big|^{1/2},\bigg|\frac{\ddot\psi}{\dot\psi}\bigg|,\cdots\bigg\}\,. \label{eq: Eback}
\end{equation}
We find $E_\mathrm{back}(t)\sim |H(t)|\sim 1/|t|$ before the bounce phase, that is, for $t\lesssim t_{\mathrm{b}-}\sim 1/|H_{\mathrm{b}-}|$. Thus, the maximal energy scale before the bounce is $|H_{\mathrm{b}-}|$. Using the potential \eqref{eq:U_phi} as an example, one estimates\footnote{The points where $\dot H=0$ can be found in terms of $\bar\varphi$ by finding the roots of $U_{,\varphi\varphi}$ [recall \eqref{eq:cuscConstrDeriv}]. For the potential \eqref{eq:U_phi}, those points can be expressed in terms of the Lambert $W$ function as
\[
    \bar\varphi=\pm\frac{m}{2}\sqrtb{-1+2W\left(\frac{1}{2}\exp[\frac{1}{2}+\frac{2\varphi_\infty^2}{m^2}]\right)}\,.
\]
Then, the value of $H$ at those points is found through \eqref{eq:backCuscutonConstraint}, and we assume $\varphi_\infty\gg m$.}
\begin{equation}
    |H_{\mathrm{b}-}|\approx\frac{m^2\varphi_\infty}{3\mu^2\mpl^2}\approx\frac{\mu^2}{3}\varphi_\infty\,,\label{eq:Hbmest}
\end{equation}
where the second approximation assumes $m\approx\mu^2\mpl$ [more about this later, below \eqref{eq:L2zetabIR}]. In order for the bounce to be semiclassical (sub-Planckian), we ask $|H_{\mathrm{b}-}|\ll\mpl$. Given that the potential \eqref{eq:U_phi} assumes $m/\mpl\leq\mu^2\ll\varphi_\infty/\mpl<1$ (see \cite{Boruah:2018pvq}), this is indeed the case. Under the same approximation, \eqref{eq:Upsilonpotentialparams} tells us that\footnote{This also shows that a minimal requirement for a successful bounce with $0<\Upsilon<\infty$ is generally that $0<m<\sqrtb{3/2}\mu^2\mpl$; see \cite{Boruah:2018pvq,Quintin:2019orx}. \label{footnote:breq}}
\begin{equation}
    \Upsilon\approx\frac{m^4}{\mpl^4}\frac{\varphi_\infty^2}{3\mu^4-2m^2/\mpl^2}\approx\mu^4\varphi_\infty^2\approx 9H_{\mathrm{b}-}^2\,,\label{eq:UpsilonHbmapprox}
\end{equation}
where the second approximation assumes $m\approx\mu^2\mpl$ and the third uses \eqref{eq:Hbmest}.
Hence, according to \eqref{eq:aHHpb}, we expect the relevant energy scale through the bounce, $|\dot H|^{1/2}\approx\Upsilon^{1/2}$, to remain roughly constant and of the same order as $|H_{\mathrm{b}-}|$.

\subsubsection{Linear perturbations}

\paragraph*{Before the bounce}
Recalling \eqref{eq:S2zetaS2chi} and below, deep in the contraction when $\epsilon\approx\epsphi\approx\mathrm{constant}$, we simply have $\mathcal{L}^{(2)}_\zeta\approx\epsphi(\dot\zeta^2-a^{-2}(\partial_i\zeta)^2)$ and $\mathcal{L}^{(2)}_\chi=(F/2)(\dot\chi^2-a^{-2}(\partial_i\chi)^2)$, and this is true on all scales. One can check the estimated coarse-grained size of $\mathcal{L}^{(2)}_\zeta$ and $\mathcal{L}^{(2)}_\chi$ as outlined in Sec.~\ref{sec:strongCouplingStrategy}. First in the IR, if we have ekpyrosis, then $|\mathcal{L}^{(2)}_\zeta|\sim k_\mathrm{p}^4/\mpl^2$ [using \eqref{eq:vk-zetak-ek-superHubble} and \eqref{eq:zetakpek} in the large $\bar\epsilon$ limit, as well as \eqref{eq:xiW} and \eqref{eq:pxiW}; also recall $k_\mathrm{p}\equiv k/a$ is the physical momentum]. If we have kination ($\psi$ is massless), then using \eqref{eq:zeta_super}
\begin{equation}
    |\mathcal{L}^{(2,\textrm{IR})}_\zeta|\sim\frac{k_\mathrm{p}^3H^2}{\mpl^2|H_{\mathrm{b}-}|}\,.\label{eq:L2zetaIRkincontr}
\end{equation}
And in either case in the IR using \eqref{eq:IR_est_s-i} we have
\begin{equation}
    |\mathcal{L}^{(2,\textrm{IR})}_\chi|\sim\frac{k_\mathrm{p}^2H^2}{\mpl^2}\,.\label{eq:L2chiIRcontr}
\end{equation}
In all cases, this confirms that $|\mathcal{L}^{(2)}|\ll|\mathcal{L}^{(0)}|\sim H^2$ before the bounce phase in the IR. In the UV for both $\zeta$ and $\chi$ (collectively denoted $\xi$), one finds using \eqref{eq:UV_est_xi}
\begin{equation}
    |\mathcal{L}_\xi^{(2,\textrm{UV})}|\sim\frac{k_\mathrm{p}^4}{\mpl^2}\,,\label{eq:L2UV}
\end{equation}
which upon comparing with $|\mathcal{L}^{(0)}|$ suggests a strong coupling scale of the order of $\Lambda_\mathrm{strong}\sim\sqrtb{|H|\mpl}$. Note that this is, in essence, only a statement about semiclassical GR since we assume that the Cuscuton is negligible before the bounce phase. However, as it is the Cuscuton that sets the maximal Hubble scale $|H_{\mathrm{b}-}|$ right at the onset of the bounce phase, this is already informative of the Cuscuton's strong coupling scale; using \eqref{eq:Hbmest}, it suggests
\begin{equation}
    \Lambda_\mathrm{strong}^\textrm{(Cuscuton)}\sim\sqrtb{|H_{\mathrm{b}-}|\mpl}\sim\frac{m}{\mu}\sqrtb{\frac{\varphi_\infty}{\mpl}}\sim\mu\mpl\sqrtb{\frac{\varphi_\infty}{\mpl}}\,,\label{eq:Estrong1}
\end{equation}
where we use $m\approx\mu^2\mpl$ in the last estimate.
Unsurprisingly, the resulting scale is proportional to $\mu\mpl=\mL$, which is the new mass scale that the Cuscuton introduced from the start in \eqref{eq:Lcuscintro}.
If we set $\varphi_\infty\sim\mpl$ for the sake of simplicity, then the strong coupling scale is solely controlled by $\mL$ as $\Lambda_\mathrm{strong}^\textrm{(Cuscuton)}\sim\mL$.
Generally we want $\mu^2\ll\varphi_\infty/\mpl<1$, so assuredly $\Lambda_\mathrm{strong}^\textrm{(Cuscuton)}<\mpl$, but the strong coupling scale is not necessarily orders of magnitude smaller than the Planck scale at this point. However, if $|H_{\mathrm{b}-}|$ is sufficiently small compared to $\mpl$, then we have the hierarchy $|H|\leq |H_{\mathrm{b}-}|\ll\Lambda_\mathrm{strong}^\textrm{(Cuscuton)}\sim\sqrtb{|H_{\mathrm{b}-}|\mpl}\ll\mpl$. This is indeed assured whenever $\sqrtb{m/\mpl}\lesssim\mu\ll\sqrtb{\varphi_\infty/\mpl}<1$.

\paragraph*{Bounce phase, isocurvature perturbations}
Let us now focus on the bounce phase, starting with $\chi$. We again have $\mathcal{L}^{(2)}_\chi=(F/2)(\dot\chi^2-a^{-2}(\partial_i\chi)^2)$, but now $F$ essentially becomes a constant through the bounce phase [recall the discussion below \eqref{eq:phipbasUb}]. As we saw around \eqref{eq:chiktaubp}, $\chi_k$ also remains more or less constant through the bounce phase on superhorizon scales, so one can estimate the size of $\mathcal{L}^{(2)}_\chi$ in the IR from its value at the onset of the bounce. Thus, we infer from \eqref{eq:L2chiIRcontr} that $|\mathcal{L}^{(2,\mathrm{IR})}_\chi|\sim (k_{\mathrm{p}}/\mpl)^2H_{\mathrm{b}-}^2$. Since $|\mathcal{L}^{(0)}|\sim H_{\mathrm{b}-}^2$, this confirms $|\mathcal{L}^{(2,\mathrm{IR})}_\chi|\ll|\mathcal{L}^{(0)}|$ through the bounce in the IR. Note that this is true regardless of whether the prebounce phase was ekpyrosis or kination. Likewise in the UV, $|\mathcal{L}^{(2)}_\chi|$ is the same as at the onset of the bounce, \eqref{eq:L2UV}, and the resulting strong coupling scale is given by \eqref{eq:Estrong1}, which remains constant through the bounce.

\paragraph*{Bounce phase, curvature perturbations}
Moving on to $\zeta$ through the bounce, let us start with the analysis on subhorizon scales. Since the action for $\zeta$ in the UV reduces to the usual `canonical' form [recall \eqref{eq:S2zetaUV}], one finds the same result for $\zeta$ as for $\chi$ regarding the UV strong coupling scale derived above. One may be concerned about the point where $H=0$ [recall the discussion after \eqref{eq:S2zetaUV}], but by doing the limits carefully, we confirm that the result is unchanged. In fact, taking $H\to 0$ before $k\to\infty$ in \eqref{eq:z-cs2-zs}, one finds $z\simeq\sqrtb{2/\sigma}(k/H)$, so the amplitude of the UV fluctuations at the bounce point is not given by \eqref{eq:UV_est_xi}, but rather by
\begin{equation}
    |\zeta^\mathrm{(UV)}|\stackrel{H\approx 0}{\sim}\frac{\sqrtb{\sigma}|H|}{\mpl}\,,\label{eq:zetaUVbouncePoint}
\end{equation}
where we recall that $\sqrtb{\sigma}|H|$ is $\mathcal{O}(H^0)$. Combining this and \eqref{eq:S2zetaUVH0}, which is the UV action about $H=0$, one recovers \eqref{eq:L2UV}. In passing, we have to recall from \eqref{eq:eps_app} and \eqref{eq:mplt} that $\sigma/\epsphi=1-\epsilon/\epsphi\approx 1-\mpl^2/\mplt^2$ is $\mathcal{O}(1)$ about $H\approx 0$.

Away from the large-$k$ limit, the Lagrangian for $\zeta$ is generally quite complicated in the bounce phase, so let us explore the opposite regime, i.e., the IR limit given by \eqref{eq:S2zetaIR}, which is relevant for superhorizon modes of observational relevance. To approximate $\mathcal{L}^{(2,\mathrm{IR})}_\zeta$, we make use of the fact that we have previously carried out the series expansion for most of the required background quantities around the bounce point, as presented in \eqref{eq:aHHpb}, \eqref{eq:phib}, \eqref{eq:1oepspsib}, and \eqref{eq:epsoepsphib}. Now, let us evaluate the series expansion for the other necessary `slow-roll' parameters:
\begin{equation}
    \epsilon\stackrel{t\approx 0}{\simeq} -\frac{1}{\Upsilon t^2}+\mathcal{O}(t^0)\,,\qquad\eta\stackrel{t\approx 0}{\simeq}-\frac{2}{\Upsilon t^2}+\mathcal{O}(t^0)\,,\qquad\eta_\psi\stackrel{t\approx 0}{\simeq}-\frac{2}{\Upsilon t^2}+\mathcal{O}(t^0)\,.\label{eq:epsetaetapsiH0}
\end{equation}
Applying this to \eqref{eq:S2zetaIR}, one finds to leading order
\begin{align}
    \mathcal{L}^{(2,\mathrm{IR})}_\zeta\stackrel{H\approx 0}{\simeq}\frac{3\dot\psi^{2}_\mathrm{b}}{2\mpl^2\Upsilon+\dot\psi^{2}_\mathrm{b}}\dot\zeta^2-\frac{4\mpl^2\Upsilon-\dot\psi^{2}_\mathrm{b}}{2\mpl^2\Upsilon+\dot\psi^{2}_\mathrm{b}}\frac{(\partial_i\zeta)^2}{a^2}
    &\approx\left(3-\frac{2m^2}{\mu^4\mpl^2}\right)\left(\dot\zeta^2-\frac{2m^2-\mu^4\mpl^2}{3\mu^4\mpl^2-2m^2}\frac{(\partial_i\zeta)^2}{a^2}\right)\nonumber\\
    &\approx\dot\zeta^2-\frac{(\partial_i\zeta)^2}{a^2}\,,\label{eq:L2zetabIR}
\end{align}
where the first approximation is for IR perturbations about $t\approx 0$ (equivalently $H\approx 0$), the second approximation assumes the potential \eqref{eq:U_phi} with $m/\mpl\leq\mu^2\ll\varphi_\infty/\mpl$ [and uses \eqref{eq:phipbasUb} and \eqref{eq:Upsilonpotentialparams}], and the last approximation is when $m\approx\mu^2\mpl$. Note that \eqref{eq:L2zetabIR} makes it clear why $m\approx\mu^2\mpl$ is a sensible assumption since the Lagrangian becomes canonical with unit sound speed; indeed, \cite{Quintin:2019orx} showed that the theory behaves best when $m\approx\mu^2\mpl$, that is, the sound speed remains closest to unity across time and $k$-space. In \cite{Quintin:2019orx}, it was also observed that to maintain a positive coefficient for the gradient term (i.e., the sound speed\footnote{While a negative sound speed squared would not necessarily be a sign of a gradient instability since this is in the IR as opposed to the UV, it may still be an undesirable feature. We will in fact see later that $m\to 0^+$ could lead to the production of arbitrarily large non-Gaussianities through the bounce.\label{foot:negcs2IRm0}} squared) and to ensure that the sound speed remains subluminal, the condition
\begin{equation}
    \frac{1}{2}<\frac{m^2}{\mu^2\mpl^2}\leq 1\label{eq:gammaRange}
\end{equation}
must be satisfied. Therefore, \cite{Quintin:2019orx} anticipated that similar constraints should likely be applied to prevent strong coupling issues.

From \eqref{eq:L2zetabIR}, if the superhorizon near-scale-invariant $\chi$ perturbations are converted to $\zeta$ perturbations just before the onset of the bounce, then as was the case for $\chi$ before, we conclude that $|\mathcal{L}_\zeta^{(2,\mathrm{IR})}|\sim (k_{\mathrm{p}}/\mpl)^2H_{\mathrm{b}-}^2$ on superhorizon scales through the bounce. If the conversion occurs after the bounce one finds on superhorizon scales through the bounce $|\mathcal{L}_\zeta^{(2,\mathrm{IR})}|\sim k_{\mathrm{p}}^4/\mpl^2$ or $|\mathcal{L}_\zeta^{(2,\mathrm{IR})}|\sim k_{\mathrm{p}}^3|H_{\mathrm{b}-}|/\mpl^2$ depending on whether the prebounce phase is ekpyrosis or kination, respectively. In all cases, $|\mathcal{L}_\zeta^{(2,\mathrm{IR})}|\ll|\mathcal{L}^{(0)}|\sim H_{\mathrm{b}-}^2$ for $k_\mathrm{p}\ll|H_{\mathrm{b}-}|$.

\subsubsection{Interactions}\label{sec:strongCouplingInteractions}

Moving on to evaluating $\mathcal{L}^{(3)}/\mathcal{L}^{(2)}$, we first notice from the derivation of the third-order action in Sec.~\ref{sec:3rdorder} that $\mathcal{L}^{(3)}$ separates into terms that represent $\zeta\chi^2$ interactions [what we called $\mathcal{L}^{(3)}_{(\chi)}$ in \eqref{eq:L3chir}; equivalently $\mathcal{L}_{\zeta\chi\chi}$, $\mathcal{L}_{\zeta\chi^2}$ or $\mathcal{L}_{\chi\chi\zeta}$ henceforth] and terms that represent $\zeta^3$ interactions [which corresponds to $\mathcal{L}^{(3)}_{(\delta\varphi)}+\mathcal{L}^{(3)}_\textrm{(rest)}$ of \eqref{eq:L3deltavarphiallr} and \eqref{eq:L3restr}; $\mathcal{L}_{\zeta^3}$ or $\mathcal{L}_{\zeta\zeta\zeta}$ henceforth]. Therefore, there are two meaningful ratios to evaluate: the ratio of the most dominant terms in $\mathcal{L}_{\zeta\chi\chi}$ over the most dominant terms in $\mathcal{L}^{(2)}_{\chi}$ ($\mathcal{L}_{\chi\chi}$ henceforth); and the ratio of the most dominant terms in $\mathcal{L}_{\zeta\zeta\zeta}$ over the most dominant terms in $\mathcal{L}^{(2)}_{\zeta}$ ($\mathcal{L}_{\zeta\zeta}$ henceforth).

\paragraph*{Before the bounce}
Before the bounce, the Cuscuton is negligible, so from \eqref{eq:L3chir} we have as an order-of-magnitude estimate of the size of the terms involved in $\mathcal{L}_{\zeta\chi\chi}$ (ignoring signs and numerical constants),
\begin{align}
    \mathcal{L}_{\zeta\chi\chi}\sim\left(\left(1-\frac{XF_{,X}}{F}\right)\frac{\dot\zeta}{H}+\zeta\right)\mathcal{L}_{\chi\chi}+F\left(\frac{\partial_i\zeta}{aH}-a\partial_i\partial^{-2}\dot\zeta\right)\frac{\partial^i\chi}{a}\dot\chi\,,\label{eq:Lzetachichiorder}
\end{align}
and likewise $\mathcal{L}_{\zeta\zeta\zeta}$ is approximately given by \eqref{eq:L3GR}. In the UV, one can verify that $|\mathcal{L}_{\zeta\zeta\zeta}/\mathcal{L}_{\zeta\zeta}|\sim |\zeta|$ and therefore the strong coupling scale at which $|\mathcal{L}_{\zeta\zeta\zeta}|\sim|\mathcal{L}_{\zeta\zeta}|$ corresponds to the Planck scale (when $|\zeta|\sim 1$). Meanwhile, $|\mathcal{L}_{\zeta\chi\chi}/\mathcal{L}_{\chi\chi}|$ is dominated by terms such as $\dot\zeta/H\sim k_\mathrm{p}^2/(|H|\mpl)$, so the strong coupling scale, similar to the case of $\mathcal{L}^{(2)}/\mathcal{L}^{(0)}$ discussed after \eqref{eq:L2UV}, is expected to be of the order of $\sqrtb{|H|\mpl}$. In the IR, using \eqref{eq:zeta_super} for $\zeta$, we estimate
\begin{equation}
    |\zeta|\sim\frac{k_\mathrm{p}^{3/2}}{\mpl\sqrtb{|H_{\mathrm{b}-}|}}\ln\left(\frac{|H|}{k_\mathrm{p}}\right)\,,\qquad|\dot\zeta|\sim\frac{k_\mathrm{p}^{3/2}|H|}{\mpl\sqrtb{|H_{\mathrm{b}-}|}}\,,\qquad\frac{|\partial\zeta|}{a}\sim k_\mathrm{p}|\zeta|\,,\label{eq:zetaIRpb}
\end{equation}
and similarly using \eqref{eq:chi_anysol} for $\chi$, \eqref{eq:IR_est_s-i} follows. This implies $|\mathcal{L}_{\zeta\chi\chi}/\mathcal{L}_{\chi\chi}|\sim|\mathcal{L}_{\zeta\zeta\zeta}/\mathcal{L}_{\zeta\zeta}|\sim|\zeta|\ll 1$ as desired.

\paragraph*{Through the bounce in the IR}
The IR limit ($k/a\ll |H|$) of the Cuscuton constraint \eqref{eq:deltacusc} gives
\begin{equation}
    \delta\varphi\stackrel{\textrm{(IR)}}{\simeq}\frac{2\sigma}{\mu^2(\sigma+3)}\dot\zeta\,.
\end{equation}
Upon substituting in \eqref{eq:L3deltavarphiallr} and \eqref{eq:L3restr}, the dominant terms in $\mathcal{L}_{\zeta\zeta\zeta}$ are of the form $\zeta\dot\zeta^2$ and $\dot\zeta^3$, so one finds
\begin{equation}
    \frac{\mathcal{L}_{\zeta\zeta\zeta}}{\mathcal{L}_{\zeta\zeta}}\stackrel{\textrm{(IR)}}{\simeq}\frac{9H(2\epsilon^2-\epsilon(12+5\epsphi)+3(6+5\epsphi+\epsphi^2))\zeta-2(2\epsilon^2+9(3+\epsphi)-\epsilon(15+\eta))\dot\zeta}{6H(3-\epsilon+\epsphi)^2}\,.\label{eq:Lzeta3oLzeta2IRb}
\end{equation}
Near the beginning or end of the bounce phase when $|H|\approx|H_{\mathrm{b}-}|$, $\epsilon\approx 0$, and $\sqrtb{\epsphi}\sim\mathcal{O}(1)$, this results in $|\mathcal{L}_{\zeta\zeta\zeta}/\mathcal{L}_{\zeta\zeta}|\sim|\zeta|\sim\sqrtb{A_s}\ll 1$, assuming $\chi$ converts to curvature perturbations before the bounce. If the conversion occurs after the bounce, it results in
\begin{equation}
    \left|\frac{\mathcal{L}_{\zeta\zeta\zeta}}{\mathcal{L}_{\zeta\zeta}}\right|\stackrel{\textrm{(IR)}}{\sim}|\zeta|\sim\frac{k_\mathrm{p}^{3/2}}{\mpl\sqrtb{|H_{\mathrm{b}-}|}}\ln\left(\frac{|H_{\mathrm{b}-}|}{k_\mathrm{p}}\right)\,,\label{eq:L3zoL2zbIRe}
\end{equation}
which is also very small in the IR. In both cases, we make use of the fact that $\zeta$ receives marginal growth through the bounce phase in accordance with \eqref{eq:zetakscaleinvariantthroughbounce} and \eqref{eq:zetakbouncemassless}, respectively. Furthermore, if entropy perturbations pass through the bounce before being converted, then they also receive marginal growth according to \eqref{eq:chiktaubp}, and we find $|\mathcal{L}_{\zeta\chi\chi}/\mathcal{L}_{\chi\chi}|\sim|\zeta|$, which is suppressed according to \eqref{eq:L3zoL2zbIRe}.

Note that \eqref{eq:Lzeta3oLzeta2IRb} is not applicable too close to the bounce point where $H\approx 0$ since, firstly, it appears to diverge in that region and since, secondly, superhorizon IR modes (while still satisfying $k\ll E_{\textrm{back}}$ [see \eqref{eq: Eback}]) have crossed into the $k/a \gtrsim |H|$ regime. Therefore, as mentioned before, extra caution is necessary when taking any limits in the vicinity of $H=0$. Recalling the expansions \eqref{eq:deltavarphiH0} and \eqref{eq:alphabetaH0} near the bounce point, substituting them in \eqref{eq:L30}, using \eqref{eq:epsetaetapsiH0} and \eqref{eq:1oepspsib}, and taking an IR limit yields instead
\begin{equation}
    \mathcal{L}^\mathrm{(IR)}_{\zeta\zeta\zeta}\stackrel{H\approx 0}{\simeq}\frac{9\dot\psi_\mathrm{b}^2(4\mpl^2\Upsilon+3\dot\psi_\mathrm{b}^2)}{2(2\mpl^2\Upsilon+\dot\psi_\mathrm{b}^2)^2}\zeta\dot\zeta^2\approx\left(\frac{27}{2}-\frac{12m^2}{\mu^4\mpl^2}+\frac{2m^4}{\mu^8\mpl^4}\right)\zeta\dot\zeta^2\,.\label{eq:Lzeta3IRH0}
\end{equation}
Upon comparing with the dominant term of \eqref{eq:L2zetabIR}, we conclude that
\begin{equation}
    \left|\frac{\mathcal{L}_{\zeta\zeta\zeta}}{\mathcal{L}_{\zeta\zeta}}\right|\sim\left(\frac{9}{2}-\frac{m^2}{\mu^4\mpl^2}\right)|\zeta|\ll 1\qquad(H\approx 0,\ \textrm{IR})
\end{equation}
holds when $H$ passes through zero.

\paragraph*{Through the bounce in the UV}
Similar to the IR, we have to be careful when taking limits through the bounce phase in the UV. For example, as long as $H\neq 0$, the UV limit of the Cuscuton constraint \eqref{eq:deltacusc} gives $\mu^2\delta\varphi\simeq 2\sigma H\zeta$, so the dominant terms in the respective third-order Lagrangians are schematically of the form
\begin{equation}
\label{eq:dom-term-s3}
    \mathcal{L}_{\zeta\zeta\zeta}\stackrel{\textrm{(UV)}}{\sim}\frac{\epsilon_\psi}{H^2}\nabla^4\zeta^3\,,\qquad\mathcal{L}_{\zeta\chi\chi}\stackrel{\textrm{(UV)}}{\sim}\frac{F}{|H|}\nabla^3(\zeta\chi^2)\,,
\end{equation}
where we recall $\nabla$ indicates either a spatial or time derivative here (as they scale equally in the UV).
For $\mathcal{L}_{\zeta\chi^2}$, such terms can be seen in \eqref{eq:L3chir}, and for $\mathcal{L}_{\zeta\zeta\zeta}$ they can be seen in \eqref{eq:L3deltavarphi2} and \eqref{eq:L3restr}, although recall that in this case they may amount to boundary terms in the formal UV limit $k\to\infty$. Since these expressions are invalid in the vicinity of the bounce point (indeed, they appear to diverge as $H\to 0$), they are only applicable at the very beginning (or very end) of the bounce phase, when $\sqrtb{\epsphi}$ is of order unity and $|H|\approx|H_{\mathrm{b}-}|$. Correspondingly,
\begin{equation}
\label{eq:3/2ratio_subH}
    \left|\frac{\mathcal{L}_{\zeta\zeta\zeta}}{\mathcal{L}_{\zeta\zeta}}\right|\stackrel{\textrm{(UV)}}{\sim}\frac{|\nabla^2\zeta|}{H_{\mathrm{b}-}^2}\sim\frac{k_\mathrm{p}^3}{H_{\mathrm{b}-}^2\mpl}\,,\qquad\left|\frac{\mathcal{L}_{\zeta\chi\chi}}{\mathcal{L}_{\chi\chi}}\right|\stackrel{\textrm{(UV)}}{\sim}\frac{|\nabla\zeta|}{|H_{\mathrm{b}-}|}\sim\frac{k_\mathrm{p}^2}{|H_{\mathrm{b}-}|\mpl}\,.
\end{equation}
The latter ratio yields the same strong coupling scale as \eqref{eq:Estrong1}, but the former ratio points toward a smaller strong coupling scale of the order of
\begin{equation}
    \Lambda_\mathrm{strong}^\textrm{(Cuscuton)}\sim(H_{\mathrm{b}-}^2\mpl)^{1/3}\sim\left(\frac{m}{\mu\mpl}\right)^{4/3}\left(\frac{\varphi_\infty}{\mpl}\right)^{2/3}\mpl\sim\left(\frac{\mL}{\mpl}\right)^{4/3}\mpl\,,\label{eq:Lambdas23}
\end{equation}
where the last estimate uses $m\sim\mu^2\mpl$, $\varphi_\infty\sim\mpl$, and $\mu=\mL/\mpl$. While $(H_{\mathrm{b}-}^2\mpl)^{1/3}<(|H_{\mathrm{b}-}|\mpl)^{1/2}$, we still have $|H_{\mathrm{b}-}|\ll(H_{\mathrm{b}-}^2\mpl)^{1/3}$ provided $|H_{\mathrm{b}-}|$ is sufficiently small compared to $\mpl$, so such a strong coupling scale would not necessarily represent an immediate strong coupling issue for the model.

Examining the vicinity of $H\approx 0$ as discussed before [using \eqref{eq:deltavarphiH0}, \eqref{eq:epsetaetapsiH0}, and \eqref{eq:1oepspsib} in \eqref{eq:L30}], one finds no divergent terms in $\mathcal{L}_{\zeta\zeta\zeta}$ and $\mathcal{L}_{\zeta\chi\chi}$, and a full estimate with the same approximations yields
\begin{align}
    \left|\frac{\mathcal{L}_{\zeta\zeta\zeta}}{\mathcal{L}_{\zeta\zeta}}\right|&\sim\frac{\mu^6\mpl^3k_\mathrm{p}^2}{m\varphi_\infty(2m^2+3\mu^4\mpl^2)\sqrtb{3\mu^4\mpl^2-2m^2}}\,,\qquad(H\approx 0,\ \textrm{UV})\nonumber\\
    \left|\frac{\mathcal{L}_{\zeta\chi\chi}}{\mathcal{L}_{\chi\chi}}\right|&\sim\frac{\mu^2\mpl k_\mathrm{p}^2}{m\varphi_\infty\sqrtb{3\mu^4\mpl^2-2m^2}}\,.\qquad(H\approx 0,\ \textrm{UV})
\end{align}
The corresponding strong coupling scales are
\begin{align}
    \Lambda_\mathrm{strong}^\textrm{(Cuscuton)}&\sim\sqrtb{\frac{m\varphi_\infty}{\mu^6\mpl^3}(3\mu^4\mpl^2+2m^2)\sqrtb{3\mu^4\mpl^2-2m^2}}\,,\nonumber\\
    \Lambda_\mathrm{strong}^\textrm{(Cuscuton)}&\sim\sqrtb{\frac{m\varphi_\infty}{\mu^2\mpl}\sqrtb{3\mu^4\mpl^2-2m^2}}\,,
\end{align}
and using \eqref{eq:UpsilonHbmapprox}, they can be further expressed as
\begin{align}
    \frac{\Lambda_\mathrm{strong}^\textrm{(Cuscuton)}}{\sqrtb{\Upsilon}}&\sim\sqrtb{\frac{3\mu^4\mpl^2+2m^2}{\mu^6\mpl m^3\varphi_\infty}}(3\mu^4\mpl^2-2m^2)^{3/4}\,,\nonumber\\
    \frac{\Lambda_\mathrm{strong}^\textrm{(Cuscuton)}}{\sqrtb{\Upsilon}}&\sim\sqrtb{\frac{\mpl}{\mu^2m^3\varphi_\infty}}(3\mu^4\mpl^2-2m^2)^{3/4}\,.\label{eq:scCf}
\end{align}
Since $E_\mathrm{back}\sim\sqrtb{\Upsilon}$ when $H\approx 0$, we must ask for the above ratios to be very large for there to be no strong coupling issue. This cannot be achieved in the limit $m^2\to(3/2)\mu^4\mpl^2$, but in the same limit no bounce can occur at the background level in the first place --- recall either Refs.~\cite{Boruah:2018pvq,Quintin:2019orx}, eq.~\eqref{eq:UpsilonHbmapprox} or footnote \ref{footnote:breq}. Therefore, if we instead demand $m^2\leq\mu^4\mpl^2$ as before, we find in both cases
\begin{equation}
    \frac{\Lambda_\mathrm{strong}^\textrm{(Cuscuton)}}{\sqrtb{\Upsilon}}\gtrsim\frac{1}{\mu}\sqrtb{\frac{\mpl}{\varphi_\infty}}\gg 1\,,
\end{equation}
hence there is no strong coupling issue provided $\mu\ll\sqrtb{\varphi_\infty/\mpl}<1$ as before. In fact making use of \eqref{eq:Hbmest}, the above implies a strong coupling scale of the same order as \eqref{eq:Estrong1} again, which goes as $\Lambda_\mathrm{strong}^\textrm{(Cuscuton)}/|H_{\mathrm{b}-}|\sim\sqrtb{\mpl/|H_{\mathrm{b}-}|}$.

\paragraph*{Summary}
To summarize this subsection, we showed that whenever the Cuscuton parameters satisfy $\sqrtb{m/\mpl}\leq\mu\ll\sqrtb{\varphi_\infty/\mpl}<1$, the Cuscuton model remains properly weakly coupled in the IR throughout time, notably through the bounce phase. The same is true in the UV ($k_\mathrm{p}\gg |H|$): the strong coupling scale is always larger than the background energy scale, as desired ($|H|\ll k_\mathrm{p}\ll\Lambda_\mathrm{strong}$). Indeed, in the contracting phase, $\Lambda_\mathrm{strong}\sim\sqrtb{|H|\mpl}$, which by the time the Cuscuton becomes important at the onset of the bounce is $\sqrtb{|H_{\mathrm{b}-}|\mpl}\sim\mL$. Through the bounce, though, the additional interactions the Cuscuton provides lower the strong coupling scale to $\Lambda_\mathrm{strong}^\textrm{(Cuscuton)}\sim(H_{\mathrm{b}-}^2\mpl)^{1/3}\sim(\mL/\mpl)^{4/3}\mpl$. However, as commented above, provided that $|H_{\mathrm{b}-}|$ is relatively small compared to $\mpl$, such a strong coupling scale does not imply an immediate strong coupling problem for the model. Additionally, as discussed in the previous subsection, the operators yielding this lower strong coupling scale might be removable by integration by parts perturbatively in $1/k$ in the UV. Nevertheless, this scale could be taken as a conservative lower bound, while $\Lambda_\mathrm{strong}\sim\sqrtb{|H_{\mathrm{b}-}|\mpl}$ might be more robust.

\subsection{Bispectrum calculations}\label{sec:bispectrum}

\subsubsection{Setup}

The strong coupling order-of-magnitude estimate from the previous subsection was rough in the sense that we used the 2-point correlation function to approximate the coarse-grained amplitude of the real-space fluctuations, and we used that to compare coefficients of the second- and third-order Lagrangians. We can improve this estimate by performing an actual computation of the higher correlation functions. Specifically in this subsection, we will be interested in the 3-point function, represented as the bispectrum in Fourier space.

At the same time as being of theoretical relevance for the strong coupling issue, the bispectrum (like other correlation functions in Fourier space) is an observable that we can measure or constrain through cosmological observations (see, e.g., \cite{Bartolo:2004if,Planck:2019kim}), which serves as additional motivation for its computation. A nonzero 3-point function is synonymous with a deviation from Gaussianity, which if local in real space can parametrized as (we omit the time dependence here)
\begin{align}
    \zeta(\bm{x})&=\zeta_\G(\bm{x})+f_\NL^\zeta\left(\zeta_\G(\bm{x})^2-\langle\zeta_\G(\bm{x})^2\rangle\right)+C_\NL^\zeta\zeta_\G(\bm{x})\chi_\G(\bm{x})+\ldots\,,\nonumber\\
    \chi(\bm{x})&=\chi_\G(\bm{x})+f_\NL^\chi\left(\chi_\G(\bm{x})^2-\langle\chi_\G(\bm{x})^2\rangle\right)+C_\NL^\chi\chi_\G(\bm{x})\zeta_\G(\bm{x})+\ldots\,,\label{eq:realSpaceLNonGdef}
\end{align}
where the subscript {\small `G'} indicates that this is the Gaussian part of the full perturbation (assuming no interaction) and where the ellipses denote further (third-order) deviations from Gaussianity (which would start appearing at the level of the four-point function). Note that there are different conventions in the literature, which may differ from \eqref{eq:realSpaceLNonGdef}, the more standard one being of the form $\zeta=\zeta_\G-(3/5)f_\NL^\zeta\zeta_\G^2-\ldots$ (e.g., \cite{Komatsu:2001rj,Maldacena:2002vr}). In the above, $f_\NL^\zeta$, $f_\NL^\chi$, $C_\NL^\zeta$, and $C_\NL^\chi$ have the interpretation of local non-Gaussianity parameters, but shortly we will consider non-Gaussianities beyond the locality assumption. From \eqref{eq:realSpaceLNonGdef}, one can compute the various 3-point functions in Fourier space involving $\zeta$ and $\chi$ (see Appendix \ref{app:3-pt} for a derivation). We define the general bispectrum $B$ and the dimensionless bispectrum $\tilde B$ according to
\begin{align}
\label{eq:bispec-def}
    \langle\xi^I_{\bm{k}_1}\xi^J_{\bm{k}_2}\xi^K_{\bm{k}_3}\rangle&=(2\pi)^3\delta^{(3)}(\bm{k}_1+\bm{k}_2+\bm{k}_3)B_{\xi^I\xi^J\xi^K}(\bm{k}_1,\bm{k}_2,\bm{k}_3)\,,\nonumber\\
    \qquad\tilde B_{\xi^I\xi^J\xi^K}(\bm{k}_1,\bm{k}_2,\bm{k}_3)&=\frac{k_1^3k_2^3k_3^3}{k_1^3+k_2^3+k_3^3}B_{\xi^I\xi^J\xi^K}(\bm{k}_1,\bm{k}_2,\bm{k}_3)\,,
\end{align}
where $\xi^I=(\zeta,\chi)$, $I,J,K\in\{1,2\}$. We then use the calculation of the 3-point functions for local non-Gaussianities to define the general non-Gaussianity parameters according to
{\allowdisplaybreaks
\begin{align}
    f_\NL^\zeta(\bm{k}_1,\bm{k}_2,\bm{k}_3)&=\frac{B_{\zeta\zeta\zeta}(\bm{k}_1,\bm{k}_2,\bm{k}_3)}{2\left(P_\zeta(k_1)P_\zeta(k_2)+P_\zeta(k_2)P_\zeta(k_3)+P_\zeta(k_1)P_\zeta(k_3)\right)}\,,\nonumber\\
    C_\NL^\zeta(\bm{k}_1,\bm{k}_2,\bm{k}_3)&=\frac{B_{\zeta\zeta\chi}(\bm{k}_1,\bm{k}_2,\bm{k}_3)}{\left(P_\zeta(k_1)+P_\zeta(k_2)\right)P_\chi(k_3)}\,,\nonumber\\
    f_\NL^\chi(\bm{k}_1,\bm{k}_2,\bm{k}_3)&=\frac{B_{\chi\chi\chi}(\bm{k}_1,\bm{k}_2,\bm{k}_3)}{2\left(P_\chi(k_1)P_\chi(k_2)+P_\chi(k_2)P_\chi(k_3)+P_\chi(k_1)P_\chi(k_3)\right)}\,,\nonumber\\
    C_\NL^\chi(\bm{k}_1,\bm{k}_2,\bm{k}_3)&=\frac{B_{\chi\chi\zeta}(\bm{k}_1,\bm{k}_2,\bm{k}_3)}{\left(P_\chi(k_1)+P_\chi(k_2)\right)P_\zeta(k_3)}\,,\label{eq:deffCNL}
\end{align}
}%
where
\begin{equation}
    \langle\xi^I_{\bm{k}}\xi^J_{\bm{q}}\rangle=(2\pi)^3\delta^{(3)}(\bm{k}+\bm{q})\delta^{IJ}P_{\xi^I}(k)\,,\qquad P_{\xi^I}(k)=|\xi_k^I|^2=\frac{2\pi^2}{k^3}\mathcal{P}_{\xi^I}(k)\,, \label{eq:powersdef}
\end{equation}
are the relations between the 2-point function in Fourier space, the power spectrum, and its dimensionless form. Note that $f_\NL^\zeta$ above reduces to the standard local expression, up to a factor of $-3/5$ due to the convention \cite{Komatsu:2001rj}.

The standard approach to compute the bispectrum at some time $\tau_\mathrm{e}$ is to use the in-in formalism with the interaction picture, according to which to leading order (recall our unconventional dimensions for Lagrangian densities)
\begin{equation}
    \langle\xi^I_{\bm{k}_1}(\tau_\mathrm{e})\xi^J_{\bm{k}_2}(\tau_\mathrm{e})\xi^K_{\bm{k}_3}(\tau_\mathrm{e})\rangle=-2\mpl^2\mathrm{Im}\langle\xi^I_{\bm{k}_1}(\tau_\mathrm{e})\xi^J_{\bm{k}_2}(\tau_\mathrm{e})\xi^K_{\bm{k}_3}(\tau_\mathrm{e})\int_{-\infty(1+i\varepsilon)}^{\tau_\mathrm{e}}\dd\tau\int\dd^3\bm{x}\,a^4\mathcal{L}^{(3)}_{\xi^I\xi^J\xi^K}\rangle\,;\label{eq:inindef}
\end{equation}
see, e.g., \cite{Chen:2010xka,Wang:2013zva} for reviews.
The notation above makes it explicit that, in order to compute, e.g., $\langle\zeta^3\rangle$, only terms in $\mathcal{L}_{\zeta\zeta\zeta}$ should matter; likewise to compute $\langle\zeta\chi^2\rangle$, we only need $\mathcal{L}_{\zeta\chi^2}$. The underlying reason is that, when expanding the quantum operators into mode functions and raising/lowering operators,
\begin{equation} 
    \hat\xi^I(\tau,\bm{x})=\int_{\bm{k}}\hat\xi_{\bm{k}}^I(\tau)e^{i\bm{k}\cdot\bm{x}}=\int_{\bm{k}}\left(\xi_k^I(\tau)\mathfrak{\hat a}_{\bm{k}}^Ie^{i\bm{k}\cdot\bm{x}}+\xi_k^I(\tau)^*\mathfrak{\hat a}_{\bm{k}}^{I\dagger}e^{-i\bm{k}\cdot\bm{x}}\right)\,,\label{eq:expanded_operators}
\end{equation}
the only nonvanishing commutators are
\begin{equation}
    [\mathfrak{\hat a}_{\bm{k}}^I,\mathfrak{\hat a}_{\bm{q}}^{J\dagger}]=(2\pi)^3\delta^{IJ}\delta^{(3)}(\bm{k}-\bm{q})\,.\label{eq:quantization}
\end{equation}
We temporarily used hats to explicitly denote quantum operators in the last two equations, but we omit them elsewhere.

As our third-order Lagrangian only contains $\zeta^3$ and $\zeta\chi^2$ interactions, the only nonvanishing 3-point correlators will be $\langle\zeta^3\rangle$ and $\langle\zeta\chi^2\rangle$; in other words, $\langle\chi^3\rangle=\langle\zeta^2\chi\rangle=0$. One could wonder how robust the vanishing of those bispectra is. As long as $\chi$ solely enters the action at the perturbation level with no contribution to the background evolution, with a standard kinetic term, and as long as no potential is generated for $\chi$, this seems to be robust, because the perturbation of the $\chi$ Lagrangian to $n$th order will always be of the form $\mathcal{L}_\chi^{(n)}\sim\mathcal{O}(\zeta^{n-2}\chi^2)$. However, the stability of $\chi$ to higher-order corrections and whether a potential term could appear due to, e.g., quantum loop corrections (which are intrinsically linked to the question of stability) is beyond the scope of this work. Conservatively, it appears $B_{\chi\chi\chi}\approx B_{\zeta\zeta\chi}\approx 0$ (hence $f_\NL^\chi\approx C_\NL^\zeta\approx 0$) hold at least to leading order in our approximations, so our focus shall be on evaluating $B_{\zeta\zeta\zeta}$ and $B_{\chi\chi\zeta}$ (and the corresponding $f_\NL^\zeta$ and $C_\NL^\chi$ functions).\footnote{Let us stress that all of this applies ignoring the eventual conversion of isocurvature perturbations into curvature perturbations. We will comment on this separately later.}

One can intuitively understand how computing the remaining nonvanishing bispectra assesses strong coupling by inspecting the local-type non-Gaussianities as \eqref{eq:realSpaceLNonGdef}, which with $f_\NL^\chi\approx C_\NL^\zeta\approx 0$ suggest
\begin{equation}
    \frac{\zeta}{\zeta_\G}\sim 1+f_\NL^\zeta\zeta_\G\,,\qquad\frac{\chi}{\chi_\G}\sim 1+C_\NL^\chi\zeta_\G\,.
\end{equation}
Therefore, remaining under perturbative control (being weakly coupled in other words) amounts to having $|f_\NL^\zeta\zeta_\G|\ll 1$ and $|C_\NL^\chi\zeta_\G|\ll 1$. If we use our definition of $f_\NL^\zeta$ and $C_\NL^\chi$ from \eqref{eq:deffCNL} and evaluate them for $k_1=k_2=k_3\equiv k$, then the ratios we wish to check can be expressed as
\begin{equation}
\label{eq:fNLl-CNl_eqL}
    |f_\NL^\zeta\zeta_\G|\approx\frac{k^6|B_{\zeta\zeta\zeta}|}{24\pi^4\mathcal{P}_\zeta^{3/2}}=\frac{|\tilde B_{\zeta\zeta\zeta}|}{8
\pi^4\mathcal{P}_\zeta^{3/2}}\,,\qquad|C_\NL^\chi\zeta_\G|\approx\frac{k^6|B_{\chi\chi\zeta}|}{8\pi^4\mathcal{P}_\chi\mathcal{P}_\zeta^{1/2}}=\frac{3|\tilde B_{\chi\chi\zeta}|}{8\pi^4\mathcal{P}_\chi\mathcal{P}_\zeta^{1/2}}\,,
\end{equation}
where we approximated as before $|\zeta_\G|\approx\mathcal{P}_\zeta^{1/2}$.

Let us proceed with the calculation of $B_{\zeta\chi^2}(\bm{k}_1,\bm{k}_2,\bm{k}_3)$, where the Lagrangian that enters the calculation, $\mathcal{L}_{\zeta\chi^2}$, is given by \eqref{eq:L3chir}. Expanding all the terms in the Lagrangian and using \eqref{eq:deltacusc} to substitute $\delta\varphi$, the bispectrum as a function of the momenta $\bm{k}_n$, $n\in\{1,2,3\}$, and at time $\tau$ becomes
{\allowdisplaybreaks
\begin{align}
    B_{\zeta\chi^2}(\bm{k}_1,\bm{k}_2,\bm{k}_3,\tau)=&-\mpl^2\Im\bigg[\zeta_{k_1}(\tau)\chi_{k_2}(\tau)\chi_{k_3}(\tau)\nonumber\\
    &\times\int_{-\infty(1+i\varepsilon)}^{\tau}\dd\tilde\tau\,a(\tilde\tau)^2F(\tilde\tau)\Big(2\mathcal{I}_{\zeta\chi^2}^\textrm{1-perm}(\bm{k}_1,\bm{k}_2,\bm{k}_3,\tilde\tau)^*+\mathcal{I}_{\zeta\chi^2}^\textrm{2-perm}(\bm{k}_1,\bm{k}_2,\bm{k}_3,\tilde\tau)^*\Big)\bigg]\,,\label{eq:Bzetachi2i}
\end{align}
}%
where
{\small
{\allowdisplaybreaks
\begin{align}
    \mathcal{I}_{\zeta\chi^2}^\textrm{1-perm}(\bm{k}_1,\bm{k}_2,\bm{k}_3,\tau)=&~
    A_1^{\zeta\chi^2}(\bm{k}_1,\bm{k}_2,\bm{k}_3,\tau)\zeta_{k_1}(\tau)\chi_{k_2}'(\tau)\chi_{k_3}'(\tau)
    +A_2^{\zeta\chi^2}(\bm{k}_1,\bm{k}_2,\bm{k}_3,\tau)\zeta_{k_1}'(\tau)\chi_{k_2}'(\tau)\chi_{k_3}'(\tau)\nonumber\\
    &+A_3^{\zeta\chi^2}(\bm{k}_1,\bm{k}_2,\bm{k}_3,\tau)\zeta_{k_1}(\tau)\chi_{k_2}(\tau)\chi_{k_3}(\tau)
    +A_4^{\zeta\chi^2}(\bm{k}_1,\bm{k}_2,\bm{k}_3,\tau)\zeta_{k_1}'(\tau)\chi_{k_2}(\tau)\chi_{k_3}(\tau)\,,\nonumber\\
    \mathcal{I}_{\zeta\chi^2}^\textrm{2-perm}(\bm{k}_1,\bm{k}_2,\bm{k}_3,\tau)=&~\left[A_5^{\zeta\chi^2}(\bm{k}_1,\bm{k}_2,\bm{k}_3,\tau)\zeta_{k_1}(\tau)\chi_{k_2}(\tau)\chi_{k_3}'(\tau)
    +A_6^{\zeta\chi^2}(\bm{k}_1,\bm{k}_2,\bm{k}_3,\tau)\zeta_{k_1}'(\tau)\chi_{k_2}(\tau)\chi_{k_3}'(\tau)\right]\nonumber\\
    &+[\bm{k}_2\leftrightarrow \bm{k}_3]\,.\label{eq:Azetachi2}
\end{align}
}}%
In the above, the $\bm{k}_n$- and $\tau$-dependent coefficients are
{\allowdisplaybreaks
\begin{align}
    A_1^{\zeta\chi^2}(\bm{k}_1,\bm{k}_2,\bm{k}_3)&=3\left(1+\sigma\frac{k_1^2}{\mathcal{K}_1^2}\right)\,,
    & A_2^{\zeta\chi^2}(\bm{k}_1,\bm{k}_2,\bm{k}_3)&=-\frac{3}{\mathcal{H}}\left(1-\sigma\epsphi\frac{\mathcal{H}^2}{\mathcal{K}_1^2}\right)\,,\nonumber\\
    A_3^{\zeta\chi^2}(\bm{k}_1,\bm{k}_2,\bm{k}_3)&=\frac{A_1^{\zeta\chi^2}(\bm{k}_1,\bm{k}_2,\bm{k}_3)}{3}\bm{k}_2\cdot\bm{k}_3\,,
    & A_4^{\zeta\chi^2}(\bm{k}_1,\bm{k}_2,\bm{k}_3)&=\frac{A_2^{\zeta\chi^2}(\bm{k}_1,\bm{k}_2,\bm{k}_3)}{3}\bm{k}_2\cdot\bm{k}_3\,,\nonumber\\
    A_5^{\zeta\chi^2}(\bm{k}_1,\bm{k}_2,\bm{k}_3)&=\frac{2A_2^{\zeta\chi^2}(\bm{k}_1,\bm{k}_2,\bm{k}_3)}{3}\bm{k}_1\cdot\bm{k}_2\,,
    & A_6^{\zeta\chi^2}(\bm{k}_1,\bm{k}_2,\bm{k}_3)&=\frac{\epsphi\mathcal{H}A_5^{\zeta\chi^2}(\bm{k}_1,\bm{k}_2,\bm{k}_3)}{k_1^2}\,,\label{eq:Azetachi22}
\end{align}
}%
where we define the momenta $\mathcal{K}_n^2\equiv k_n^2+(3+\sigma)\epsphi\mathcal{H}^2$ related to the Fourier transform of the modified Laplace operator defined in \eqref{eq:deltacusc}. Note that $\mathcal{K}_n$ is both $k_n$ and $\tau$ dependent.
In the above, we are assuming $XF_{,X}\approx F$ to simplify some of the coefficients, which is a good approximation taking, e.g., \eqref{eq:Fofphi-massless2} as the coupling function when $\delta\ll 1$.

The calculation of $B_{\zeta^3}(\bm{k}_1,\bm{k}_2,\bm{k}_3)$ is found in a similar fashion, where the Lagrangian that enters the in-in formula \eqref{eq:inindef} is now $\mathcal{L}_{\zeta^3}$, which corresponds to \eqref{eq:L3restr} together with \eqref{eq:L3deltavarphiallr} where $\delta\varphi$ is explicitly replaced by its $\zeta$-dependent expression \eqref{eq:deltacusc}.
The resulting bispectrum is of the same form as before,
{\allowdisplaybreaks
\begin{align}
    B_{\zeta^3}(\bm{k}_1,\bm{k}_2,\bm{k}_3,\tau)=&-2\mpl^2\Im\bigg[\zeta_{k_1}(\tau)\zeta_{k_2}(\tau)\zeta_{k_3}(\tau)\nonumber\\
    &\times\int_{-\infty(1+i\varepsilon)}^{\tau}\dd\tilde\tau\,a(\tilde\tau)^2
    \Big(6\mathcal{I}^\textrm{1-perm}_{\zeta^3}(\bm{k}_1,\bm{k}_2,\bm{k}_3,\tilde\tau)^*+2\mathcal{I}^\textrm{3-perm}_{\zeta^3}(\bm{k}_1,\bm{k}_2,\bm{k}_3,\tilde\tau)^*\Big)\bigg]\,,\label{eq:Bzeta3i}
\end{align}
}%
with
{\allowdisplaybreaks
\begin{align}
    \mathcal{I}^\textrm{1-perm}_{\zeta^3}(\bm{k}_1,\bm{k}_2,\bm{k}_3,\tau)=&~
    A_1^{\zeta^3}(\bm{k}_1,\bm{k}_2,\bm{k}_3,\tau)\zeta_{k_1}(\tau)\zeta_{k_2}(\tau)\zeta_{k_3}(\tau)
    +A_2^{\zeta^3}(\bm{k}_1,\bm{k}_2,\bm{k}_3,\tau)\zeta_{k_1}'(\tau)\zeta_{k_2}'(\tau)\zeta_{k_3}'(\tau)\,,\nonumber\\
    \mathcal{I}^\textrm{3-perm}_{\zeta^3}(\bm{k}_1,\bm{k}_2,\bm{k}_3,\tau)=&~
    \left[A_3^{\zeta^3}(\bm{k}_1,\bm{k}_2,\bm{k}_3,\tau)\zeta_{k_1}(\tau)\zeta_{k_2}(\tau)\zeta_{k_3}'(\tau)\right]
    +[\bm{k}_1\leftrightarrow \bm{k}_3]+[\bm{k}_2\leftrightarrow \bm{k}_3]\nonumber\\
    &+\left[A_4^{\zeta^3}(\bm{k}_1,\bm{k}_2,\bm{k}_3,\tau)\zeta_{k_1}(\tau)\zeta_{k_2}'(\tau)\zeta_{k_3}'(\tau)\right]
    +[\bm{k}_1\leftrightarrow \bm{k}_2]+[\bm{k}_1\leftrightarrow \bm{k}_3]\,.\label{eq:Azeta3}
\end{align}
}%
While the coefficients $A^{\zeta^3}$ are straightforward to compute from the Fourier transform of the Lagrangian, they are extremely long, so we omit them here. (They are also not particularly insightful.) Nevertheless, we use the full coefficients in our calculations below.

\subsubsection{Analytical estimates}\label{sec:bianalycial}

Before numerically computing the full bispectra outlined above, aiming to minimize approximations, let us examine a few regimes of interest. In these cases, we can obtain some analytical insight by applying analogous applicable approximations.

\paragraph*{Before the bounce in the IR}
Starting with the prebounce phase in the IR regime, we can assume $\sigma\simeq 0$, $\epsphi\simeq 3$, $\mathcal{H}\simeq 1/(2\tau)$, $a^2F\simeq 1/(4\pi^2 A_s\mpl^2\tau^2)$, and use \eqref{eq:zeta_super} and \eqref{eq:chi_anysol} to estimate $ \tilde B_{\zeta\chi^2}$. Keeping only leading-order terms in the IR results in [the $A_3^{\zeta\chi^2}$, $A_4^{\zeta\chi^2}$, and $A_6^{\zeta\chi^2}$ terms turn out to dominate in \eqref{eq:Azetachi2}]
\begin{equation}\label{eq:Bzetachi2-analytic}
    \tilde B_{\zeta\chi^2}(\bm{k}_1,\bm{k}_2,\bm{k}_3,\tau_{\mathrm{b}-})\simeq-\frac{\pi^2A_sk_1(k_1^2\bm{k}_2\cdot\bm{k}_3+3k_2^2\bm{k}_1\cdot\bm{k}_3+3k_3^2\bm{k}_1\cdot\bm{k}_2)}{6\mpl^2(k_1^3+k_2^3+k_3^3)}\,,
\end{equation}
which is $\mathcal{O}(k^2/\mpl^2)$, hence highly suppressed.\footnote{This is to be contrasted with the dimensionless bispectrum corresponding to a scale-invariant perturbation (e.g., inflation or the matter bounce scenario), which depends on the angles of the $\bm{k}_n$ configuration but does not have an overall power-law dependence on the magnitude of the wavenumbers.} Indeed, the deviation from Gaussianity can be quantified as in \eqref{eq:fNLl-CNl_eqL} by
\begin{equation}
    |C_\NL^\chi\zeta_\G|\simeq\frac{7}{8\mpl}\sqrtb{\frac{|H_{\mathrm{b}-}|k_\mathrm{p}}{6\pi}}\bigg/\left|\ln\left(\frac{k_\mathrm{p}}{2|H_{\mathrm{b}-}|}\right)\right|\,,\label{eq:CNLIRpb}
\end{equation}
where IR modes of observational relevance have $k_\mathrm{p}\ll|H_{\mathrm{b}-}|\ll\mpl$, so this is a very small number.
Let us stress that this is before any conversion of $\chi$ into $\zeta$, hence $C_\NL^\chi$ is not straightforwardly mapped onto observational constraints on $f_\NL$. Nevertheless, since the correlation $\langle\zeta\chi^2\rangle$ involves $\zeta$ (which is suppressed on cosmological scales [its spectrum is blue]), the contribution to the final $f_\NL$ will be severely IR suppressed, even though $C_\NL^\chi$ itself can be greater than one.

We can similarly evaluate $\tilde B_{\zeta^3}$ in the IR. In the equilateral limit, to leading order the integrand of \eqref{eq:Bzeta3i} is $9i(-3\tau_{\mathrm{b}-}/2)^{1/2}(\gammaE +i\pi/2+\ln[-k\tau/2])/(4\pi^{3/2}\mpl^3\tau)$ --- the Lagrangian is essentially dominated by its $\zeta\zeta'^2$ term --- so we find
\begin{equation}
    \tilde B_{\zeta^3}(k,\tau_{\mathrm{b}-})\simeq\frac{k^6}{256\pi^2\mpl^4\mathcal{H}_{\mathrm{b}-}^2}\left(\ln\left[\frac{k}{2|\mathcal{H}_{\mathrm{b}-}|}\right]\right)^4\,.
\end{equation}
Following \eqref{eq:fNLl-CNl_eqL}, we can express the corresponding deviation from Gaussianity as
\begin{equation}\label{eq:fNLIRpb}
    |f_\NL^\zeta\zeta_\G|\simeq\frac{(3k_\mathrm{p})^{3/2}}{64\mpl\sqrtb{2\pi^3|H_{\mathrm{b}-}|}}\left|\ln\left(\frac{k_\mathrm{p}}{2|H_{\mathrm{b}-}|}\right)\right|\,,
\end{equation}
which is again highly suppressed. In fact, the right-hand side of the above is the same as $|\zeta_\G|$ up to an $\mathcal{O}(1)$ number [cf.~\eqref{eq:zetaIRpb}], in agreement with the estimate $|\mathcal{L}_{\zeta^3}/\mathcal{L}_{\zeta^2}|\sim|\zeta|$ from Sec.~\ref{sec:strongCouplingInteractions}. This shows that $|\mathcal{L}_{\zeta^3}/\mathcal{L}_{\zeta^2}|\sim |f_\NL^\zeta\zeta_\G|$, but we note that \eqref{eq:CNLIRpb} implies $|\mathcal{L}_{\zeta\chi^2}/\mathcal{L}_{\chi^2}|\sim|\zeta|\nsim|C_\NL^\chi\zeta_\G|$. This simply indicates that the ratio of the third- and second-order Lagrangians can sometimes be a good proxy for non-Gaussianities (or, for that matter, corrections to any other observable due to any nonlinear interactions), but they are not necessarily (formally) the same quantities. The bispectrum and corresponding quantities such as $f_\NL$ and $C_\NL$ are observable (at least in principle), while the ratio of the Lagrangians is really only a rough \emph{theoretical} estimate of the strong coupling scale. In particular, there can sometimes be cancellations in the full calculation of the bispectrum.

\paragraph*{Before the bounce in the UV}
Previously, we calculated $\Lambda_\mathrm{strong}$ in Sec.~\ref{sec:strong-coupling} by comparing the size of the third- and second-order terms in the action. Here, we show how $\Lambda_\mathrm{strong}$ can be obtained through calculations of the bispectrum. Indeed, a more robust way to quantify this scale is to determine the scale where the non-Gaussianities become order unity. As mentioned previously, $|f_\NL^\zeta\zeta_\G|$ and $|C_\NL^\chi\zeta_\G|$ represent the level of non-Gaussianities. Focusing on $\langle\zeta^3\rangle$, we first compute the $B_{\zeta^3}$ bispectrum using \eqref{eq:Bzeta3i} and \eqref{eq:Azeta3} with all the terms from the third-order action [cf.~\eqref{eq:L3deltavarphiallr} and \eqref{eq:L3restr}] for the equilateral shape. Assuming $\sigma\simeq 0$, $\epsphi\simeq 3$, and $\mathcal{H}\simeq 1/(2\tau)$ for the background contraction again and taking a UV limit, where in particular the mode function is approximated by $\zeta_k\simeq\sqrtb{\tau_{\mathrm{b}-}/(3k\tau)}e^{-ik\tau}/(2\mpl)$, one finds
the integral in \eqref{eq:Bzeta3i} to be dominated by
\begin{equation}
    \frac{k^{5/2}}{2\mpl^3}\sqrtb{\frac{-\tau_{\mathrm{b}-}}{3}}\int^{\tau_{\mathrm{b}-}}\dd\tau\,(-\tau)^{3/2}e^{3ik\tau}=\frac{(-1)^{3/4}\sqrtb{-\tau_{\mathrm{b}-}}}{54\mpl^3}\Gamma\Big(\frac{5}{2},-3ik\tau_{\mathrm{b}-}\Big)\stackrel{-k\tau_{\mathrm{b}-}\gg 1}{\simeq}-\frac{ik^{3/2}\tau_{\mathrm{b}-}^2e^{3ik\tau_{\mathrm{b}-}}}{6\sqrtb{3}\mpl^3}\,,
\end{equation}
where $\Gamma$ is the incomplete gamma function here,
hence the dimensionless bispectrum by the end of the contracting phase is found to be (also keeping subleading terms)
\begin{equation}
    |\tilde B_{\zeta^3}(k,\tau_{\mathrm{b}-})|\simeq\frac{k^6\tau_{\mathrm{b}-}^2}{1296\mpl^4}\left|1-\frac{14}{(k\tau_{\mathrm{b}-})^2}\right|\,.
\end{equation}
Correspondingly,
\begin{equation}
    |f_\NL^\zeta\zeta_\G|\simeq\frac{7}{18\sqrtb{6}\pi}\left|\frac{k_\mathrm{p}^3}{56H_{\mathrm{b}-}^2\mpl}-\frac{k_\mathrm{p}}{\mpl}\right|\,,\label{eq:fNLUV}
\end{equation}
in agreement with \eqref{eq:3/2ratio_subH}, suggesting here $|\mathcal{L}_{\zeta^3}/\mathcal{L}_{\zeta^2}|\sim|f_\NL^\zeta\zeta_\G|$.
From \eqref{eq:fNLUV}, the strong coupling scale appears to be
\begin{equation}
    \Lambda_\mathrm{strong}^\textrm{(Cuscuton)}\simeq 2^{3/2}3^{5/6}\pi^{1/3}\left(H_{\mathrm{b}-}^2\mpl\right)^{1/3}\,,
\end{equation}
again in agreement with \eqref{eq:Lambdas23}, up to the numerical prefactor here that is around $10.35$, so about an order of magnitude larger than the estimated $(H_{\mathrm{b}-}^2\mpl)^{1/3}$. However, recall that the higher-order operators yielding the above strong coupling scale could in principle be removed by integration by parts perturbatively in $1/k$ in the UV where it matches GR. Accordingly, the subleading term in \eqref{eq:fNLUV} would actually imply a strong coupling scale at the Planck scale.

Performing a similar calculation for $B_{\zeta\chi^2}$ using \eqref{eq:Bzetachi2i}, \eqref{eq:Azetachi2}, and \eqref{eq:Azetachi22}, together with the same approximations as before, where in particular in the UV now $\chi_k\simeq\sqrtb{2A_s/k}(-\pi\tau)e^{-ik\tau}$, we find the integral in \eqref{eq:Bzetachi2i} to be dominated by
\begin{align}
    -\frac{3i\sqrtb{-3k^3\tau_{\mathrm{b}-}}}{2\mpl^3}\int^{\tau_{\mathrm{b}-}}\dd\tau\,\sqrtb{-\tau}e^{3ik\tau}=&~\frac{(-1)^{3/4}\sqrtb{-\tau_{\mathrm{b}-}}}{2\mpl^3}\Gamma\Big(\frac{3}{2},-3ik\tau_{\mathrm{b}-}\Big)\nonumber\\
    &\stackrel{-k\tau_{\mathrm{b}-}\gg 1}{\simeq}\frac{e^{3ik\tau_{\mathrm{b}-}}}{4\sqrtb{3k}\mpl^3}(6k\tau_{\mathrm{b}-}+i+\ldots)\,.\label{eq:Bzetachi2iprebUV}
\end{align}
We keep the subleading term since when computing \eqref{eq:Bzetachi2i}, we have to multiply the above integral by $\zeta_k(\tau_{\mathrm{b}-})\chi_k(\tau_{\mathrm{b}-})^2$ --- whose sole complex contribution goes as $e^{-3ik\tau_{\mathrm{b}-}}$ --- and take the imaginary part of the result, hence only the subleading term in \eqref{eq:Bzetachi2iprebUV} survives. In the end, we find
\begin{equation}
    |C_\NL^\chi\zeta_\G|\simeq\frac{k_\mathrm{p}}{8\sqrtb{6}\pi\mpl}\,,
\end{equation}
which becomes strongly coupled only near the Planck scale.
We note that the leading-order term that canceled is the one that would otherwise have yielded $|C_\NL^\chi\zeta_\G|\sim k_\mathrm{p}^2/(|H_{\mathrm{b}-}|\mpl)$ as initially expected in the discussion below \eqref{eq:Lzetachichiorder}. Thus, this is another situation where the leading-order terms in our previous order-of-magnitude estimate of the strong coupling scale cancel in the full calculation of the bispectrum, hence $|\mathcal{L}_{\zeta\chi^2}/\mathcal{L}_{\chi^2}|\nsim|C_\NL^\chi\zeta_\G|$.

\paragraph{Through the bounce in the IR}
We now evaluate the relevant bispectra $B_{\zeta\chi^2}$ and $B_{\zeta^3}$ in the IR at the time $\tau_{\mathrm{b}+}$, i.e., once the bounce phase has ended. In the case where all three modes $\bm{k}_1$, $\bm{k}_2$, and $\bm{k}_3$ exit the horizon in the prebounce phase, one can split the conformal-time integrations in \eqref{eq:Bzetachi2i} and \eqref{eq:Bzeta3i} as $\int^{\tau_{\mathrm{b}-}}+\int_{\tau_{\mathrm{b}-}}^{\tau_{\mathrm{b}+}}$; the first integral was already evaluated in the IR calculation above, so the new contribution comes from the second integral. To get a rough approximation of the integral from $\tau_{\mathrm{b}-}$ to $\tau_{\mathrm{b}+}$, as before, we analytically expand and estimate the background and linear perturbations about the bounce point (i.e., about $\tau=0$), according to what was derived in Sec.~\ref{eq:bounceanalytical}. Starting with the evaluation of $\langle\zeta\chi^2\rangle$, after substitution for those quantities we find
\begin{align}\label{eq:Bzetachi2-analytic-afterB}
    \tilde B_{\zeta\chi^2}(\bm{k}_1,\bm{k}_2,\bm{k}_3,\tau_{\mathrm{b}+})\simeq&-\frac{\pi^2A_sk_1}{6\mpl^2(k_1^3+k_2^3+k_3^3)}\Bigg(-k_1^2\bm{k}_2\cdot\bm{k}_3\left[1-\frac{\mpl^2}{\mplt^2}\left(2+\frac{\mu^4\varphi_\infty^2}{2\mathcal{H}_{\mathrm{b}-}^2}\right)\right]\nonumber\\
    &+\left(k_2^2\bm{k}_1\cdot\bm{k}_3+k_3^2\bm{k}_1\cdot\bm{k}_2\right)\left[3+\frac{m^2\varphi_\infty^2}{\mpl^2\mathcal{H}_{\mathrm{b}-}^2}\right]\Bigg)\,.
\end{align}
Upon comparing with \eqref{eq:Bzetachi2-analytic}, the overall $\mathcal{O}(k^2/\mpl^2)$ dependence before and after the bounce remains the same, but the coefficients of $k_1^2\bm{k}_2\cdot\bm{k}_3$ and $k_2^2\bm{k}_1\cdot\bm{k}_3+k_3^2\bm{k}_1\cdot\bm{k}_2$ now depend on the parameters of the bouncing phase (parameters of the Cuscuton potential), hence the precise shape changes. It is informative to compare the size of non-Gaussianities through $C_\NL^\chi\zeta_\G$ before and after the bounce as a ratio,
\begin{equation}
    \left|\frac{C_\NL^\chi(\tau_{\mathrm{b}+})\zeta_\G(\tau_{\mathrm{b}+})}{C_\NL^\chi(\tau_{\mathrm{b}-})\zeta_\G(\tau_{\mathrm{b}-})}\right|\simeq\left|1+\frac{15}{7\gamma^2}-\frac{12}{7(3-2\gamma^2)}\right|\,,\label{eq:CNLratio}
\end{equation}
where we made use of the first approximation in \eqref{eq:Hbmest} and defined $\gamma\equiv m/(\mu^2\mpl)$. We first note that the above ratio diverges as $\gamma\to 0$ or as $\gamma\to\sqrtb{3/2}$, indicating the theory becomes strongly coupled (and breaks down) in those limits. However, these correspond to the boundaries of the range over which the bounce phase can occur in the first place [recall again the footnotes \ref{footnote:breq} and \ref{foot:negcs2IRm0}, Refs.~\cite{Boruah:2018pvq,Quintin:2019orx}, eq.~\eqref{eq:UpsilonHbmapprox}, or the discussion below \eqref{eq:scCf}]. Now, let us explore this ratio within that range. For $\gamma\approx 1$, the bounce phase does not generate any significant enhancement of non-Gaussianities from $\chi$ interactions since the above ratio is $\approx 1.43$. In fact, the ratio \eqref{eq:CNLratio} is $\lesssim 10$ for $0.471\lesssim \gamma \lesssim 1.196$, and it is $\lesssim 100$ for $0.147\lesssim\gamma\lesssim 1.221$. Thus, we confirm that the theory is not strongly coupled if $1/2<m/(\mu^2\mpl)\leq 1$; recall the discussion below \eqref{eq:L2zetabIR}. However, note that this is a conservative range as \eqref{eq:Bzetachi2-analytic} and \eqref{eq:Bzetachi2-analytic-afterB} indicate $|C_\NL^\chi(\tau_{\mathrm{b}+})\zeta_\G(\tau_{\mathrm{b}+})|$ is suppressed in the IR, so $\gamma$ would have to be very close to $0$ or $\sqrtb{3/2}$ before it reaches the strong coupling limit.

The same calculation as above for $\langle\zeta^3\rangle$ yields
\begin{equation}
    \left|\frac{f_\NL^\zeta(\tau_{\mathrm{b}+})\zeta_\G(\tau_{\mathrm{b}+})}{f_\NL^\zeta(\tau_{\mathrm{b}-})\zeta_\G(\tau_{\mathrm{b}-})}\right|\simeq 1+\frac{4(36+\gamma^2)}{3(3-2\gamma^2)\left|\ln(k_\mathrm{p}/|H_{\mathrm{b}-}|)\right|}\,,\label{eq:fNLratio1}
\end{equation}
indicating once again that the bounce phase has little effect on the deviation from Gaussianity in the IR provided $\gamma$ is not close to $\sqrtb{3/2}$. For $\gamma\lesssim 1$, the above ratio is between $1$ and $\sim 1.5$. Once again one finds a divergence (and thus a strong coupling issue) if $\gamma$ is close to $\sqrtb{3/2}$, but then the same discussion as in the previous paragraph applies.

Lastly, if $\chi$ converts into $\zeta$ instantaneously at the onset of the bounce, so that $\zeta$ becomes scale invariant at $\tau_{\mathrm{b}-}$, then $\zeta$ evolves through the bounce according to \eqref{eq:zeta_b_approx}, but with its initial conditions approximately set at $\tau_{\mathrm{b}-}$ as \eqref{eq:chichiptaubmek}. Using that in the calculation of $B_{\zeta^3}$ through the bounce in the IR, we can quantify the enhancement of the non-Gaussianity from $\langle\zeta^3\rangle$ as
\begin{equation}
    \left|\frac{f_\NL^\zeta(\tau_{\mathrm{b}+})\zeta_\G(\tau_{\mathrm{b}+})}{f_\NL^\zeta(\tau_{\mathrm{b}-})\zeta_\G(\tau_{\mathrm{b}-})}\right|\simeq 1+\frac{16(5-\gamma^2)}{(3-2\gamma^2)\left|\ln(k_\mathrm{p}/|H_{\mathrm{b}-}|)\right|}\sqrtb{\frac{6\pi^7A_s^3k_\mathrm{p}\mpl^6}{|H_{\mathrm{b}-}|^7}}\,,
\end{equation}
where we simply assumed $f_\NL^\zeta(\tau_{\mathrm{b}-})\zeta_\G(\tau_{\mathrm{b}-})\approx C_\NL^\chi(\tau_{\mathrm{b}-})\chi_\G(\tau_{\mathrm{b}-})$. Other than the same apparent strong coupling issue as before if $\gamma\to\sqrtb{3/2}$, we note the additional $k$ dependence in the above enhancement indicating the spectral index of the bispectrum changes after conversion, through the bounce. This dependence implies that the bispectrum is either enhanced or suppressed through the bounce depending on whether $k_\mathrm{p}/|H_{\mathrm{b}-}|$ is greater or smaller than $(|H_{\mathrm{b}-}|/\mpl)^6$. However, in magnitude, it remains the case that the contribution to $f_\NL^\zeta$ from the bounce phase is suppressed in the IR. The dominant contribution, if there is any, would then come from the conversion process itself, which was ignored in the above.

\subsubsection{Numerical results}

Let us confirm the above analytical IR estimates of $B_{\zeta\chi^2}$ and $B_{\zeta^3}$ by performing full numerical computations of the bispectra. We note that the three $\bm{k}_n$ modes of the bispectra relevant to observations first evolve from their subhorizon regime, cross the horizon, evolve on superhorizon scales in the prebounce regime, and then through the bounce. Thus, when evaluating the bispectra at late times (after the bounce phase), the integrals involved in computing \eqref{eq:Bzetachi2i} and \eqref{eq:Bzeta3i} go through all of these regimes. Until the post horizon-crossing regime, our model resembles other contracting cosmologies, for which non-Gaussianities have been studied and shown to be negligible (e.g., \cite{Cai:2009fn,Li:2016xjb}) --- this is in contrast to inflation, where contributions to the bispectrum mainly come from the horizon-crossing regime. Our numerical computations further confirm that the contribution from the superhorizon regime dominates over the preceding regimes. Thus, we focus on the superhorizon regime during the prebounce phase and then through the bounce phase. We evaluate the bispectrum using the exact numerical solutions presented in Sec.~\ref{sec:numerical} for both the background and the mode functions $\chi_k(\tau)$ and $\zeta_k(\tau)$.

\begin{figure}[t]
    \centering
    \includegraphics[width=.45\columnwidth]{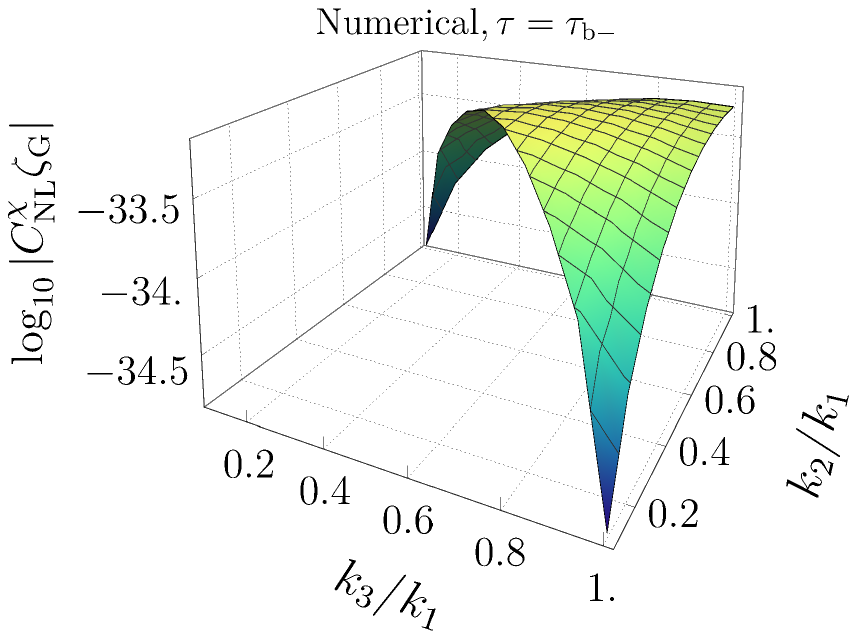}
    \hspace*{5pt}
    \includegraphics[width=.45\columnwidth]{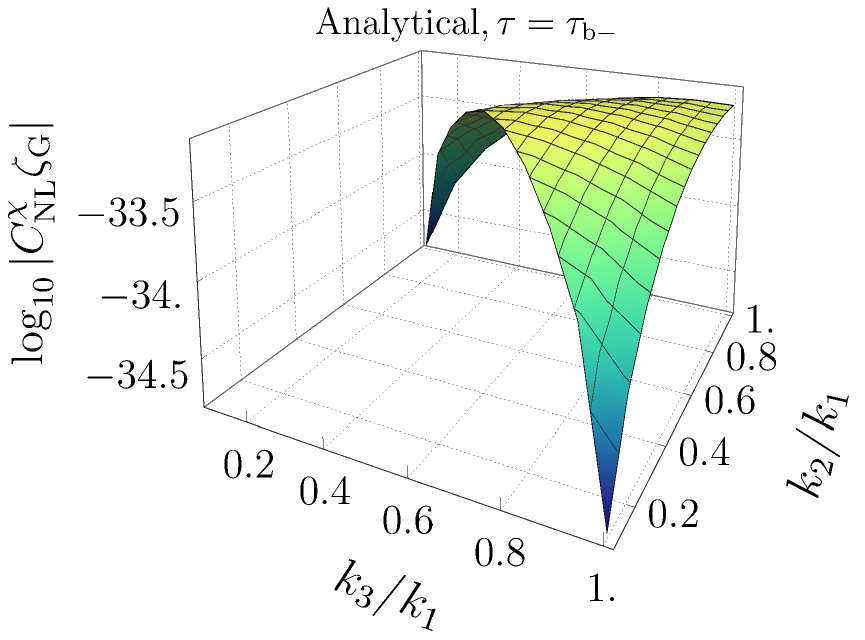}
    \includegraphics[width=.45\columnwidth]{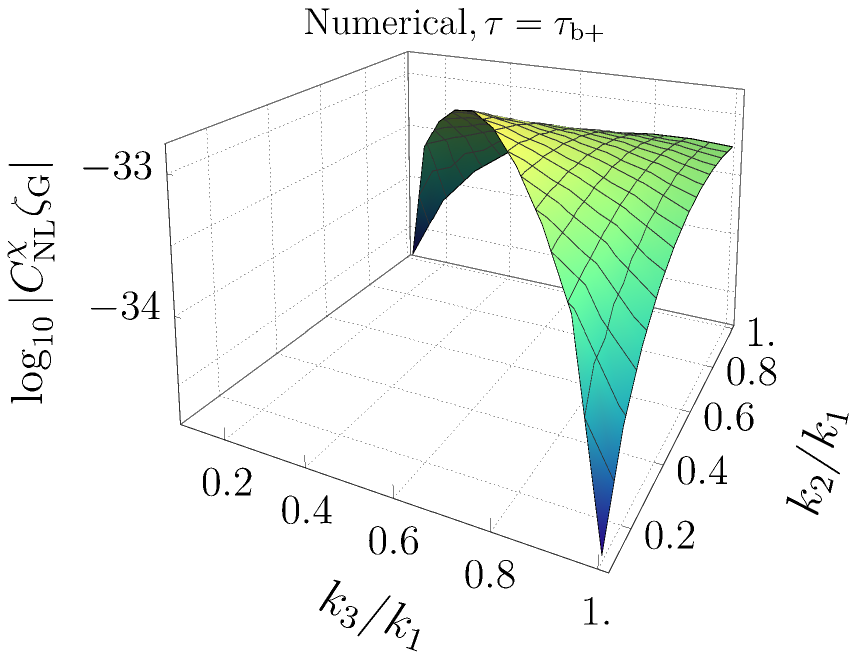}
    \hspace*{5pt}
    \includegraphics[width=.45\columnwidth]{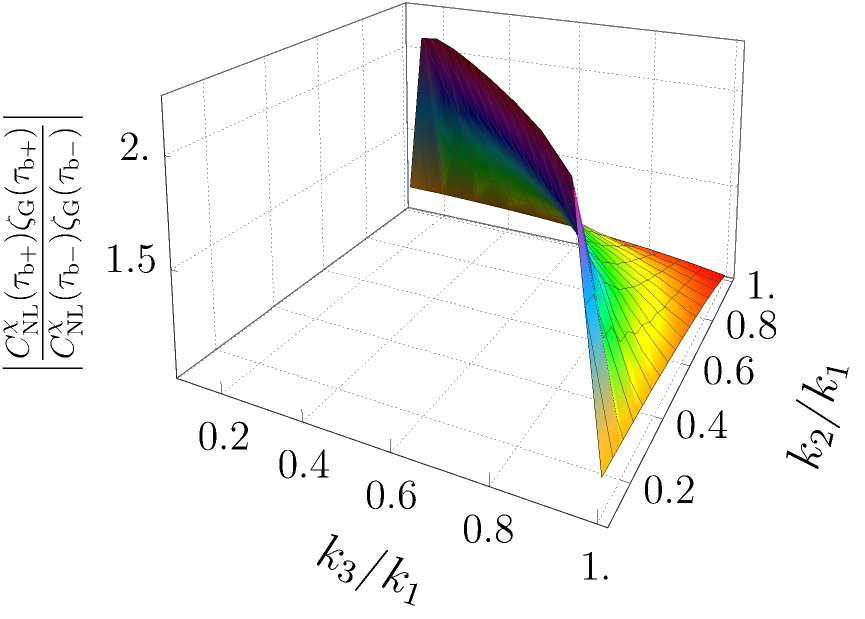}
    \caption{{\footnotesize Results for the nonlinear corrections from $\zeta\chi^2$ bispectrum. The pivot value of the first wavenumber is set to be $k_1=10^{-58}\mpl$, and the value of $\zeta_\G$ is approximated by $|\zeta_\G|\approx\mathcal{P}_\zeta(k_1)^{1/2}$. The bispectrum is computed numerically in the two panels on the left, ending the integration at $\tau=\tau_{\mathrm{b}-}$ in the top-left plot and at $\tau=\tau_{\mathrm{b}+}$ in the bottom-left plot. In the top-right panel, the analytical estimate of \eqref{eq:Bzetachi2-analytic} is plotted, which closely matches the corresponding numerical results in the top-left panel. The ratio of $|C_\NL^\chi\zeta_\G|$ numerical values at $\tau=\tau_{\mathrm{b}+}$ and $\tau=\tau_{\mathrm{b}-}$ is plotted in the bottom-right panel.}}
    \label{fig:cnlzeta}
\end{figure}
 
The numerical results for the strength of nonlinear effects expressed through the quantity $|C_\NL^\chi(\bm{k}_1,\bm{k}_2,\bm{k}_3,\tau)\zeta_\G|$ --- recall \eqref{eq:deffCNL} --- are shown in Fig.~\ref{fig:cnlzeta}. The pivot value of the first wavenumber is set to $k_1=10^{-58}\mpl$, and the value of $\zeta_\G$ is approximated by $\zeta_\G(\tau;1/k_1)\approx\mathcal{P}_\zeta(k_1)^{1/2}$ as before [recall \eqref{eq:xiW}]. We initiate the integral of \eqref{eq:Bzetachi2i} deep in the contracting phase, but the precise starting point is unimportant, as the contribution from the lower integration bound is negligible. In the top-left panel of Fig.~\ref{fig:cnlzeta}, we show the result when the integration is stopped at $\tau=\tau_{\mathrm{b}-}$, which is to say that the bounce phase is not included in the computation.
The top-right panel comparatively shows the analytical result obtained from \eqref{eq:Bzetachi2-analytic}; the top-left and -right panels are nearly identical, thus demonstrating the validity of the many approximations used in deriving \eqref{eq:Bzetachi2-analytic}.
The bottom-left panel then shows the result when the integration is stopped at $\tau=\tau_{\mathrm{b}+}$, which is to say that the bounce phase is included in the computation. In these three panels, note that $|C_\NL^\chi\zeta_\G|$ is shown on a logarithmic scale, and as such, it indicates that nonlinear effects are strongly suppressed, as expected from the analytical estimates.

In the bottom-right panel of Fig.~\ref{fig:cnlzeta}, we further plot the ratio of $|C_\NL^\chi\zeta_\G|$ after the bounce (at $\tau=\tau_{\mathrm{b}+}$) to its value before the bounce (at $\tau=\tau_{\mathrm{b}-}$). Again, we checked that this closely matches the ratio of the analytical estimates \eqref{eq:Bzetachi2-analytic-afterB} and \eqref{eq:Bzetachi2-analytic}. The plot further shows that the bounce phase (with reasonable Cuscuton parameters such as those used in the numerical computations) has only a small effect on the size and shape of the bispectrum.

\begin{figure}[t]
    \centering
    \includegraphics[width=.44\columnwidth]{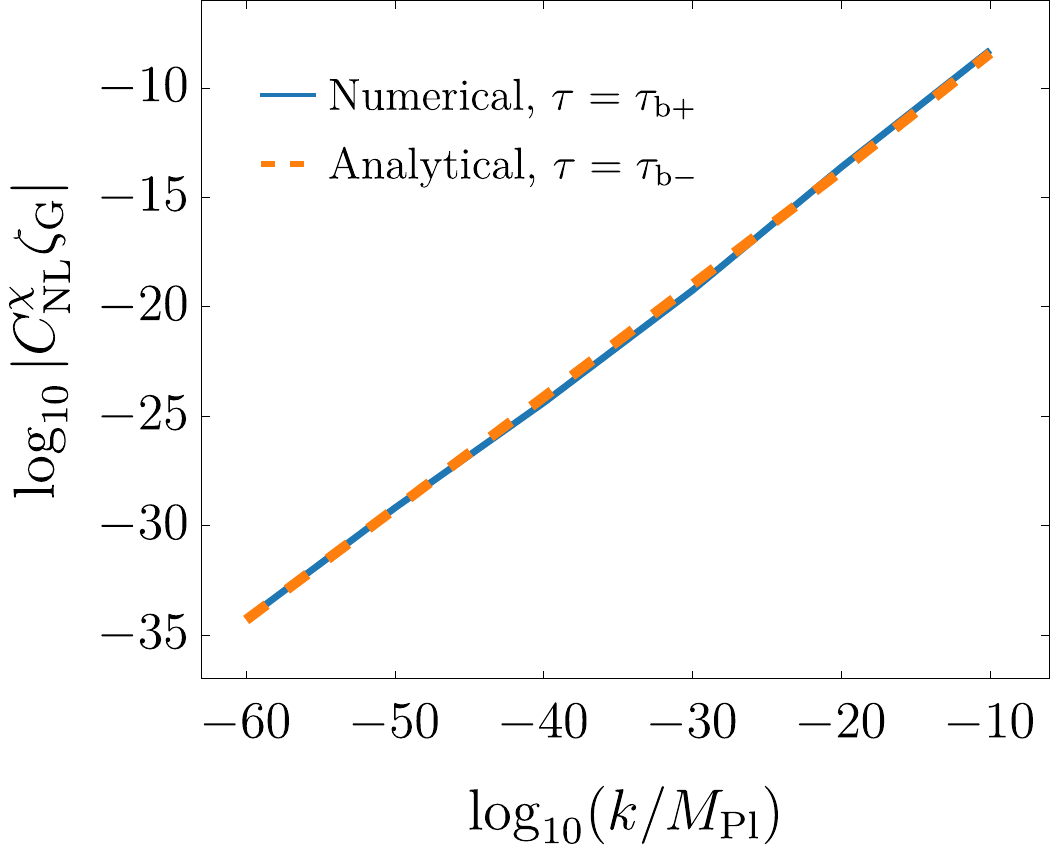}
    \hspace{3pt}
    \includegraphics[width=.45\columnwidth]{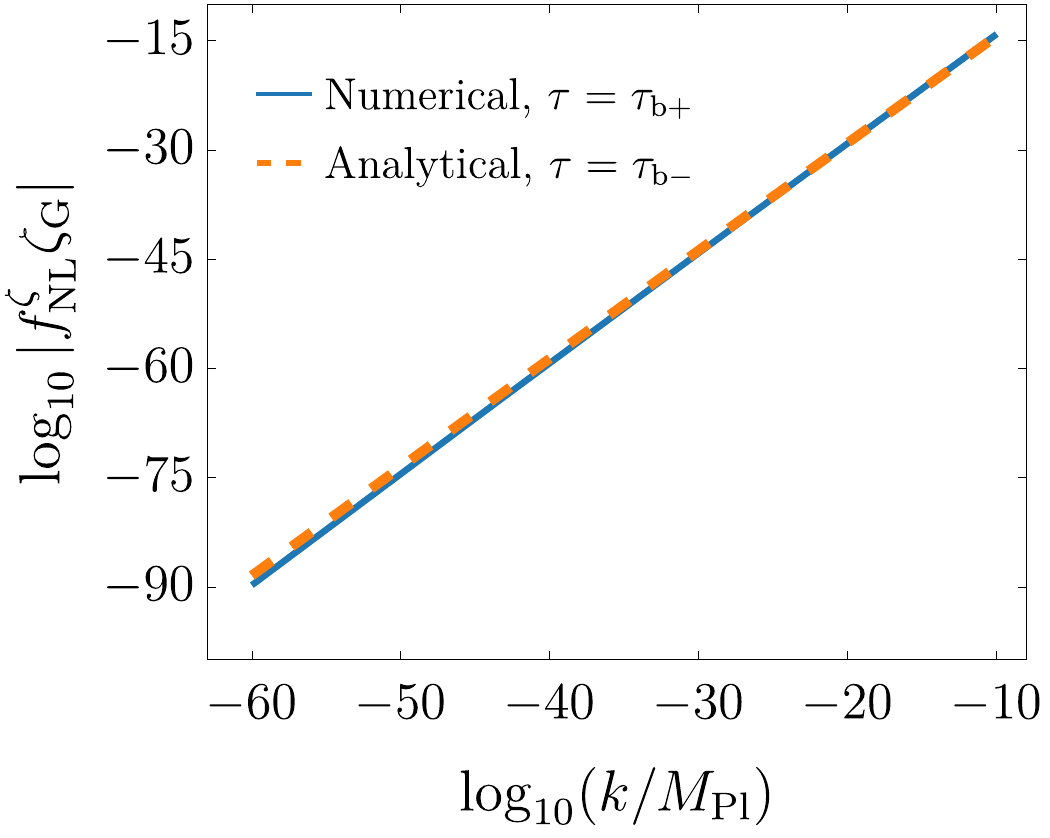}
    \caption{{\footnotesize Numerical and analytical estimates of $|C_\NL^\chi\zeta_\G|$ (left) and $|f_\NL^\zeta\zeta_\G|$ (right) as functions of $k\equiv k_1=k_2=k_3$.}}
    \label{fig:cnlzeta_eq}
\end{figure}

To examine the overall scale dependence of the $\zeta\chi^2$ and $\zeta^3$ bispectra, we compute the quantities $|C_\NL^\chi\zeta_\G|$ and $|f_\NL^\zeta\zeta_\G|$ at $\tau_{\mathrm{b}+}$ for the simple case of $k_1=k_2=k_3\equiv k$ over a range of $k$ values. The results are respectively plotted in the left and right plots of Fig.~\ref{fig:cnlzeta_eq}. Additionally, the analytical estimates from \eqref{eq:CNLIRpb} and \eqref{eq:fNLIRpb} are included as dashed curves, which correspond to the bispectra evaluated at $\tau_{\mathrm{b}-}$.
The plots thus show that the bounce has only a marginal effect on the size of non-Gaussianities [as expected from \eqref{eq:CNLratio} and \eqref{eq:fNLratio1}] and that the analytical estimates closely match the full numerical computations. In both cases, non-Gaussianities are suppressed in the IR due to the `blue' $k$ dependence; indeed, ignoring logarithmic dependence, \eqref{eq:CNLIRpb} and \eqref{eq:fNLIRpb} imply $|C_\NL^\chi\zeta_\G|\sim k^{1/2}$ and $|f_\NL^\zeta\zeta_\G|\sim k^{3/2}$. Note that these power-laws are good approximations (coming from the IR limit) for the whole $k$ range used in Fig.~\ref{fig:cnlzeta_eq}. In other words, $k$ modes even as high as $10^{-10}\mpl$ safely remain superhorizon (as defined in footnote \ref{foot:horizonDef}) for the time intervals of interest when computing the bispectra. This can be easily seen in the prebounce phase as the Hubble radius is at its smallest at $\tau_{\mathrm{b}-}$ and $|H_{\mathrm{b}-}|\sim 10^{-3}\mpl$ in the numerical example used. The same is true post bounce. Through the bounce, the horizon scale differs significantly from the Hubble radius, but \eqref{eq:kIRregimes} still holds for these modes.

\begin{figure}[t]
    \centering
    \includegraphics[width=.44\columnwidth]{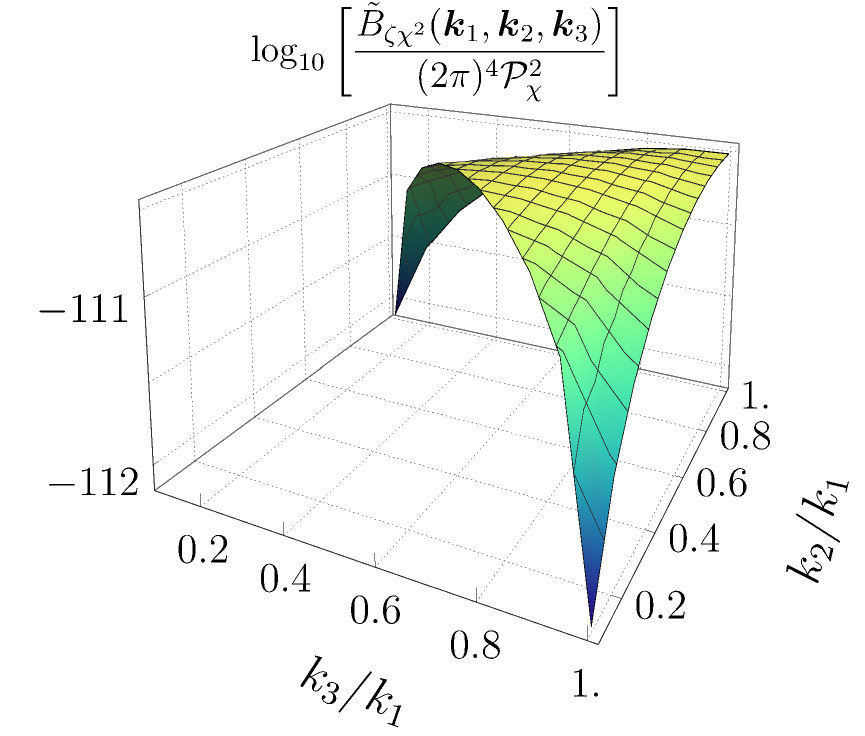}
    \hspace{3pt}
    \includegraphics[width=.45\columnwidth]{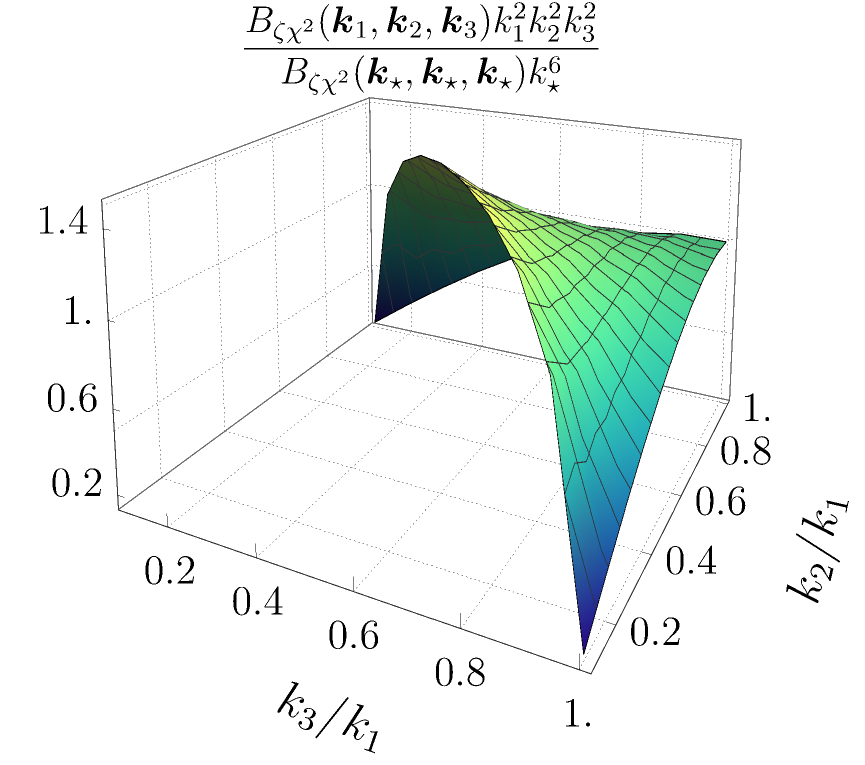}
    \caption{{\footnotesize The left panel represents the level of non-Gaussianities, assuming that $\chi$ converted into $\zeta$, ignoring potential enhancement due to nonlinear interactions needed for this process to happen. The right panel shows the shape function normalized by its value at $k_1=k_2=k_3\equiv k_\star=10^{-58}\mpl$.}}
    \label{fig:fnl-shape}
\end{figure}

In the previous plots, we demonstrated that the theory is not strongly coupled and that the perturbations remain highly Gaussian. However, the current observational constraints on non-Gaussianities \cite{Planck:2019kim} are more stringent than the strong coupling constraints. In principle, if non-Gaussianities are observed in future experiments (e.g., \cite{SimonsObservatory:2018koc,Meerburg:2019qqi}), they can be used to constrain model parameters. However, the quantities shown in Fig.~\ref{fig:cnlzeta_eq} cannot be directly mapped to CMB observables. Therefore, for completeness, we also estimate the non-Gaussianity `amplitude' assuming that $\chi$ perturbations directly convert into $\zeta$, ignoring any additional enhancements of non-Gaussianities from this conversion process. That is, we estimate $f_\NL^\zeta$ post conversion as $\tilde{B}_{\zeta\chi^2}(\bm{k}_1,\bm{k}_2,\bm{k}_3)/((2\pi)^4\mathcal{P}_\chi^2)$. This essentially corresponds to our definition of $f_\NL^\zeta$ provided in \eqref{eq:deffCNL}, with $B_{\zeta\zeta\zeta}$ replaced by $B_{\zeta\chi\chi}$ and $P_\zeta$ by $P_\chi$. The numerical results are presented on a logarithmic scale in the left panel of Fig.~\ref{fig:fnl-shape}, indicating extremely low values, substantially beneath current observational limits for observationally relevant modes. This agrees with our analytical estimate using \eqref{eq:Bzetachi2-analytic}, where we anticipate, for a fixed triangle shape, the overall amplitude to scale as $k^2/(\mpl^2A_s)$, thereby resulting in significant suppression.

Finally, we end this section by presenting in the right panel of Fig.~\ref{fig:fnl-shape} the bispectrum's shape function, $B_{\zeta\chi^2}(\bm{k}_1,\bm{k}_2,\bm{k}_3)k_1^2k_2^2k_3^2$, normalized by its value at $k_1=k_2=k_3\equiv k_\star=10^{-58}\mpl$. The shape does not resemble any of the well-known shapes (equilateral, folded, or local); therefore, the bispectrum from the Cuscuton bounce, while highly suppressed, has a distinct shape of its own.

\section{Discussion and conclusions}\label{sec:conclusions}

In this work, we revisited the Cuscuton bounce model of \cite{Boruah:2018pvq,Kim:2020iwq}, and besides the initial scenario where the matter field is massless, we also expanded our study to include the case where the matter field may be an ekpyrotic field. We provided new insight regarding the evolution of the perturbations through the bounce, reconfirming stability at the linear level, and extending both the numerical and analytical scope of the previous analyses (e.g., \cite{Boruah:2018pvq,Kim:2020iwq,Quintin:2019orx}).

The main novelty then lies in going beyond the linear level. We derived the full third-order action for the system consisting of the Cuscuton and a massless scalar. As far as we are aware, third-order actions involving the Cuscuton (or related theories) have only been derived in \cite{Bartolo:2021wpt} under slow-roll approximations and in a related but different modified gravity based bouncing model in \cite{Ganz:2024ihb}. We used the third-order action with two goals in mind: to check the consistency of the models (i.e., find the range of validity for a perturbative approach and weak coupling) and to see if they produced sizable non-Gaussianities (which satisfy observational constraints or could lead to observable predictions).

Our strong coupling analysis showed that the Cuscuton bounce has a reasonable regime of validity. Two conditions are needed: the second derivative of the Cuscuton potential at the bounce point, $U_{,\varphi\varphi}(\bar\varphi=0)$, must not be larger than $\mL^4/\mpl^2$ [see \eqref{eq:gammaRange}]; and the mass coefficient of the Cuscuton's `kinetic' term $\mL$ must be much smaller than the Planck scale. Under these conditions, we find the strong coupling scale to be at least of the order $\mL^{4/3}/\mpl^{1/3}$ [see \eqref{eq:Lambdas23}], while the background energy scale is at most of the order of $\mL^2/\mpl$ [see \eqref{eq:Hbmest}]. Such a hierarchy of scales encourages us that the Cuscuton bounce may remain weakly coupled over cosmological scales of relevance, something other nonsingular bouncing cosmologies can hardly realize (see again, e.g., \cite{Koehn:2015vvy,deRham:2017aoj,Dobre:2017pnt}). While we reach this conclusion by comparing the third- and second-order actions and by computing the bispectra, it may be interesting to check its robustness using alternative methods in future work, such as through scattering amplitudes and positivity bounds (see again, e.g., \cite{deRham:2017aoj,Cai:2022ori}).

Our analysis of the bispectra revealed non-Gaussianities are very suppressed, in contrast to other bouncing models such as \cite{Gao:2014hea,Gao:2014eaa,Quintin:2015rta,Delgado:2021mxu,vanTent:2022vgy,Ageeva:2024knc,Akama:2025ows} and notably \cite{Ganz:2024ihb}, which has a Cuscuton-like modification to GR that enables the bounce \cite{Ganz:2022zgs} (although the mechanism for generating the perturbations before the bounce is rather different from ours since it is purely adiabatic). In particular, we found barely any enhancement of nonlinearities as perturbations pass through the bounce in the IR, provided the Cuscuton parameters are in the ranges described above (i.e., within the regime of validity of the model) and provided $U_{,\varphi\varphi}(\bar\varphi=0)$ is not too close to zero either [see \eqref{eq:CNLratio}], something foreseen in \cite{Quintin:2019orx}.

Although finding no new observable non-Gaussian signature may be a disappointment, it is also a good thing in terms of the validity of the models under study and within the present observational bounds. Let us emphasize that this study focused on multifield models where a near-scale-invariant power spectrum was generated through an isocurvature perturbation.\footnote{We did not consider purely adiabatic scenarios such as the matter bounce and varying equations of state \cite{Wands:1998yp,Finelli:2001sr,Brandenberger:2012zb,Khoury:2009my,Khoury:2011ii,Akama:2025ows}. However, we believe our conclusions regarding the bouncing phase not contributing any significant changes to non-Gaussianities as the perturbations evolve through the bounce would still hold. This may be a good thing for the models that produce sizable non-Gaussianities on their own \cite{Cai:2009fn,Baumann:2011dt,Li:2016xjb,Akama:2025ows}.} As such, our models are not fully developed yet in the sense that there still needs to be some form of transition (e.g., through a reheating surface after the bounce phase) through which the isocurvature modes are converted into curvature perturbations. This has been analyzed in various other works such as \cite{Lehners:2007wc,Lehners:2008my,Lehners:2009qu,Lehners:2009ja,Lehners:2010fy,Ijjas:2014fja,Fertig:2015ola,Fertig:2016czu,Ijjas:2020cyh,Ijjas:2021ewd}, so we omitted this in our analysis. Still, from those references, we know that $\mathcal{O}(1-10)$ non-Gaussianities can be generated during the conversion process, already intriguingly close to current observational constraints such as from \textit{Planck} \cite{Planck:2019kim}. Therefore, any additional significant non-Gaussianities from the preconversion phase would imply that these bouncing models are ruled out entirely. As we have shown in this work, this is not the case in Cuscuton bounce models since preconversion non-Gaussianities are extremely suppressed, but such $\mathcal{O}(1-10)$ non-Gaussianities from the conversion may be within observational reach by future cosmological surveys.

Another aspect of the Cuscuton bounce that we left aside in this work is primordial gravitational waves. Indeed, we ignored tensorial metric perturbations $\gamma_{ij}$ from the start. At the level of the power spectrum, the evolution and spectrum of the tensor modes have been explored in \cite{Kim:2020iwq}, and they have been shown to be stable and nonsingular through the bounce phase. Their prebounce power spectrum can be understood quite intuitively. Tensor modes behave in a similar way to adiabatic perturbations (so schematically $|\gamma_{ij}|\sim|\zeta|$). In particular in the models of this work, this means that the tensor 2-point function $\langle\gamma_{ij}^2\rangle\sim\langle\zeta^2\rangle$ has a blue (vacuum) spectrum, which is highly suppressed on cosmologically relevant scales.\footnote{Typically, contracting models that produce a near-scale-invariant power spectrum through isocurvature perturbations, like ekpyrotic cosmology, do not forecast a detectable tensor-to-scalar ratio. This contrasts with models based on purely adiabatic perturbations, such as the matter bounce ($\epsilon=3/2$) for which superhorizon modes grow significantly. The reason has to do with the fact that within contracting models for which $\epsilon\gtrsim 3$, superhorizon perturbations freeze (or at most grow logarithmically), ensuring that the model remains stable against anisotropies. However, as modes freeze out on relatively large physical scales during the contracting phase (in contrast to inflation), their amplitude is also very suppressed (see Sec.~4.1 of \cite{Geshnizjani:2014bya} for more details).} A similar statement is expected to hold for the 3-point functions, i.e., $\langle\gamma_{ij}^3\rangle\sim\langle\zeta^3\rangle$ and $\langle\gamma_{ij}\chi^2\rangle\sim\langle\zeta\chi^2\rangle$, and as such we would not expect any observable bispectra involving graviton legs. (For an alternative bouncing model that predicts an interesting signal in the tensor 3-point function, see \cite{Akama:2024vtu}.) Moreover, it is important to stress that the presence of the Cuscuton does not affect the action/EoM governing tensor perturbations at the linear level (so while the $\zeta$ evolution may be `nontrivial' due to the Cuscuton, that of $\gamma_{ij}$ would be exactly as in GR \cite{Kim:2020iwq}), but it could be interesting to explore more carefully what happens when tensor perturbations are included at higher order.

In general, one of the holy grails for any early universe scenario would be to have a model that predicts sizable non-Gaussianities, yet within the observational bounds, at the same time as being distinguishable from other scenarios and detectable in the future. Although the bispectrum does not achieve this for the current Cuscuton bounce models, there could potentially be a distinguishable signature in the trispectrum (see, e.g., \cite{Lehners:2009ja,Fertig:2015ola} for trispectrum computations in ekpyrosis). Exploring this would involve going to next order in perturbations again, which would represent a significant computational challenge; hence we leave it for future work. Still, it would also provide further scrutinization in terms of the robustness and validity of the model to higher-order nonlinearities, and thanks to computational advances, we may be able to have interesting observational constraints on the trispectrum in the future (see, e.g., \cite{Philcox:2025bvj,Philcox:2025lrr,Philcox:2025wts}). In passing, we wish to mention that while the overall size of the correlations may be relevant predictions for various models of primordial cosmology, specific features (such as shapes or oscillations in the amplitude) within the correlations could have an even greater constraining power in terms of having model-independent signatures of inflation versus bouncing alternatives (see, e.g., \cite{Chen:2011zf,Chen:2011tu,Chen:2015lza,Chen:2018cgg,Domenech:2020qay,Quintin:2024boj}). Yet another observable signature of bouncing cosmology that deserves further exploration is the formation of primordial black holes (see, e.g., \cite{Carr:2011hv,Quintin:2016qro,Chen:2016kjx,Clifton:2017hvg,Corman:2022rqo,Chen:2022usd,Cai:2023ptf}). The Cuscuton bounce may provide an interesting framework for investigating primordial black holes in the context of bouncing cosmology.

Let us end by mentioning that the Cuscuton, which is still not widely studied, has revealed lots of interesting properties in recent years (e.g., \cite{Chagoya:2016inc,deRham:2016ged,DeFelice:2018ewo,Pajer:2018egx,Grall:2019qof,Faraoni:2022doe,Miranda:2022wkz}). While we treated the Cuscuton as an effective field theory as in \cite{Mylova:2023ddj}, the interesting question remains whether this theory admits a UV completion, which still eludes us (see \cite{Afshordi:2006ad,Afshordi:2009tt,Bhattacharyya:2016mah} for a few potential avenues, in particular the connection with Ho\v{r}ava-Lifshitz gravity \cite{Horava:2009uw}). Various directions could be explored in the future, such as those of \cite{Mukohyama:2020lsu,Lara:2021piy} for $k$-essence theories or discretization as alluded to in \cite{Afshordi:2006ad,Mylova:2023ddj}. Certainly, the Cuscuton still has lots of secrets to unveil.

\vskip23pt
\subsection*{Acknowledgments}
{\small We thank Niayesh Afshordi, Chunshan Lin, Maria Mylova, and Ryo Namba for stimulating discussions about the Cuscuton over the years. We also thank the Fields Institute for a stimulating atmosphere in the early stages of this work. This research was supported by a Discovery Grant from the Natural Science and Engineering Research Council of Canada (NSERC). This research was also supported in part by the Perimeter Institute for Theoretical Physics. Research at the Perimeter Institute is supported by the Government of Canada through the Department of Innovation, Science and Economic Development and by the Province of Ontario through the Ministry of Colleges and Universities. J.Q.~further acknowledges financial support from the University of Waterloo's Faculty of Mathematics William T.~Tutte Postdoctoral Fellowship. This research made use of \texttt{MathGR} \cite{Wang:2013mea}. The codes underlying this article may be shared upon reasonable request to the corresponding authors.}
\vskip23pt

\appendix

\section{Three-point functions for two locally non-Gaussian scalar fields}\label{app:3-pt}

Here we show how \eqref{eq:deffCNL} is derived from \eqref{eq:realSpaceLNonGdef}. In order to make the derivation easy to follow, we take into account the terms in \eqref{eq:realSpaceLNonGdef} one by one and derive the final expression of \eqref{eq:deffCNL} step by step.

\subsection{One field}

Let us consider the following\footnote{Note that $\langle\zeta_\G(\bm{x})^2\rangle$ is essentially a constant ensuring that $\langle\zeta(\bm{x})\rangle = 0$ to quadratic order, which fixes the gauge freedom in defining $\zeta$. However, when $\zeta_\G$ has a scale-invariant or red power spectrum, the real-space 2-point function presents an IR divergence, resulting in $\langle\zeta_\G(\bm{x})^2\rangle$ being infinite. As shown below in the bispectrum calculation, this term matters only for $k=0$. Yet, both theoretically and observationally, we always have to consider an IR cut-off.},
\begin{equation}
    \zeta(\bm{x})=\zeta_\G(\bm{x})+f_\NL^\zeta\left(\zeta_\G(\bm{x})^2-\langle\zeta_\G(\bm{x})^2\rangle\right)+ \cdots\,,
\end{equation}
and evaluate $\zeta_{\bm{k}}$, the Fourier transform of $\zeta$, 
\begin{equation}
   \zeta_{\bm{k}}=\int\dd^3\bm{x}\,\zeta(\bm{x})e^{-i\bm{k}\cdot\bm{x}}
   =\int\dd^3\bm{x}\,\zeta_\G(\bm{x})e^{-i\bm{k}\cdot\bm{x}}
   +f_\NL^\zeta\bigg(\int\dd^3\bm{x}\,\zeta_\G(\bm{x})^2e^{-i\bm{k}\cdot\bm{x}}
   -\int\dd^3\bm{x}\,\langle\zeta_\G(\bm{x})^2\rangle e^{-i\bm{k}\cdot\bm{x}}\bigg)\,.
\end{equation}
The first term simply evaluates to $\zeta^\G_{\bm{k}}$, the Fourier transform of $\zeta_\G(\bm{x})$.
For the second term, after inserting $\zeta_\G(\bm{x})=\int_{\bm{p}}\zeta^\G_{\bm{p}}\exp(i\bm{p}\cdot\bm{x})$, we have
\begin{align}
    \int\dd^3\bm{x}\, \zeta_\G(\bm{x})^2 e^{-i\bm{k}\cdot\bm{x}}&=\int\dd^3\bm{x}\,
    \int_{\bm{p}}\,\zeta^\G_{\bm{p}} e^{i\bm{p}\cdot\bm{x}} \int_{\bm{q}}\,\zeta^\G_{\bm{q}} e^{i\bm{q}\cdot\bm{x}}e^{-i\bm{k}\cdot\bm{x}}
    = \int_{\bm{p}}\, \int_{\bm{q}}\,\zeta^\G_{\bm{p}}\zeta^\G_{\bm{q}}\int\dd^3\bm{x}\, e^{i(\bm{p}+\bm{q}-\bm{k})\cdot\bm{x}} \nonumber\\ 
    &= \int_{\bm{p}}\, \int_{\bm{q}}\,\zeta^\G_{\bm{p}}\zeta^\G_{\bm{q}} (2\pi)^3\delta^{(3)}(\bm{p}+\bm{q}-\bm{k})= \int_{\bm{p}}\,\zeta^\G_{\bm{p}}\zeta^\G_{\bm{k}-\bm{p}}\,,
\end{align}
and for the last term,
\begin{align}
    \int\dd^3\bm{x}\, \langle\zeta_\G(\bm{x})^2\rangle e^{-i\bm{k}\cdot\bm{x}}=\langle\zeta_\G(\bm{x})^2\rangle (2\pi)^3\delta^{(3)}(\bm{k})\,,   
\end{align}
which as mentioned before drops out for all values of $k$ except for $k=0$.
Putting all the terms together, we can write
\begin{align}
    \zeta_{\bm{k}}=\zeta^\G_{\bm{k}}+\zeta^{\NL}_{\bm{k}}\,, \qquad \zeta^\NL_{\bm{k}}\equiv f_\NL^\zeta\int_{\bm{p}}\,\zeta^\G_{\bm{p}}\zeta^\G_{\bm{k}-\bm{p}}-(2\pi)^3f_\NL^\zeta\langle\zeta_\G(\bm{x})^2\rangle\delta^{(3)}(\bm{k})\,. \label{eq:zetakNldef}
\end{align}

Now let us calculate $\langle\zeta_{\bm{k}_1}\zeta_{\bm{k}_2}\zeta_{\bm{k}_3}\rangle$. Inserting the above, we have
\begin{align}
    \langle\zeta_{\bm{k}_1}\zeta_{\bm{k}_2}\zeta_{\bm{k}_3}\rangle&=\langle(\zeta^\G_{\bm{k}_1}+\zeta^{\NL}_{\bm{k}_1})(\zeta^\G_{\bm{k}_2}+\zeta^{\NL}_{\bm{k}_2})(\zeta^\G_{\bm{k}_3}+\zeta^{\NL}_{\bm{k}_3})\rangle \nonumber \\ 
    &=\langle\zeta^\G_{\bm{k}_1}\zeta^\G_{\bm{k}_2}\zeta^\NL_{\bm{k}_3}\rangle+[\bm{k}_2\leftrightarrow \bm{k}_3]+[\bm{k}_1\leftrightarrow \bm{k}_3]+\mathcal{O}\left((\zeta^\NL_{\bm{k}})^2\right) \,, \label{eq:Imoutoflabels}
\end{align}
where we set $\langle\zeta^\G_{\bm{k}_1}\zeta^\G_{\bm{k}_2}\zeta^\G_{\bm{k}_3}\rangle=0$ based on Wick's theorem. Let us then focus on evaluating $\langle\zeta^\G_{\bm{k}_1}\zeta^\G_{\bm{k}_2}\zeta^\NL_{\bm{k}_3}\rangle$. Inserting \eqref{eq:zetakNldef}, we have
{\allowdisplaybreaks
\begin{align}
    \langle\zeta^\G_{\bm{k}_1}\zeta^\G_{\bm{k}_2}\zeta^\NL_{\bm{k}_3}\rangle=&~
    f_\NL^\zeta\int_{\bm{p}}\, \langle\zeta^\G_{\bm{k}_1}\zeta^\G_{\bm{k}_2}\zeta^\G_{\bm{p}}\zeta^\G_{\bm{k}_3-\bm{p}}\rangle -\langle\zeta^\G_{\bm{k}_1}\zeta^\G_{\bm{k}_2}\rangle(2\pi)^3f_\NL^\zeta\langle\zeta_\G(\bm{x})^2\rangle\delta^{(3)}(\bm{k}_3)\nonumber \\ 
    =&~f_\NL^\zeta\int_{\bm{p}}\,\left(  
    \langle\zeta^\G_{\bm{k}_1}\zeta^\G_{\bm{p}}\rangle\langle\zeta^\G_{\bm{k}_2}\zeta^\G_{\bm{k}_3-\bm{p}}\rangle 
    +\langle\zeta^\G_{\bm{k}_1}\zeta^\G_{\bm{k}_3-\bm{p}}\rangle\langle\zeta^\G_{\bm{k}_2}\zeta^\G_{\bm{p}}\rangle 
    +\langle\zeta^\G_{\bm{k}_1}\zeta^\G_{\bm{k}_2}\rangle\langle\zeta^\G_{\bm{p}}\zeta^\G_{\bm{k}_3-\bm{p}}\rangle
    \right) \nonumber\\
    &-\langle\zeta^\G_{\bm{k}_1}\zeta^\G_{\bm{k}_2}\rangle(2\pi)^3f_\NL^\zeta\langle\zeta_\G(\bm{x})^2\rangle\,\delta^{(3)}(\bm{k}_3)\nonumber\\
    =&~2(2\pi)^3f_\NL^\zeta P_\zeta(k_1)P_\zeta(k_2)\,\delta^{(3)}(\bm{k}_1+\bm{k}_2+\bm{k}_3)\nonumber\\
    &+(2\pi)^6 f_\NL^\zeta P_\zeta(k_1)\, \delta^{(3)}
    (\bm{k}_1+\bm{k}_2)\delta^{(3)}
    (\bm{k}_3)\bigg(\int_{\bm{p}}P_\zeta(p)-\langle\zeta_\G(\bm{x})^2\rangle\bigg) \,, \label{eq:longderivationofggN}
\end{align}
}%
where the second line follows from Wick's theorem, and in the last two lines, the definition of the power spectrum from \eqref{eq:powersdef} is used. Noting that $\int_{\bm{p}}P_\zeta(p) =\langle\zeta_\G(\bm{x})^2\rangle$ is the Fourier transform of the 2-point function, the last term of \eqref{eq:longderivationofggN} vanishes regardless of the value of $\bm{k}_3$. Together with \eqref{eq:Imoutoflabels}, we thus have
\begin{align}
    \langle\zeta_{\bm{k}_1}\zeta_{\bm{k}_2}\zeta_{\bm{k}_3}\rangle=2(2\pi)^3f_\NL^\zeta \left[
    P_\zeta(k_1)P_\zeta(k_2)+P_\zeta(k_1)P_\zeta(k_3)+P_\zeta(k_2)P_\zeta(k_3)\right]\delta^{(3)}\bigg(\sum_{n=1}^3\bm{k}_n\bigg)\,.\label{eq:finalfor<zzz>}
\end{align}
This gives the first line of \eqref{eq:deffCNL}.

\subsection{Two fields}

Next, we consider the full definition of $\zeta(\bm{x})$ and $\chi(\bm{x})$ with local-type non-Gaussianities considered in \eqref{eq:realSpaceLNonGdef}. This updates \eqref{eq:zetakNldef} as
\begin{align}
    \zeta_{\bm{k}}&=\zeta^\G_{\bm{k}}+\zeta^{\NL}_{\bm{k}}\,, & \zeta^\NL_{\bm{k}}&\equiv f_\NL^\zeta\int_{\bm{p}}\,\zeta^\G_{\bm{p}}\zeta^\G_{\bm{k}-\bm{p}}-(2\pi)^3f_\NL^\zeta\langle\zeta_\G(\bm{x})^2\rangle\delta^{(3)}(\bm{k})+C_\NL^\zeta\int_{\bm{p}}\,\zeta^\G_{\bm{p}}\chi^\G_{\bm{k}-\bm{p}}\nonumber\\
    \chi_{\bm{k}}&=\chi^\G_{\bm{k}}+\chi^{\NL}_{\bm{k}}\,, & \chi^\NL_{\bm{k}}&\equiv f_\NL^\chi\int_{\bm{p}}\,\chi^\G_{\bm{p}}\chi^\G_{\bm{k}-\bm{p}}-(2\pi)^3f_\NL^\chi\langle\chi_\G(\bm{x})^2\rangle\delta^{(3)}(\bm{k})+C_\NL^\chi\int_{\bm{p}}\,\chi^\G_{\bm{p}}\zeta^\G_{\bm{k}-\bm{p}}\,.\label{eq:zetachiNLfulldef}
\end{align}
We then need to repeat the derivation of $\langle\zeta^\G_{\bm{k}_1}\zeta^\G_{\bm{k}_2}\zeta^\NL_{\bm{k}_3}\rangle$ in \eqref{eq:longderivationofggN}. The presence of $C_\NL^\zeta$ brings an extra term compared to \eqref{eq:longderivationofggN},
\begin{align}
    \langle\zeta^\G_{\bm{k}_1}\zeta^\G_{\bm{k}_2}\zeta^\NL_{\bm{k}_3}\rangle \supset &~
    C_\NL^\zeta\int_{\bm{p}}\, \langle\zeta^\G_{\bm{k}_1}\zeta^\G_{\bm{k}_2}\zeta^\G_{\bm{p}}\chi^\G_{\bm{k}_3-\bm{p}}\rangle \nonumber \\ 
    =&~C_\NL^\zeta\int_{\bm{p}}\,\left(  
    \langle\zeta^\G_{\bm{k}_1}\zeta^\G_{\bm{p}}\rangle\langle\zeta^\G_{\bm{k}_2}\chi^\G_{\bm{k}_3-\bm{p}}\rangle 
    +\langle\zeta^\G_{\bm{k}_1}\chi^\G_{\bm{k}_3-\bm{p}}\rangle\langle\zeta^\G_{\bm{k}_2}\zeta^\G_{\bm{p}}\rangle 
    +\langle\zeta^\G_{\bm{k}_1}\zeta^\G_{\bm{k}_2}\rangle\langle\zeta^\G_{\bm{p}}\chi^\G_{\bm{k}_3-\bm{p}}\rangle
    \right)=0 \,, \label{eq:4point=0}
\end{align}
since $\langle\zeta^\G_{\bm{k}}\chi^\G_{\bm{p}}\rangle=0$ for any $\bm{k}$ and $\bm{p}$; recall \eqref{eq:quantization} and \eqref{eq:powersdef}. Therefore, \eqref{eq:finalfor<zzz>} still holds.

To derive the second line of \eqref{eq:deffCNL}, we should start from the correlator $\langle\zeta_{\bm{k}_1}\zeta_{\bm{k}_2}\chi_{\bm{k}_3}\rangle$:
\begin{align}
    \langle\zeta_{\bm{k}_1}\zeta_{\bm{k}_2}\chi_{\bm{k}_3}\rangle&=\langle(\zeta^\G_{\bm{k}_1}+\zeta^{\NL}_{\bm{k}_1})(\zeta^\G_{\bm{k}_2}+\zeta^{\NL}_{\bm{k}_2})(\chi^\G_{\bm{k}_3}+\chi^{\NL}_{\bm{k}_3})\rangle \nonumber \\ &=\langle\zeta^\G_{\bm{k}_1}\zeta^\NL_{\bm{k}_2}\chi^\G_{\bm{k}_3}\rangle+[\bm{k}_1\leftrightarrow \bm{k}_2]+\langle\zeta^\G_{\bm{k}_1}\zeta^\G_{\bm{k}_2}\chi^\NL_{\bm{k}_3}\rangle+\mathcal{O}\left((\zeta_{\bm{k}}^\NL)^2,\zeta_{\bm{k}}^\NL\chi_{\bm{k}}^\NL\right) \,.
\end{align}
First, we show that $\langle\zeta^\G_{\bm{k}_1}\zeta^\G_{\bm{k}_2}\chi^\NL_{\bm{k}_3}\rangle=0$, hence only the first two terms will contribute. Indeed, using \eqref{eq:zetachiNLfulldef},
{\allowdisplaybreaks
\begin{align}
    \langle\zeta^\G_{\bm{k}_1}\zeta^\G_{\bm{k}_2}\chi^\NL_{\bm{k}_3}\rangle=&~
    f_\NL^\chi\int_{\bm{p}}\, \langle\zeta^\G_{\bm{k}_1}\zeta^\G_{\bm{k}_2}\chi^\G_{\bm{p}}\chi^\G_{\bm{k}_3-\bm{p}}\rangle -\langle\zeta^\G_{\bm{k}_1}\zeta^\G_{\bm{k}_2}\rangle(2\pi)^3f_\NL^\chi\langle\chi_\G(\bm{x})^2\rangle\delta^{(3)}(\bm{k}_3)\nonumber\\
    &+C_\NL^\chi\int_{\bm{p}}\, \langle\zeta^\G_{\bm{k}_1}\zeta^\G_{\bm{k}_2}\chi^\G_{\bm{p}}\zeta^\G_{\bm{k}_3-\bm{p}}\rangle\nonumber \\ 
    =&~f_\NL^\chi\langle\zeta^\G_{\bm{k}_1}\zeta^\G_{\bm{k}_2}\rangle \left(  \int_{\bm{p}}\,
    \langle\chi^\G_{\bm{p}}\chi^\G_{\bm{k}_3-\bm{p}}\rangle-(2\pi)^3\langle\chi_\G(\bm{x})^2\rangle\delta^{(3)}(\bm{k}_3)\right)\nonumber\\
    =&~(2\pi)^3f_\NL^\chi\langle\zeta^\G_{\bm{k}_1}\zeta^\G_{\bm{k}_2}\rangle\delta^{(3)}(\bm{k}_3) \left(  \int_{\bm{p}}\,
    P_\chi(p)-\langle\chi_\G(\bm{x})^2\rangle\right)=0\,, 
\end{align}
}%
where in the first equality the last term vanishes in a similar fashion to \eqref{eq:4point=0}. Lastly by similar arguments,
{\allowdisplaybreaks
\begin{align}
    \langle\zeta^\G_{\bm{k}_1}\zeta^\NL_{\bm{k}_2}\chi^\G_{\bm{k}_3}\rangle=&~
    f_\NL^\zeta\int_{\bm{p}}\, \langle\zeta^\G_{\bm{k}_1}\zeta^\G_{\bm{p}}\zeta^\G_{\bm{k}_2-\bm{p}}\chi^\G_{\bm{k}_3}\rangle -(2\pi)^3f_\NL^\zeta\langle\zeta_\G(\bm{x})^2\rangle\delta^{(3)}(\bm{k}_3)\langle\zeta^\G_{\bm{k}_1}\chi^\G_{\bm{k}_3}\rangle\nonumber\\
    &+C_\NL^\zeta\int_{\bm{p}}\, \langle\zeta^\G_{\bm{k}_1}\zeta^\G_{\bm{p}}\chi^\G_{\bm{k}_2-\bm{p}}\chi^\G_{\bm{k}_3}\rangle \nonumber \\ 
    =&~C_\NL^\zeta\int_{\bm{p}}\, \langle\zeta^\G_{\bm{k}_1}\zeta^\G_{\bm{p}}\rangle\langle\chi^\G_{\bm{k}_2-\bm{p}}\chi^\G_{\bm{k}_3}\rangle =
    (2\pi)^3 C_\NL^\zeta P_\zeta(k_1)P_\chi(k_3)\delta^{(3)}\bigg(\sum_{n=1}^3\bm{k}_n\bigg)\,.
\end{align}
}%
Therefore, 
\begin{align}
    \langle\zeta_{\bm{k}_1}\zeta_{\bm{k}_2}\chi_{\bm{k}_3}\rangle=(2\pi)^3 C_\NL^\zeta\left [P_\zeta(k_1)P_\chi(k_3)+P_\zeta(k_2)P_\chi(k_3)\right ]\delta^{(3)}\bigg(\sum_{n=1}^3\bm{k}_n\bigg)\,.
\end{align}
This gives the second line of \eqref{eq:deffCNL}. By the symmetry $\zeta \leftrightarrow \chi$, the last two lines of \eqref{eq:deffCNL} also follow.


\addcontentsline{toc}{section}{References}

\let\oldbibliography\thebibliography
\renewcommand{\thebibliography}[1]{
  \oldbibliography{#1}
  \setlength{\itemsep}{0pt}
  \footnotesize
}

\bibliographystyle{JHEP2}
\bibliography{refs}

\providecommand{\url}[1]{#1}\providecommand{\href}[2]{#2}\begingroup\raggedright\begin{thebibliography}{100}

\bibitem{Guth:1980zm}
A.H.~Guth, \emph{{The Inflationary Universe: A Possible Solution to the Horizon
  and Flatness Problems}},
  \href{https://doi.org/10.1103/PhysRevD.23.347}{\emph{Phys. Rev. D} {\bfseries
  23} (1981) 347}.

\bibitem{Linde:1981mu}
A.D.~Linde, \emph{{A New Inflationary Universe Scenario: A Possible Solution of
  the Horizon, Flatness, Homogeneity, Isotropy and Primordial Monopole
  Problems}}, \href{https://doi.org/10.1016/0370-2693(82)91219-9}{\emph{Phys.
  Lett. B} {\bfseries 108} (1982) 389}.

\bibitem{Mukhanov:1981xt}
V.F.~Mukhanov and G.V.~Chibisov, \emph{{Quantum Fluctuations and a Nonsingular
  Universe}}, {\emph{JETP Lett.} {\bfseries 33} (1981) 532},
  \url{http://www.jetpletters.ru/ps/1510/article_23079.shtml}.

\bibitem{Bardeen:1983qw}
J.M.~Bardeen, P.J.~Steinhardt and M.S.~Turner, \emph{{Spontaneous Creation of
  Almost Scale-Free Density Perturbations in an Inflationary Universe}},
  \href{https://doi.org/10.1103/PhysRevD.28.679}{\emph{Phys. Rev. D} {\bfseries
  28} (1983) 679}.

\bibitem{Brandenberger:2012uj}
R.~Brandenberger, \emph{{Do we have a Theory of Early Universe Cosmology?}},
  \href{https://doi.org/10.1016/j.shpsb.2013.09.008}{\emph{Stud. Hist. Phil.
  Sci. B} {\bfseries 46} (2014) 109}
  [\href{https://arxiv.org/abs/1204.6108}{{\ttfamily arXiv:1204.6108}}].

\bibitem{Ijjas:2013vea}
A.~Ijjas, P.J.~Steinhardt and A.~Loeb, \emph{{Inflationary paradigm in trouble
  after Planck2013}},
  \href{https://doi.org/10.1016/j.physletb.2013.05.023}{\emph{Phys. Lett. B}
  {\bfseries 723} (2013) 261}
  [\href{https://arxiv.org/abs/1304.2785}{{\ttfamily arXiv:1304.2785}}].

\bibitem{Ijjas:2014nta}
A.~Ijjas, P.J.~Steinhardt and A.~Loeb, \emph{{Inflationary schism}},
  \href{https://doi.org/10.1016/j.physletb.2014.07.012}{\emph{Phys. Lett. B}
  {\bfseries 736} (2014) 142}
  [\href{https://arxiv.org/abs/1402.6980}{{\ttfamily arXiv:1402.6980}}].

\bibitem{Agrawal:2018own}
P.~Agrawal, G.~Obied, P.J.~Steinhardt and C.~Vafa, \emph{{On the Cosmological
  Implications of the String Swampland}},
  \href{https://doi.org/10.1016/j.physletb.2018.07.040}{\emph{Phys. Lett. B}
  {\bfseries 784} (2018) 271}
  [\href{https://arxiv.org/abs/1806.09718}{{\ttfamily arXiv:1806.09718}}].

\bibitem{DiTucci:2019xcr}
A.~Di~Tucci, J.~Feldbrugge, J.L.~Lehners and N.~Turok, \emph{{Quantum
  Incompleteness of Inflation}},
  \href{https://doi.org/10.1103/PhysRevD.100.063517}{\emph{Phys. Rev. D}
  {\bfseries 100} (2019) 063517}
  [\href{https://arxiv.org/abs/1906.09007}{{\ttfamily arXiv:1906.09007}}].

\bibitem{Bedroya:2019tba}
A.~Bedroya, R.~Brandenberger, M.~Loverde and C.~Vafa, \emph{{Trans-Planckian
  Censorship and Inflationary Cosmology}},
  \href{https://doi.org/10.1103/PhysRevD.101.103502}{\emph{Phys. Rev. D}
  {\bfseries 101} (2020) 103502}
  [\href{https://arxiv.org/abs/1909.11106}{{\ttfamily arXiv:1909.11106}}].

\bibitem{Borde:2001nh}
A.~Borde, A.H.~Guth and A.~Vilenkin, \emph{{Inflationary space-times are
  incomplete in past directions}},
  \href{https://doi.org/10.1103/PhysRevLett.90.151301}{\emph{Phys. Rev. Lett.}
  {\bfseries 90} (2003) 151301}
  [\href{https://arxiv.org/abs/gr-qc/0110012}{{\ttfamily gr-qc/0110012}}].

\bibitem{Yoshida:2018ndv}
D.~Yoshida and J.~Quintin, \emph{{Maximal extensions and singularities in
  inflationary spacetimes}},
  \href{https://doi.org/10.1088/1361-6382/aacf4b}{\emph{Class. Quant. Grav.}
  {\bfseries 35} (2018) 155019}
  [\href{https://arxiv.org/abs/1803.07085}{{\ttfamily arXiv:1803.07085}}].

\bibitem{Geshnizjani:2023hyd}
G.~Geshnizjani, E.~Ling and J.~Quintin, \emph{{On the initial singularity and
  extendibility of flat quasi-de Sitter spacetimes}},
  \href{https://doi.org/10.1007/JHEP10(2023)182}{\emph{JHEP} {\bfseries 10}
  (2023) 182} [\href{https://arxiv.org/abs/2305.01676}{{\ttfamily
  arXiv:2305.01676}}].

\bibitem{Battefeld:2014uga}
D.~Battefeld and P.~Peter, \emph{{A Critical Review of Classical Bouncing
  Cosmologies}},
  \href{https://doi.org/10.1016/j.physrep.2014.12.004}{\emph{Phys. Rept.}
  {\bfseries 571} (2015) 1} [\href{https://arxiv.org/abs/1406.2790}{{\ttfamily
  arXiv:1406.2790}}].

\bibitem{Brandenberger:2016vhg}
R.~Brandenberger and P.~Peter, \emph{{Bouncing Cosmologies: Progress and
  Problems}}, \href{https://doi.org/10.1007/s10701-016-0057-0}{\emph{Found.
  Phys.} {\bfseries 47} (2017) 797}
  [\href{https://arxiv.org/abs/1603.05834}{{\ttfamily arXiv:1603.05834}}].

\bibitem{Gasperini:1996fu}
M.~Gasperini, M.~Maggiore and G.~Veneziano, \emph{{Towards a nonsingular
  pre-big bang cosmology}},
  \href{https://doi.org/10.1016/S0550-3213(97)00149-1}{\emph{Nucl. Phys. B}
  {\bfseries 494} (1997) 315}
  [\href{https://arxiv.org/abs/hep-th/9611039}{{\ttfamily hep-th/9611039}}].

\bibitem{Gasperini:2003pb}
M.~Gasperini, M.~Giovannini and G.~Veneziano, \emph{{Perturbations in a
  nonsingular bouncing universe}},
  \href{https://doi.org/10.1016/j.physletb.2003.07.028}{\emph{Phys. Lett. B}
  {\bfseries 569} (2003) 113}
  [\href{https://arxiv.org/abs/hep-th/0306113}{{\ttfamily hep-th/0306113}}].

\bibitem{Gasperini:2004ss}
M.~Gasperini, M.~Giovannini and G.~Veneziano, \emph{{Cosmological perturbations
  across a curvature bounce}},
  \href{https://doi.org/10.1016/j.nuclphysb.2004.06.020}{\emph{Nucl. Phys. B}
  {\bfseries 694} (2004) 206}
  [\href{https://arxiv.org/abs/hep-th/0401112}{{\ttfamily hep-th/0401112}}].

\bibitem{Quintin:2018loc}
J.~Quintin, R.H.~Brandenberger, M.~Gasperini and G.~Veneziano, \emph{{Stringy
  black-hole gas in $\alpha^{\prime}$-corrected dilaton gravity}},
  \href{https://doi.org/10.1103/PhysRevD.98.103519}{\emph{Phys. Rev. D}
  {\bfseries 98} (2018) 103519}
  [\href{https://arxiv.org/abs/1809.01658}{{\ttfamily arXiv:1809.01658}}].

\bibitem{Hohm:2019jgu}
O.~Hohm and B.~Zwiebach, \emph{{Duality invariant cosmology to all orders in
  $\alpha^{\prime}$}},
  \href{https://doi.org/10.1103/PhysRevD.100.126011}{\emph{Phys. Rev. D}
  {\bfseries 100} (2019) 126011}
  [\href{https://arxiv.org/abs/1905.06963}{{\ttfamily arXiv:1905.06963}}].

\bibitem{Wang:2019kez}
P.~Wang, H.~Wu, H.~Yang and S.~Ying, \emph{{Non-singular string cosmology via
  $\alpha^{\prime}$ corrections}},
  \href{https://doi.org/10.1007/JHEP10(2019)263}{\emph{JHEP} {\bfseries 10}
  (2019) 263} [\href{https://arxiv.org/abs/1909.00830}{{\ttfamily
  arXiv:1909.00830}}].

\bibitem{Wang:2019dcj}
P.~Wang, H.~Wu, H.~Yang and S.~Ying, \emph{{Construct $\alpha^{\prime}$
  corrected or loop corrected solutions without curvature singularities}},
  \href{https://doi.org/10.1007/JHEP01(2020)164}{\emph{JHEP} {\bfseries 01}
  (2020) 164} [\href{https://arxiv.org/abs/1910.05808}{{\ttfamily
  arXiv:1910.05808}}].

\bibitem{Quintin:2021eup}
J.~Quintin, H.~Bernardo and G.~Franzmann, \emph{{Cosmology at the top of the
  \ensuremath{\alpha'} tower}},
  \href{https://doi.org/10.1007/JHEP07(2021)149}{\emph{JHEP} {\bfseries 07}
  (2021) 149} [\href{https://arxiv.org/abs/2105.01083}{{\ttfamily
  arXiv:2105.01083}}].

\bibitem{Gasperini:2023tus}
M.~Gasperini and G.~Veneziano, \emph{{Non-singular pre-big bang scenarios from
  all-order \ensuremath{\alpha'} corrections}},
  \href{https://doi.org/10.1007/JHEP07(2023)144}{\emph{JHEP} {\bfseries 07}
  (2023) 144} [\href{https://arxiv.org/abs/2305.00222}{{\ttfamily
  arXiv:2305.00222}}].

\bibitem{Date:2004fj}
G.~Date and G.M.~Hossain, \emph{{Genericity of big bounce in isotropic loop
  quantum cosmology}},
  \href{https://doi.org/10.1103/PhysRevLett.94.011302}{\emph{Phys. Rev. Lett.}
  {\bfseries 94} (2005) 011302}
  [\href{https://arxiv.org/abs/gr-qc/0407074}{{\ttfamily gr-qc/0407074}}].

\bibitem{Singh:2006im}
P.~Singh, K.~Vandersloot and G.V.~Vereshchagin, \emph{{Non-singular bouncing
  universes in loop quantum cosmology}},
  \href{https://doi.org/10.1103/PhysRevD.74.043510}{\emph{Phys. Rev. D}
  {\bfseries 74} (2006) 043510}
  [\href{https://arxiv.org/abs/gr-qc/0606032}{{\ttfamily gr-qc/0606032}}].

\bibitem{deHaro:2012xj}
J.~de~Haro, \emph{{Does loop quantum cosmology replace the big rip singularity
  by a non-singular bounce?}},
  \href{https://doi.org/10.1088/1475-7516/2012/11/037}{\emph{JCAP} {\bfseries
  11} (2012) 037} [\href{https://arxiv.org/abs/1207.3621}{{\ttfamily
  arXiv:1207.3621}}].

\bibitem{Wilson-Ewing:2012lmx}
E.~Wilson-Ewing, \emph{{The Matter Bounce Scenario in Loop Quantum Cosmology}},
  \href{https://doi.org/10.1088/1475-7516/2013/03/026}{\emph{JCAP} {\bfseries
  03} (2013) 026} [\href{https://arxiv.org/abs/1211.6269}{{\ttfamily
  arXiv:1211.6269}}].

\bibitem{Wilson-Ewing:2013bla}
E.~Wilson-Ewing, \emph{{Ekpyrotic loop quantum cosmology}},
  \href{https://doi.org/10.1088/1475-7516/2013/08/015}{\emph{JCAP} {\bfseries
  08} (2013) 015} [\href{https://arxiv.org/abs/1306.6582}{{\ttfamily
  arXiv:1306.6582}}].

\bibitem{Cai:2014zga}
Y.F.~Cai and E.~Wilson-Ewing, \emph{{Non-singular bounce scenarios in loop
  quantum cosmology and the effective field description}},
  \href{https://doi.org/10.1088/1475-7516/2014/03/026}{\emph{JCAP} {\bfseries
  03} (2014) 026} [\href{https://arxiv.org/abs/1402.3009}{{\ttfamily
  arXiv:1402.3009}}].

\bibitem{Wilson-Ewing:2017vju}
E.~Wilson-Ewing, \emph{{The loop quantum cosmology bounce as a Kasner
  transition}}, \href{https://doi.org/10.1088/1361-6382/aaab8b}{\emph{Class.
  Quant. Grav.} {\bfseries 35} (2018) 065005}
  [\href{https://arxiv.org/abs/1711.10943}{{\ttfamily arXiv:1711.10943}}].

\bibitem{Afshordi:2006ad}
N.~Afshordi, D.J.H.~Chung and G.~Geshnizjani, \emph{{Cuscuton: A Causal Field
  Theory with an Infinite Speed of Sound}},
  \href{https://doi.org/10.1103/PhysRevD.75.083513}{\emph{Phys. Rev. D}
  {\bfseries 75} (2007) 083513}
  [\href{https://arxiv.org/abs/hep-th/0609150}{{\ttfamily hep-th/0609150}}].

\bibitem{Afshordi:2007yx}
N.~Afshordi, D.J.H.~Chung, M.~Doran and G.~Geshnizjani, \emph{{Cuscuton
  Cosmology: Dark Energy meets Modified Gravity}},
  \href{https://doi.org/10.1103/PhysRevD.75.123509}{\emph{Phys. Rev. D}
  {\bfseries 75} (2007) 123509}
  [\href{https://arxiv.org/abs/astro-ph/0702002}{{\ttfamily
  astro-ph/0702002}}].

\bibitem{Bhattacharyya:2016mah}
J.~Bhattacharyya, A.~Coates, M.~Colombo, A.E.~Gumrukcuoglu and T.P.~Sotiriou,
  \emph{{Revisiting the cuscuton as a Lorentz-violating gravity theory}},
  \href{https://doi.org/10.1103/PhysRevD.97.064020}{\emph{Phys. Rev. D}
  {\bfseries 97} (2018) 064020}
  [\href{https://arxiv.org/abs/1612.01824}{{\ttfamily arXiv:1612.01824}}].

\bibitem{Gomes:2017tzd}
H.~Gomes and D.C.~Guariento, \emph{{Hamiltonian analysis of the cuscuton}},
  \href{https://doi.org/10.1103/PhysRevD.95.104049}{\emph{Phys. Rev. D}
  {\bfseries 95} (2017) 104049}
  [\href{https://arxiv.org/abs/1703.08226}{{\ttfamily arXiv:1703.08226}}].

\bibitem{Boruah:2017tvg}
S.S.~Boruah, H.J.~Kim and G.~Geshnizjani, \emph{{Theory of Cosmological
  Perturbations with Cuscuton}},
  \href{https://doi.org/10.1088/1475-7516/2017/07/022}{\emph{JCAP} {\bfseries
  07} (2017) 022} [\href{https://arxiv.org/abs/1704.01131}{{\ttfamily
  arXiv:1704.01131}}].

\bibitem{Lin:2017oow}
C.~Lin and S.~Mukohyama, \emph{{A Class of Minimally Modified Gravity
  Theories}}, \href{https://doi.org/10.1088/1475-7516/2017/10/033}{\emph{JCAP}
  {\bfseries 10} (2017) 033}
  [\href{https://arxiv.org/abs/1708.03757}{{\ttfamily arXiv:1708.03757}}].

\bibitem{Iyonaga:2018vnu}
A.~Iyonaga, K.~Takahashi and T.~Kobayashi, \emph{{Extended Cuscuton:
  Formulation}},
  \href{https://doi.org/10.1088/1475-7516/2018/12/002}{\emph{JCAP} {\bfseries
  12} (2018) 002} [\href{https://arxiv.org/abs/1809.10935}{{\ttfamily
  arXiv:1809.10935}}].

\bibitem{Mukohyama:2019unx}
S.~Mukohyama and K.~Noui, \emph{{Minimally Modified Gravity: a Hamiltonian
  Construction}},
  \href{https://doi.org/10.1088/1475-7516/2019/07/049}{\emph{JCAP} {\bfseries
  07} (2019) 049} [\href{https://arxiv.org/abs/1905.02000}{{\ttfamily
  arXiv:1905.02000}}].

\bibitem{Gao:2019twq}
X.~Gao and Z.B.~Yao, \emph{{Spatially covariant gravity theories with two
  tensorial degrees of freedom: the formalism}},
  \href{https://doi.org/10.1103/PhysRevD.101.064018}{\emph{Phys. Rev. D}
  {\bfseries 101} (2020) 064018}
  [\href{https://arxiv.org/abs/1910.13995}{{\ttfamily arXiv:1910.13995}}].

\bibitem{Aoki:2021zuy}
K.~Aoki, F.~Di~Filippo and S.~Mukohyama, \emph{{Non-uniqueness of massless
  transverse-traceless graviton}},
  \href{https://doi.org/10.1088/1475-7516/2021/05/071}{\emph{JCAP} {\bfseries
  05} (2021) 071} [\href{https://arxiv.org/abs/2103.15044}{{\ttfamily
  arXiv:2103.15044}}].

\bibitem{DeFelice:2022uxv}
A.~De~Felice, K.i.~Maeda, S.~Mukohyama and M.C.~Pookkillath, \emph{{Comparison
  of two theories of Type-IIa minimally modified gravity}},
  \href{https://doi.org/10.1103/PhysRevD.106.024028}{\emph{Phys. Rev. D}
  {\bfseries 106} (2022) 024028}
  [\href{https://arxiv.org/abs/2204.08294}{{\ttfamily arXiv:2204.08294}}].

\bibitem{Mylova:2023ddj}
M.~Mylova and N.~Afshordi, \emph{{Effective cuscuton theory}},
  \href{https://doi.org/10.1007/JHEP04(2024)144}{\emph{JHEP} {\bfseries 04}
  (2024) 144} [\href{https://arxiv.org/abs/2312.06066}{{\ttfamily
  arXiv:2312.06066}}].

\bibitem{Lin:2017fec}
C.~Lin, J.~Quintin and R.H.~Brandenberger, \emph{{Massive gravity and the
  suppression of anisotropies and gravitational waves in a matter-dominated
  contracting universe}},
  \href{https://doi.org/10.1088/1475-7516/2018/01/011}{\emph{JCAP} {\bfseries
  01} (2018) 011} [\href{https://arxiv.org/abs/1711.10472}{{\ttfamily
  arXiv:1711.10472}}].

\bibitem{Boruah:2018pvq}
S.S.~Boruah, H.J.~Kim, M.~Rouben and G.~Geshnizjani, \emph{{Cuscuton bounce}},
  \href{https://doi.org/10.1088/1475-7516/2018/08/031}{\emph{JCAP} {\bfseries
  08} (2018) 031} [\href{https://arxiv.org/abs/1802.06818}{{\ttfamily
  arXiv:1802.06818}}].

\bibitem{Quintin:2019orx}
J.~Quintin and D.~Yoshida, \emph{{Cuscuton gravity as a classically stable
  limiting curvature theory}},
  \href{https://doi.org/10.1088/1475-7516/2020/02/016}{\emph{JCAP} {\bfseries
  02} (2020) 016} [\href{https://arxiv.org/abs/1911.06040}{{\ttfamily
  arXiv:1911.06040}}].

\bibitem{Sakakihara:2020rdy}
Y.~Sakakihara, D.~Yoshida, K.~Takahashi and J.~Quintin, \emph{{Theories with
  limited extrinsic curvature and a nonsingular anisotropic universe}},
  \href{https://doi.org/10.1103/PhysRevD.102.084004}{\emph{Phys. Rev. D}
  {\bfseries 102} (2020) 084004}
  [\href{https://arxiv.org/abs/2005.10844}{{\ttfamily arXiv:2005.10844}}].

\bibitem{Lin:2010pf}
C.~Lin, R.H.~Brandenberger and L.~Perreault~Levasseur, \emph{{A Matter Bounce
  By Means of Ghost Condensation}},
  \href{https://doi.org/10.1088/1475-7516/2011/04/019}{\emph{JCAP} {\bfseries
  04} (2011) 019} [\href{https://arxiv.org/abs/1007.2654}{{\ttfamily
  arXiv:1007.2654}}].

\bibitem{Easson:2011zy}
D.A.~Easson, I.~Sawicki and A.~Vikman, \emph{{G-Bounce}},
  \href{https://doi.org/10.1088/1475-7516/2011/11/021}{\emph{JCAP} {\bfseries
  11} (2011) 021} [\href{https://arxiv.org/abs/1109.1047}{{\ttfamily
  arXiv:1109.1047}}].

\bibitem{Qiu:2011cy}
T.~Qiu, J.~Evslin, Y.F.~Cai, M.~Li and X.~Zhang, \emph{{Bouncing Galileon
  Cosmologies}},
  \href{https://doi.org/10.1088/1475-7516/2011/10/036}{\emph{JCAP} {\bfseries
  10} (2011) 036} [\href{https://arxiv.org/abs/1108.0593}{{\ttfamily
  arXiv:1108.0593}}].

\bibitem{Cai:2012va}
Y.F.~Cai, D.A.~Easson and R.~Brandenberger, \emph{{Towards a Nonsingular
  Bouncing Cosmology}},
  \href{https://doi.org/10.1088/1475-7516/2012/08/020}{\emph{JCAP} {\bfseries
  08} (2012) 020} [\href{https://arxiv.org/abs/1206.2382}{{\ttfamily
  arXiv:1206.2382}}].

\bibitem{Cai:2013kja}
Y.F.~Cai, E.~McDonough, F.~Duplessis and R.H.~Brandenberger, \emph{{Two Field
  Matter Bounce Cosmology}},
  \href{https://doi.org/10.1088/1475-7516/2013/10/024}{\emph{JCAP} {\bfseries
  10} (2013) 024} [\href{https://arxiv.org/abs/1305.5259}{{\ttfamily
  arXiv:1305.5259}}].

\bibitem{Easson:2013bda}
D.A.~Easson, I.~Sawicki and A.~Vikman, \emph{{When Matter Matters}},
  \href{https://doi.org/10.1088/1475-7516/2013/07/014}{\emph{JCAP} {\bfseries
  07} (2013) 014} [\href{https://arxiv.org/abs/1304.3903}{{\ttfamily
  arXiv:1304.3903}}].

\bibitem{Battarra:2014tga}
L.~Battarra, M.~Koehn, J.L.~Lehners and B.A.~Ovrut, \emph{{Cosmological
  Perturbations Through a Non-Singular Ghost-Condensate/Galileon Bounce}},
  \href{https://doi.org/10.1088/1475-7516/2014/07/007}{\emph{JCAP} {\bfseries
  07} (2014) 007} [\href{https://arxiv.org/abs/1404.5067}{{\ttfamily
  arXiv:1404.5067}}].

\bibitem{Ijjas:2016tpn}
A.~Ijjas and P.J.~Steinhardt, \emph{{Classically stable nonsingular
  cosmological bounces}},
  \href{https://doi.org/10.1103/PhysRevLett.117.121304}{\emph{Phys. Rev. Lett.}
  {\bfseries 117} (2016) 121304}
  [\href{https://arxiv.org/abs/1606.08880}{{\ttfamily arXiv:1606.08880}}].

\bibitem{Ijjas:2016vtq}
A.~Ijjas and P.J.~Steinhardt, \emph{{Fully stable cosmological solutions with a
  non-singular classical bounce}},
  \href{https://doi.org/10.1016/j.physletb.2016.11.047}{\emph{Phys. Lett. B}
  {\bfseries 764} (2017) 289}
  [\href{https://arxiv.org/abs/1609.01253}{{\ttfamily arXiv:1609.01253}}].

\bibitem{Libanov:2016kfc}
M.~Libanov, S.~Mironov and V.~Rubakov, \emph{{Generalized Galileons:
  instabilities of bouncing and Genesis cosmologies and modified Genesis}},
  \href{https://doi.org/10.1088/1475-7516/2016/08/037}{\emph{JCAP} {\bfseries
  08} (2016) 037} [\href{https://arxiv.org/abs/1605.05992}{{\ttfamily
  arXiv:1605.05992}}].

\bibitem{Kobayashi:2016xpl}
T.~Kobayashi, \emph{{Generic instabilities of nonsingular cosmologies in
  Horndeski theory: A no-go theorem}},
  \href{https://doi.org/10.1103/PhysRevD.94.043511}{\emph{Phys. Rev. D}
  {\bfseries 94} (2016) 043511}
  [\href{https://arxiv.org/abs/1606.05831}{{\ttfamily arXiv:1606.05831}}].

\bibitem{deRham:2017aoj}
C.~de~Rham and S.~Melville, \emph{{Unitary null energy condition violation in
  P(X) cosmologies}},
  \href{https://doi.org/10.1103/PhysRevD.95.123523}{\emph{Phys. Rev. D}
  {\bfseries 95} (2017) 123523}
  [\href{https://arxiv.org/abs/1703.00025}{{\ttfamily arXiv:1703.00025}}].

\bibitem{Dobre:2017pnt}
D.A.~Dobre, A.V.~Frolov, J.T.~G\'alvez~Ghersi, S.~Ramazanov and A.~Vikman,
  \emph{{Unbraiding the Bounce: Superluminality around the Corner}},
  \href{https://doi.org/10.1088/1475-7516/2018/03/020}{\emph{JCAP} {\bfseries
  03} (2018) 020} [\href{https://arxiv.org/abs/1712.10272}{{\ttfamily
  arXiv:1712.10272}}].

\bibitem{Ijjas:2017pei}
A.~Ijjas, \emph{{Space-time slicing in Horndeski theories and its implications
  for non-singular bouncing solutions}},
  \href{https://doi.org/10.1088/1475-7516/2018/02/007}{\emph{JCAP} {\bfseries
  02} (2018) 007} [\href{https://arxiv.org/abs/1710.05990}{{\ttfamily
  arXiv:1710.05990}}].

\bibitem{Akama:2017jsa}
S.~Akama and T.~Kobayashi, \emph{{Generalized multi-Galileons, covariantized
  new terms, and the no-go theorem for nonsingular cosmologies}},
  \href{https://doi.org/10.1103/PhysRevD.95.064011}{\emph{Phys. Rev. D}
  {\bfseries 95} (2017) 064011}
  [\href{https://arxiv.org/abs/1701.02926}{{\ttfamily arXiv:1701.02926}}].

\bibitem{Banerjee:2018svi}
S.~Banerjee, Y.F.~Cai and E.N.~Saridakis, \emph{{Evading the theoretical no-go
  theorem for nonsingular bounces in Horndeski/Galileon cosmology}},
  \href{https://doi.org/10.1088/1361-6382/ab256a}{\emph{Class. Quant. Grav.}
  {\bfseries 36} (2019) 135009}
  [\href{https://arxiv.org/abs/1808.01170}{{\ttfamily arXiv:1808.01170}}].

\bibitem{Mironov:2019fop}
S.~Mironov, \emph{{Mathematical Formulation of the No-Go Theorem in Horndeski
  Theory}}, \href{https://doi.org/10.3390/universe5020052}{\emph{Universe}
  {\bfseries 5} (2019) 52}.

\bibitem{Ye:2019sth}
G.~Ye and Y.S.~Piao, \emph{{Bounce in general relativity and higher-order
  derivative operators}},
  \href{https://doi.org/10.1103/PhysRevD.99.084019}{\emph{Phys. Rev. D}
  {\bfseries 99} (2019) 084019}
  [\href{https://arxiv.org/abs/1901.08283}{{\ttfamily arXiv:1901.08283}}].

\bibitem{Ye:2019frg}
G.~Ye and Y.S.~Piao, \emph{{Implication of GW170817 for cosmological bounces}},
  \href{https://doi.org/10.1088/0253-6102/71/4/427}{\emph{Commun. Theor. Phys.}
  {\bfseries 71} (2019) 427}
  [\href{https://arxiv.org/abs/1901.02202}{{\ttfamily arXiv:1901.02202}}].

\bibitem{Creminelli:2016zwa}
P.~Creminelli, D.~Pirtskhalava, L.~Santoni and E.~Trincherini, \emph{{Stability
  of Geodesically Complete Cosmologies}},
  \href{https://doi.org/10.1088/1475-7516/2016/11/047}{\emph{JCAP} {\bfseries
  11} (2016) 047} [\href{https://arxiv.org/abs/1610.04207}{{\ttfamily
  arXiv:1610.04207}}].

\bibitem{Cai:2016thi}
Y.~Cai, Y.~Wan, H.G.~Li, T.~Qiu and Y.S.~Piao, \emph{{The Effective Field
  Theory of nonsingular cosmology}},
  \href{https://doi.org/10.1007/JHEP01(2017)090}{\emph{JHEP} {\bfseries 01}
  (2017) 090} [\href{https://arxiv.org/abs/1610.03400}{{\ttfamily
  arXiv:1610.03400}}].

\bibitem{Cai:2017tku}
Y.~Cai, H.G.~Li, T.~Qiu and Y.S.~Piao, \emph{{The Effective Field Theory of
  nonsingular cosmology: II}},
  \href{https://doi.org/10.1140/epjc/s10052-017-4938-y}{\emph{Eur. Phys. J. C}
  {\bfseries 77} (2017) 369}
  [\href{https://arxiv.org/abs/1701.04330}{{\ttfamily arXiv:1701.04330}}].

\bibitem{Volkova:2024mbn}
V.E.~Volkova and S.A.~Mironov, \emph{{Nonsingular cosmological scenarios in
  scalar-tensor theories and their stability}},
  \href{https://doi.org/10.3367/UFNr.2024.12.039826}{\emph{Usp. Fiz. Nauk}
  {\bfseries 195} (2025) 163}
  [\href{https://arxiv.org/abs/2409.16108}{{\ttfamily arXiv:2409.16108}}].

\bibitem{Bohnenblust:2024mou}
L.~Bohnenblust, S.~Giardino, L.~Heisenberg and N.~Nussbaumer, \emph{{To bounce
  or not to bounce in generalized Proca theory and beyond}},
  \href{https://arxiv.org/abs/2412.03977}{{\ttfamily arXiv:2412.03977}}.

\bibitem{An:2025xeb}
O.S.~An, J.U.~Kang, Y.J.~Kim, U.R.~Mun and U.G.~Ri, \emph{{Fully viable DHOST
  bounce with extra scalar}},
  \href{https://doi.org/10.1007/JHEP05(2025)005}{\emph{JHEP} {\bfseries 05}
  (2025) 005} [\href{https://arxiv.org/abs/2501.09985}{{\ttfamily
  arXiv:2501.09985}}].

\bibitem{Kolevatov:2017voe}
R.~Kolevatov, S.~Mironov, N.~Sukhov and V.~Volkova, \emph{{Cosmological bounce
  and Genesis beyond Horndeski}},
  \href{https://doi.org/10.1088/1475-7516/2017/08/038}{\emph{JCAP} {\bfseries
  08} (2017) 038} [\href{https://arxiv.org/abs/1705.06626}{{\ttfamily
  arXiv:1705.06626}}].

\bibitem{Cai:2017dyi}
Y.~Cai and Y.S.~Piao, \emph{{A covariant Lagrangian for stable nonsingular
  bounce}}, \href{https://doi.org/10.1007/JHEP09(2017)027}{\emph{JHEP}
  {\bfseries 09} (2017) 027}
  [\href{https://arxiv.org/abs/1705.03401}{{\ttfamily arXiv:1705.03401}}].

\bibitem{Mironov:2018oec}
S.~Mironov, V.~Rubakov and V.~Volkova, \emph{{Bounce beyond Horndeski with GR
  asymptotics and $\gamma$-crossing}},
  \href{https://doi.org/10.1088/1475-7516/2018/10/050}{\emph{JCAP} {\bfseries
  10} (2018) 050} [\href{https://arxiv.org/abs/1807.08361}{{\ttfamily
  arXiv:1807.08361}}].

\bibitem{Kolevatov:2018mhu}
R.~Kolevatov, S.~Mironov, V.~Rubakov, N.~Sukhov and V.~Volkova,
  \emph{{Cosmological bounce in Horndeski theory and beyond}},
  \href{https://doi.org/10.1051/epjconf/201819107013}{\emph{EPJ Web Conf.}
  {\bfseries 191} (2018) 07013}.

\bibitem{Volkova:2019jlj}
V.E.~Volkova, S.A.~Mironov and V.A.~Rubakov, \emph{{Cosmological Scenarios with
  Bounce and Genesis in Horndeski Theory and Beyond}},
  \href{https://doi.org/10.1134/S1063776119100236}{\emph{J. Exp. Theor. Phys.}
  {\bfseries 129} (2019) 553}.

\bibitem{Mironov:2019mye}
S.~Mironov, V.~Rubakov and V.~Volkova, \emph{{Subluminal cosmological bounce
  beyond Horndeski}},
  \href{https://doi.org/10.1088/1475-7516/2020/05/024}{\emph{JCAP} {\bfseries
  05} (2020) 024} [\href{https://arxiv.org/abs/1910.07019}{{\ttfamily
  arXiv:1910.07019}}].

\bibitem{Mironov:2019haz}
S.~Mironov, V.~Rubakov and V.~Volkova, \emph{{Cosmological scenarios with
  bounce and Genesis in Horndeski theory and beyond: An essay in honor of I.M.
  Khalatnikov on the occasion of his 100th birthday}},
  \href{https://arxiv.org/abs/1906.12139}{{\ttfamily arXiv:1906.12139}}.

\bibitem{Ilyas:2020qja}
A.~Ilyas, M.~Zhu, Y.~Zheng, Y.F.~Cai and E.N.~Saridakis, \emph{{DHOST Bounce}},
  \href{https://doi.org/10.1088/1475-7516/2020/09/002}{\emph{JCAP} {\bfseries
  09} (2020) 002} [\href{https://arxiv.org/abs/2002.08269}{{\ttfamily
  arXiv:2002.08269}}].

\bibitem{Zhu:2021whu}
M.~Zhu, A.~Ilyas, Y.~Zheng, Y.F.~Cai and E.N.~Saridakis, \emph{{Scalar and
  tensor perturbations in DHOST bounce cosmology}},
  \href{https://doi.org/10.1088/1475-7516/2021/11/045}{\emph{JCAP} {\bfseries
  11} (2021) 045} [\href{https://arxiv.org/abs/2108.01339}{{\ttfamily
  arXiv:2108.01339}}].

\bibitem{Mironov:2022ffa}
S.~Mironov and V.~Volkova, \emph{{Stable nonsingular cosmologies in beyond
  Horndeski theory and disformal transformations}},
  \href{https://doi.org/10.1142/S0217751X22500889}{\emph{Int. J. Mod. Phys. A}
  {\bfseries 37} (2022) 2250088}
  [\href{https://arxiv.org/abs/2204.05889}{{\ttfamily arXiv:2204.05889}}].

\bibitem{Koehn:2015vvy}
M.~Koehn, J.L.~Lehners and B.~Ovrut, \emph{{Nonsingular bouncing cosmology:
  Consistency of the effective description}},
  \href{https://doi.org/10.1103/PhysRevD.93.103501}{\emph{Phys. Rev. D}
  {\bfseries 93} (2016) 103501}
  [\href{https://arxiv.org/abs/1512.03807}{{\ttfamily arXiv:1512.03807}}].

\bibitem{Baumann:2011dt}
D.~Baumann, L.~Senatore and M.~Zaldarriaga, \emph{{Scale-Invariance and the
  Strong Coupling Problem}},
  \href{https://doi.org/10.1088/1475-7516/2011/05/004}{\emph{JCAP} {\bfseries
  05} (2011) 004} [\href{https://arxiv.org/abs/1101.3320}{{\ttfamily
  arXiv:1101.3320}}].

\bibitem{Ageeva:2022byg}
Y.A.~Ageeva and P.K.~Petrov, \emph{{Unitarity relation and unitarity bounds for
  scalars with different sound speeds}},
  \href{https://doi.org/10.3367/UFNe.2022.11.039259}{\emph{Phys. Usp.}
  {\bfseries 66} (2023) 1134}
  [\href{https://arxiv.org/abs/2206.03516}{{\ttfamily arXiv:2206.03516}}].

\bibitem{Cai:2022ori}
Y.~Cai, J.~Xu, S.~Zhao and S.~Zhou, \emph{{Perturbative unitarity and NEC
  violation in genesis cosmology}},
  \href{https://doi.org/10.1007/JHEP10(2022)140}{\emph{JHEP} {\bfseries 10}
  (2022) 140} [\href{https://arxiv.org/abs/2207.11772}{{\ttfamily
  arXiv:2207.11772}}].

\bibitem{Ageeva:2022asq}
Y.~Ageeva, P.~Petrov and V.~Rubakov, \emph{{Generating cosmological
  perturbations in non-singular Horndeski cosmologies}},
  \href{https://doi.org/10.1007/JHEP01(2023)026}{\emph{JHEP} {\bfseries 01}
  (2023) 026} [\href{https://arxiv.org/abs/2207.04071}{{\ttfamily
  arXiv:2207.04071}}].

\bibitem{Joyce:2011kh}
A.~Joyce and J.~Khoury, \emph{{Strong Coupling Problem with Time-Varying Sound
  Speed}}, \href{https://doi.org/10.1103/PhysRevD.84.083514}{\emph{Phys. Rev.
  D} {\bfseries 84} (2011) 083514}
  [\href{https://arxiv.org/abs/1107.3550}{{\ttfamily arXiv:1107.3550}}].

\bibitem{Ageeva:2018lko}
Y.A.~Ageeva, O.A.~Evseev, O.I.~Melichev and V.A.~Rubakov, \emph{{Horndeski
  Genesis: strong coupling and absence thereof}},
  \href{https://doi.org/10.1051/epjconf/201819107010}{\emph{EPJ Web Conf.}
  {\bfseries 191} (2018) 07010}
  [\href{https://arxiv.org/abs/1810.00465}{{\ttfamily arXiv:1810.00465}}].

\bibitem{Ageeva:2020gti}
Y.~Ageeva, O.~Evseev, O.~Melichev and V.~Rubakov, \emph{{Toward evading the
  strong coupling problem in Horndeski genesis}},
  \href{https://doi.org/10.1103/PhysRevD.102.023519}{\emph{Phys. Rev. D}
  {\bfseries 102} (2020) 023519}
  [\href{https://arxiv.org/abs/2003.01202}{{\ttfamily arXiv:2003.01202}}].

\bibitem{Ageeva:2020buc}
Y.~Ageeva, P.~Petrov and V.~Rubakov, \emph{{Horndeski genesis: consistency of
  classical theory}},
  \href{https://doi.org/10.1007/JHEP12(2020)107}{\emph{JHEP} {\bfseries 12}
  (2020) 107} [\href{https://arxiv.org/abs/2009.05071}{{\ttfamily
  arXiv:2009.05071}}].

\bibitem{Ageeva:2021yik}
Y.~Ageeva, P.~Petrov and V.~Rubakov, \emph{{Nonsingular cosmological models
  with strong gravity in the past}},
  \href{https://doi.org/10.1103/PhysRevD.104.063530}{\emph{Phys. Rev. D}
  {\bfseries 104} (2021) 063530}
  [\href{https://arxiv.org/abs/2104.13412}{{\ttfamily arXiv:2104.13412}}].

\bibitem{Ageeva:2022fyq}
Y.~Ageeva and P.~Petrov, \emph{{On the strong coupling problem in cosmologies
  with \textquotedblleft{}strong gravity in the past\textquotedblright{}}},
  \href{https://doi.org/10.1142/S0217732322501711}{\emph{Mod. Phys. Lett. A}
  {\bfseries 37} (2022) 2250171}
  [\href{https://arxiv.org/abs/2206.10646}{{\ttfamily arXiv:2206.10646}}].

\bibitem{Akama:2022usl}
S.~Akama and S.~Hirano, \emph{{Primordial non-Gaussianity from Galilean Genesis
  without strong coupling problem}},
  \href{https://doi.org/10.1103/PhysRevD.107.063504}{\emph{Phys. Rev. D}
  {\bfseries 107} (2023) 063504}
  [\href{https://arxiv.org/abs/2211.00388}{{\ttfamily arXiv:2211.00388}}].

\bibitem{Dubovsky:2005xd}
S.~Dubovsky, T.~Gregoire, A.~Nicolis and R.~Rattazzi, \emph{{Null energy
  condition and superluminal propagation}},
  \href{https://doi.org/10.1088/1126-6708/2006/03/025}{\emph{JHEP} {\bfseries
  03} (2006) 025} [\href{https://arxiv.org/abs/hep-th/0512260}{{\ttfamily
  hep-th/0512260}}].

\bibitem{Mironov:2020mfo}
S.~Mironov, V.~Rubakov and V.~Volkova, \emph{{Superluminality in beyond
  Horndeski theory with extra scalar field}},
  \href{https://doi.org/10.1088/1402-4896/ab996a}{\emph{Phys. Scripta}
  {\bfseries 95} (2020) 084002}
  [\href{https://arxiv.org/abs/2005.12626}{{\ttfamily arXiv:2005.12626}}].

\bibitem{Mironov:2020pqh}
S.~Mironov, V.~Rubakov and V.~Volkova, \emph{{Superluminality in DHOST theory
  with extra scalar}},
  \href{https://doi.org/10.1007/JHEP04(2021)035}{\emph{JHEP} {\bfseries 04}
  (2021) 035} [\href{https://arxiv.org/abs/2011.14912}{{\ttfamily
  arXiv:2011.14912}}].

\bibitem{Wands:1998yp}
D.~Wands, \emph{{Duality invariance of cosmological perturbation spectra}},
  \href{https://doi.org/10.1103/PhysRevD.60.023507}{\emph{Phys. Rev. D}
  {\bfseries 60} (1999) 023507}
  [\href{https://arxiv.org/abs/gr-qc/9809062}{{\ttfamily gr-qc/9809062}}].

\bibitem{Finelli:2001sr}
F.~Finelli and R.~Brandenberger, \emph{{On the generation of a scale invariant
  spectrum of adiabatic fluctuations in cosmological models with a contracting
  phase}}, \href{https://doi.org/10.1103/PhysRevD.65.103522}{\emph{Phys. Rev.
  D} {\bfseries 65} (2002) 103522}
  [\href{https://arxiv.org/abs/hep-th/0112249}{{\ttfamily hep-th/0112249}}].

\bibitem{Brandenberger:2012zb}
R.H.~Brandenberger, \emph{{The Matter Bounce Alternative to Inflationary
  Cosmology}},  \href{https://arxiv.org/abs/1206.4196}{{\ttfamily
  arXiv:1206.4196}}.

\bibitem{Khoury:2001wf}
J.~Khoury, B.A.~Ovrut, P.J.~Steinhardt and N.~Turok, \emph{{The Ekpyrotic
  universe: Colliding branes and the origin of the hot big bang}},
  \href{https://doi.org/10.1103/PhysRevD.64.123522}{\emph{Phys. Rev. D}
  {\bfseries 64} (2001) 123522}
  [\href{https://arxiv.org/abs/hep-th/0103239}{{\ttfamily hep-th/0103239}}].

\bibitem{Buchbinder:2007ad}
E.I.~Buchbinder, J.~Khoury and B.A.~Ovrut, \emph{{New Ekpyrotic cosmology}},
  \href{https://doi.org/10.1103/PhysRevD.76.123503}{\emph{Phys. Rev. D}
  {\bfseries 76} (2007) 123503}
  [\href{https://arxiv.org/abs/hep-th/0702154}{{\ttfamily hep-th/0702154}}].

\bibitem{Lehners:2008vx}
J.L.~Lehners, \emph{{Ekpyrotic and Cyclic Cosmology}},
  \href{https://doi.org/10.1016/j.physrep.2008.06.001}{\emph{Phys. Rept.}
  {\bfseries 465} (2008) 223}
  [\href{https://arxiv.org/abs/0806.1245}{{\ttfamily arXiv:0806.1245}}].

\bibitem{Gasperini:1992em}
M.~Gasperini and G.~Veneziano, \emph{{Pre-big bang in string cosmology}},
  \href{https://doi.org/10.1016/0927-6505(93)90017-8}{\emph{Astropart. Phys.}
  {\bfseries 1} (1993) 317}
  [\href{https://arxiv.org/abs/hep-th/9211021}{{\ttfamily hep-th/9211021}}].

\bibitem{Gasperini:2002bn}
M.~Gasperini and G.~Veneziano, \emph{{The pre-big bang scenario in string
  cosmology}}, \href{https://doi.org/10.1016/S0370-1573(02)00389-7}{\emph{Phys.
  Rept.} {\bfseries 373} (2003) 1}
  [\href{https://arxiv.org/abs/hep-th/0207130}{{\ttfamily hep-th/0207130}}].

\bibitem{Gasperini:2007zz}
M.~Gasperini, \emph{{Elements of string cosmology}}, Cambridge University
  Press, Cambridge, UK (2007).

\bibitem{Quintin:2015rta}
J.~Quintin, Z.~Sherkatghanad, Y.F.~Cai and R.H.~Brandenberger, \emph{{Evolution
  of cosmological perturbations and the production of non-Gaussianities through
  a nonsingular bounce: Indications for a no-go theorem in single field matter
  bounce cosmologies}},
  \href{https://doi.org/10.1103/PhysRevD.92.063532}{\emph{Phys. Rev. D}
  {\bfseries 92} (2015) 063532}
  [\href{https://arxiv.org/abs/1508.04141}{{\ttfamily arXiv:1508.04141}}].

\bibitem{Li:2016xjb}
Y.B.~Li, J.~Quintin, D.G.~Wang and Y.F.~Cai, \emph{{Matter bounce cosmology
  with a generalized single field: non-Gaussianity and an extended no-go
  theorem}}, \href{https://doi.org/10.1088/1475-7516/2017/03/031}{\emph{JCAP}
  {\bfseries 03} (2017) 031}
  [\href{https://arxiv.org/abs/1612.02036}{{\ttfamily arXiv:1612.02036}}].

\bibitem{Cai:2013vm}
Y.F.~Cai, R.~Brandenberger and P.~Peter, \emph{{Anisotropy in a Nonsingular
  Bounce}}, \href{https://doi.org/10.1088/0264-9381/30/7/075019}{\emph{Class.
  Quant. Grav.} {\bfseries 30} (2013) 075019}
  [\href{https://arxiv.org/abs/1301.4703}{{\ttfamily arXiv:1301.4703}}].

\bibitem{Levy:2016xcl}
A.M.~Levy, \emph{{Fine-tuning challenges for the matter bounce scenario}},
  \href{https://doi.org/10.1103/PhysRevD.95.023522}{\emph{Phys. Rev. D}
  {\bfseries 95} (2017) 023522}
  [\href{https://arxiv.org/abs/1611.08972}{{\ttfamily arXiv:1611.08972}}].

\bibitem{Ganguly:2021pke}
C.~Ganguly and J.~Quintin, \emph{{Microphysical manifestations of viscosity and
  consequences for anisotropies in the very early universe}},
  \href{https://doi.org/10.1103/PhysRevD.105.023532}{\emph{Phys. Rev. D}
  {\bfseries 105} (2022) 023532}
  [\href{https://arxiv.org/abs/2109.11701}{{\ttfamily arXiv:2109.11701}}].

\bibitem{Akama:2019qeh}
S.~Akama, S.~Hirano and T.~Kobayashi, \emph{{Primordial non-Gaussianities of
  scalar and tensor perturbations in general bounce cosmology: Evading the
  no-go theorem}},
  \href{https://doi.org/10.1103/PhysRevD.101.043529}{\emph{Phys. Rev. D}
  {\bfseries 101} (2020) 043529}
  [\href{https://arxiv.org/abs/1908.10663}{{\ttfamily arXiv:1908.10663}}].

\bibitem{Akama:2024bav}
S.~Akama and M.~Zhu, \emph{{Parity violation in primordial tensor
  non-Gaussianities from matter bounce cosmology}},
  \href{https://doi.org/10.1088/1475-7516/2024/07/039}{\emph{JCAP} {\bfseries
  07} (2024) 039} [\href{https://arxiv.org/abs/2404.05464}{{\ttfamily
  arXiv:2404.05464}}].

\bibitem{Akama:2024vtu}
S.~Akama, G.~Orlando and P.C.M.~Delgado, \emph{{Towards testing the general
  bounce cosmology with the CMB B-mode auto-bispectrum}},
  \href{https://doi.org/10.1088/1475-7516/2024/09/055}{\emph{JCAP} {\bfseries
  09} (2024) 055} [\href{https://arxiv.org/abs/2404.14393}{{\ttfamily
  arXiv:2404.14393}}].

\bibitem{Akama:2025ows}
S.~Akama, \emph{{Primordial full bispectra from the general bounce cosmology}},
   \href{https://arxiv.org/abs/2502.14850}{{\ttfamily arXiv:2502.14850}}.

\bibitem{Fertig:2013kwa}
A.~Fertig, J.L.~Lehners and E.~Mallwitz, \emph{{Ekpyrotic Perturbations With
  Small Non-Gaussian Corrections}},
  \href{https://doi.org/10.1103/PhysRevD.89.103537}{\emph{Phys. Rev. D}
  {\bfseries 89} (2014) 103537}
  [\href{https://arxiv.org/abs/1310.8133}{{\ttfamily arXiv:1310.8133}}].

\bibitem{Ijjas:2014fja}
A.~Ijjas, J.L.~Lehners and P.J.~Steinhardt, \emph{{General mechanism for
  producing scale-invariant perturbations and small non-Gaussianity in
  ekpyrotic models}},
  \href{https://doi.org/10.1103/PhysRevD.89.123520}{\emph{Phys. Rev. D}
  {\bfseries 89} (2014) 123520}
  [\href{https://arxiv.org/abs/1404.1265}{{\ttfamily arXiv:1404.1265}}].

\bibitem{Quintin:2024boj}
J.~Quintin, X.~Chen and R.~Ebadi, \emph{{Fingerprints of a non-inflationary
  universe from massive fields}},
  \href{https://doi.org/10.1088/1475-7516/2024/09/026}{\emph{JCAP} {\bfseries
  09} (2024) 026} [\href{https://arxiv.org/abs/2405.11016}{{\ttfamily
  arXiv:2405.11016}}].

\bibitem{Kim:2020iwq}
J.L.~Kim and G.~Geshnizjani, \emph{{Spectrum of Cuscuton Bounce}},
  \href{https://doi.org/10.1088/1475-7516/2021/03/104}{\emph{JCAP} {\bfseries
  03} (2021) 104} [\href{https://arxiv.org/abs/2010.06645}{{\ttfamily
  arXiv:2010.06645}}].

\bibitem{Erickson:2003zm}
J.K.~Erickson, D.H.~Wesley, P.J.~Steinhardt and N.~Turok, \emph{{Kasner and
  mixmaster behavior in universes with equation of state $w \geq 1$}},
  \href{https://doi.org/10.1103/PhysRevD.69.063514}{\emph{Phys. Rev. D}
  {\bfseries 69} (2004) 063514}
  [\href{https://arxiv.org/abs/hep-th/0312009}{{\ttfamily hep-th/0312009}}].

\bibitem{Coley:2005pj}
A.A.~Coley and W.C.~Lim, \emph{{Asymptotic analysis of spatially inhomogeneous
  stiff and ultra-stiff cosmologies}},
  \href{https://doi.org/10.1088/0264-9381/22/14/016}{\emph{Class. Quant. Grav.}
  {\bfseries 22} (2005) 3073}
  [\href{https://arxiv.org/abs/gr-qc/0506097}{{\ttfamily gr-qc/0506097}}].

\bibitem{Lidsey:2005wr}
J.E.~Lidsey, \emph{{Cosmic no hair for collapsing universes}},
  \href{https://doi.org/10.1088/0264-9381/23/10/018}{\emph{Class. Quant. Grav.}
  {\bfseries 23} (2006) 3517}
  [\href{https://arxiv.org/abs/hep-th/0511174}{{\ttfamily hep-th/0511174}}].

\bibitem{Garfinkle:2008ei}
D.~Garfinkle, W.C.~Lim, F.~Pretorius and P.J.~Steinhardt, \emph{{Evolution to a
  smooth universe in an ekpyrotic contracting phase with $w > 1$}},
  \href{https://doi.org/10.1103/PhysRevD.78.083537}{\emph{Phys. Rev. D}
  {\bfseries 78} (2008) 083537}
  [\href{https://arxiv.org/abs/0808.0542}{{\ttfamily arXiv:0808.0542}}].

\bibitem{Barrow:2010rx}
J.D.~Barrow and K.~Yamamoto, \emph{{Anisotropic Pressures at Ultra-stiff
  Singularities and the Stability of Cyclic Universes}},
  \href{https://doi.org/10.1103/PhysRevD.82.063516}{\emph{Phys. Rev. D}
  {\bfseries 82} (2010) 063516}
  [\href{https://arxiv.org/abs/1004.4767}{{\ttfamily arXiv:1004.4767}}].

\bibitem{Heinzle:2011qj}
J.M.~Heinzle and P.~Sandin, \emph{{The initial singularity of ultrastiff
  perfect fluid spacetimes without symmetries}},
  \href{https://doi.org/10.1007/s00220-012-1496-x}{\emph{Commun. Math. Phys.}
  {\bfseries 313} (2012) 385}
  [\href{https://arxiv.org/abs/1105.1643}{{\ttfamily arXiv:1105.1643}}].

\bibitem{Barrow:2015wfa}
J.D.~Barrow and C.~Ganguly, \emph{{Evolution of initially contracting Bianchi
  Class A models in the presence of an ultra-stiff anisotropic pressure
  fluid}}, \href{https://doi.org/10.1088/0264-9381/33/12/125004}{\emph{Class.
  Quant. Grav.} {\bfseries 33} (2016) 125004}
  [\href{https://arxiv.org/abs/1510.01095}{{\ttfamily arXiv:1510.01095}}].

\bibitem{Cook:2020oaj}
W.G.~Cook, I.A.~Glushchenko, A.~Ijjas, F.~Pretorius and P.J.~Steinhardt,
  \emph{{Supersmoothing through Slow Contraction}},
  \href{https://doi.org/10.1016/j.physletb.2020.135690}{\emph{Phys. Lett. B}
  {\bfseries 808} (2020) 135690}
  [\href{https://arxiv.org/abs/2006.01172}{{\ttfamily arXiv:2006.01172}}].

\bibitem{Fertig:2015ola}
A.~Fertig and J.L.~Lehners, \emph{{The Non-Minimal Ekpyrotic Trispectrum}},
  \href{https://doi.org/10.1088/1475-7516/2016/01/026}{\emph{JCAP} {\bfseries
  01} (2016) 026} [\href{https://arxiv.org/abs/1510.03439}{{\ttfamily
  arXiv:1510.03439}}].

\bibitem{Fertig:2016czu}
A.~Fertig, J.L.~Lehners, E.~Mallwitz and E.~Wilson-Ewing, \emph{{Converting
  entropy to curvature perturbations after a cosmic bounce}},
  \href{https://doi.org/10.1088/1475-7516/2016/10/005}{\emph{JCAP} {\bfseries
  10} (2016) 005} [\href{https://arxiv.org/abs/1607.05663}{{\ttfamily
  arXiv:1607.05663}}].

\bibitem{Mukhanov:1991zn}
V.F.~Mukhanov and R.H.~Brandenberger, \emph{{A Nonsingular universe}},
  \href{https://doi.org/10.1103/PhysRevLett.68.1969}{\emph{Phys. Rev. Lett.}
  {\bfseries 68} (1992) 1969}.

\bibitem{Brandenberger:1993ef}
R.H.~Brandenberger, V.F.~Mukhanov and A.~Sornborger, \emph{{A Cosmological
  theory without singularities}},
  \href{https://doi.org/10.1103/PhysRevD.48.1629}{\emph{Phys. Rev. D}
  {\bfseries 48} (1993) 1629}
  [\href{https://arxiv.org/abs/gr-qc/9303001}{{\ttfamily gr-qc/9303001}}].

\bibitem{Yoshida:2017swb}
D.~Yoshida, J.~Quintin, M.~Yamaguchi and R.H.~Brandenberger,
  \emph{{Cosmological perturbations and stability of nonsingular cosmologies
  with limiting curvature}},
  \href{https://doi.org/10.1103/PhysRevD.96.043502}{\emph{Phys. Rev. D}
  {\bfseries 96} (2017) 043502}
  [\href{https://arxiv.org/abs/1704.04184}{{\ttfamily arXiv:1704.04184}}].

\bibitem{Levy:2015awa}
A.M.~Levy, A.~Ijjas and P.J.~Steinhardt, \emph{{Scale-invariant perturbations
  in ekpyrotic cosmologies without fine-tuning of initial conditions}},
  \href{https://doi.org/10.1103/PhysRevD.92.063524}{\emph{Phys. Rev. D}
  {\bfseries 92} (2015) 063524}
  [\href{https://arxiv.org/abs/1506.01011}{{\ttfamily arXiv:1506.01011}}].

\bibitem{Ijjas:2020dws}
A.~Ijjas, W.G.~Cook, F.~Pretorius, P.J.~Steinhardt and E.Y.~Davies,
  \emph{{Robustness of slow contraction to cosmic initial conditions}},
  \href{https://doi.org/10.1088/1475-7516/2020/08/030}{\emph{JCAP} {\bfseries
  08} (2020) 030} [\href{https://arxiv.org/abs/2006.04999}{{\ttfamily
  arXiv:2006.04999}}].

\bibitem{Ijjas:2021gkf}
A.~Ijjas, A.P.~Sullivan, F.~Pretorius, P.J.~Steinhardt and W.G.~Cook,
  \emph{{Ultralocality and slow contraction}},
  \href{https://doi.org/10.1088/1475-7516/2021/06/013}{\emph{JCAP} {\bfseries
  06} (2021) 013} [\href{https://arxiv.org/abs/2103.00584}{{\ttfamily
  arXiv:2103.00584}}].

\bibitem{Ijjas:2021wml}
A.~Ijjas, F.~Pretorius, P.J.~Steinhardt and A.P.~Sullivan, \emph{{The effects
  of multiple modes and reduced symmetry on the rapidity and robustness of slow
  contraction}},
  \href{https://doi.org/10.1016/j.physletb.2021.136490}{\emph{Phys. Lett. B}
  {\bfseries 820} (2021) 136490}
  [\href{https://arxiv.org/abs/2104.12293}{{\ttfamily arXiv:2104.12293}}].

\bibitem{Ijjas:2021zyf}
A.~Ijjas, F.~Pretorius, P.J.~Steinhardt and D.~Garfinkle, \emph{{Dynamical
  attractors in contracting spacetimes dominated by kinetically coupled scalar
  fields}}, \href{https://doi.org/10.1088/1475-7516/2021/12/030}{\emph{JCAP}
  {\bfseries 12} (2021) 030}
  [\href{https://arxiv.org/abs/2109.09768}{{\ttfamily arXiv:2109.09768}}].

\bibitem{Ijjas:2022qsv}
A.~Ijjas, \emph{{Numerical Relativity as a New Tool for Fundamental
  Cosmology}}, \href{https://doi.org/10.3390/physics4010021}{\emph{MDPI
  Physics} {\bfseries 4} (2022) 301}
  [\href{https://arxiv.org/abs/2201.03752}{{\ttfamily arXiv:2201.03752}}].

\bibitem{Kist:2022mew}
T.~Kist and A.~Ijjas, \emph{{The robustness of slow contraction and the shape
  of the scalar field potential}},
  \href{https://doi.org/10.1088/1475-7516/2022/08/046}{\emph{JCAP} {\bfseries
  08} (2022) 046} [\href{https://arxiv.org/abs/2205.01519}{{\ttfamily
  arXiv:2205.01519}}].

\bibitem{Ijjas:2023dnb}
A.~Ijjas, \emph{{Slow Contraction and the Weyl Curvature Hypothesis}},
  \href{https://arxiv.org/abs/2304.10030}{{\ttfamily arXiv:2304.10030}}.

\bibitem{Ijjas:2024oqn}
A.~Ijjas, P.J.~Steinhardt, D.~Garfinkle and W.G.~Cook, \emph{{Smoothing and
  flattening the universe through slow contraction versus inflation}},
  \href{https://doi.org/10.1088/1475-7516/2024/07/077}{\emph{JCAP} {\bfseries
  07} (2024) 077} [\href{https://arxiv.org/abs/2404.00867}{{\ttfamily
  arXiv:2404.00867}}].

\bibitem{Lehners:2011kr}
J.L.~Lehners, \emph{{Cosmic Bounces and Cyclic Universes}},
  \href{https://doi.org/10.1088/0264-9381/28/20/204004}{\emph{Class. Quant.
  Grav.} {\bfseries 28} (2011) 204004}
  [\href{https://arxiv.org/abs/1106.0172}{{\ttfamily arXiv:1106.0172}}].

\bibitem{Planck:2018vyg}
{\scshape Planck} collaboration, N.~Aghanim et~al., \emph{{Planck 2018 results.
  VI. Cosmological parameters}},
  \href{https://doi.org/10.1051/0004-6361/201833910}{\emph{Astron. Astrophys.}
  {\bfseries 641} (2020) A6}
  [\href{https://arxiv.org/abs/1807.06209}{{\ttfamily arXiv:1807.06209}}].

\bibitem{Liddle:2000cg}
A.R.~Liddle and D.H.~Lyth, \emph{{Cosmological inflation and large scale
  structure}}, Cambridge University Press, Cambridge, UK (2000),
  \href{https://doi.org/10.1017/CBO9781139175180}{10.1017/CBO9781139175180}.

\bibitem{Lehners:2018vgi}
J.L.~Lehners, \emph{{Small-Field and Scale-Free: Inflation and Ekpyrosis at
  their Extremes}},
  \href{https://doi.org/10.1088/1475-7516/2018/11/001}{\emph{JCAP} {\bfseries
  11} (2018) 001} [\href{https://arxiv.org/abs/1807.05240}{{\ttfamily
  arXiv:1807.05240}}].

\bibitem{Bernardo:2021wnv}
H.~Bernardo and R.~Brandenberger, \emph{{Contracting cosmologies and the
  swampland}}, \href{https://doi.org/10.1007/JHEP07(2021)206}{\emph{JHEP}
  {\bfseries 07} (2021) 206}
  [\href{https://arxiv.org/abs/2104.00630}{{\ttfamily arXiv:2104.00630}}].

\bibitem{Shiu:2023yzt}
G.~Shiu, F.~Tonioni and H.V.~Tran, \emph{{Collapsing universe before time}},
  \href{https://doi.org/10.1088/1475-7516/2024/05/124}{\emph{JCAP} {\bfseries
  05} (2024) 124} [\href{https://arxiv.org/abs/2312.06772}{{\ttfamily
  arXiv:2312.06772}}].

\bibitem{Takamizu:2004rq}
Y.i.~Takamizu and K.i.~Maeda, \emph{{Collision of domain walls and reheating of
  the brane universe}},
  \href{https://doi.org/10.1103/PhysRevD.70.123514}{\emph{Phys. Rev. D}
  {\bfseries 70} (2004) 123514}
  [\href{https://arxiv.org/abs/hep-th/0406235}{{\ttfamily hep-th/0406235}}].

\bibitem{Lehners:2007wc}
J.L.~Lehners and P.J.~Steinhardt, \emph{{Non-Gaussian density fluctuations from
  entropically generated curvature perturbations in Ekpyrotic models}},
  \href{https://doi.org/10.1103/PhysRevD.77.063533}{\emph{Phys. Rev. D}
  {\bfseries 77} (2008) 063533}
  [\href{https://arxiv.org/abs/0712.3779}{{\ttfamily arXiv:0712.3779}}].

\bibitem{Battefeld:2007st}
T.~Battefeld, \emph{{Modulated Perturbations from Instant Preheating after new
  Ekpyrosis}}, \href{https://doi.org/10.1103/PhysRevD.77.063503}{\emph{Phys.
  Rev. D} {\bfseries 77} (2008) 063503}
  [\href{https://arxiv.org/abs/0710.2540}{{\ttfamily arXiv:0710.2540}}].

\bibitem{Lehners:2008my}
J.L.~Lehners and P.J.~Steinhardt, \emph{{Intuitive understanding of
  non-gaussianity in ekpyrotic and cyclic models}},
  \href{https://doi.org/10.1103/PhysRevD.78.023506}{\emph{Phys. Rev. D}
  {\bfseries 78} (2008) 023506}
  [\href{https://arxiv.org/abs/0804.1293}{{\ttfamily arXiv:0804.1293}}].

\bibitem{Lehners:2009qu}
J.L.~Lehners and P.J.~Steinhardt, \emph{{Non-Gaussianity Generated by the
  Entropic Mechanism in Bouncing Cosmologies Made Simple}},
  \href{https://doi.org/10.1103/PhysRevD.80.103520}{\emph{Phys. Rev. D}
  {\bfseries 80} (2009) 103520}
  [\href{https://arxiv.org/abs/0909.2558}{{\ttfamily arXiv:0909.2558}}].

\bibitem{Lehners:2009ja}
J.L.~Lehners and S.~Renaux-Petel, \emph{{Multifield Cosmological Perturbations
  at Third Order and the Ekpyrotic Trispectrum}},
  \href{https://doi.org/10.1103/PhysRevD.80.063503}{\emph{Phys. Rev. D}
  {\bfseries 80} (2009) 063503}
  [\href{https://arxiv.org/abs/0906.0530}{{\ttfamily arXiv:0906.0530}}].

\bibitem{Lehners:2010fy}
J.L.~Lehners, \emph{{Ekpyrotic Non-Gaussianity: A Review}},
  \href{https://doi.org/10.1155/2010/903907}{\emph{Adv. Astron.} {\bfseries
  2010} (2010) 903907} [\href{https://arxiv.org/abs/1001.3125}{{\ttfamily
  arXiv:1001.3125}}].

\bibitem{Quintin:2014oea}
J.~Quintin, Y.F.~Cai and R.H.~Brandenberger, \emph{{Matter creation in a
  nonsingular bouncing cosmology}},
  \href{https://doi.org/10.1103/PhysRevD.90.063507}{\emph{Phys. Rev. D}
  {\bfseries 90} (2014) 063507}
  [\href{https://arxiv.org/abs/1406.6049}{{\ttfamily arXiv:1406.6049}}].

\bibitem{Hipolito-Ricaldi:2016kqq}
W.S.~Hipolito-Ricaldi, R.~Brandenberger, E.G.M.~Ferreira and L.L.~Graef,
  \emph{{Particle Production in Ekpyrotic Scenarios}},
  \href{https://doi.org/10.1088/1475-7516/2016/11/024}{\emph{JCAP} {\bfseries
  11} (2016) 024} [\href{https://arxiv.org/abs/1605.04670}{{\ttfamily
  arXiv:1605.04670}}].

\bibitem{Ijjas:2020cyh}
A.~Ijjas and R.~Kolevatov, \emph{{Sourcing curvature modes with entropy
  perturbations in non-singular bouncing cosmologies}},
  \href{https://doi.org/10.1088/1475-7516/2021/06/012}{\emph{JCAP} {\bfseries
  06} (2021) 012} [\href{https://arxiv.org/abs/2012.08249}{{\ttfamily
  arXiv:2012.08249}}].

\bibitem{Ijjas:2021ewd}
A.~Ijjas and R.~Kolevatov, \emph{{Nearly scale-invariant curvature modes from
  entropy perturbations during the graceful exit phase}},
  \href{https://doi.org/10.1103/PhysRevD.103.L101302}{\emph{Phys. Rev. D}
  {\bfseries 103} (2021) L101302}
  [\href{https://arxiv.org/abs/2102.03818}{{\ttfamily arXiv:2102.03818}}].

\bibitem{Chen:2006nt}
X.~Chen, M.x.~Huang, S.~Kachru and G.~Shiu, \emph{{Observational signatures and
  non-Gaussianities of general single field inflation}},
  \href{https://doi.org/10.1088/1475-7516/2007/01/002}{\emph{JCAP} {\bfseries
  01} (2007) 002} [\href{https://arxiv.org/abs/hep-th/0605045}{{\ttfamily
  hep-th/0605045}}].

\bibitem{Langlois:2008qf}
D.~Langlois, S.~Renaux-Petel, D.A.~Steer and T.~Tanaka, \emph{{Primordial
  perturbations and non-Gaussianities in DBI and general multi-field
  inflation}}, \href{https://doi.org/10.1103/PhysRevD.78.063523}{\emph{Phys.
  Rev. D} {\bfseries 78} (2008) 063523}
  [\href{https://arxiv.org/abs/0806.0336}{{\ttfamily arXiv:0806.0336}}].

\bibitem{Planck:2019kim}
{\scshape Planck} collaboration, Y.~Akrami et~al., \emph{{Planck 2018 results.
  IX. Constraints on primordial non-Gaussianity}},
  \href{https://doi.org/10.1051/0004-6361/201935891}{\emph{Astron. Astrophys.}
  {\bfseries 641} (2020) A9}
  [\href{https://arxiv.org/abs/1905.05697}{{\ttfamily arXiv:1905.05697}}].

\bibitem{Camera:2014bwa}
S.~Camera, M.G.~Santos and R.~Maartens, \emph{{Probing primordial
  non-Gaussianity with SKA galaxy redshift surveys: a fully relativistic
  analysis}}, \href{https://doi.org/10.1093/mnras/stv040}{\emph{Mon. Not. Roy.
  Astron. Soc.} {\bfseries 448} (2015) 1035}
  [\href{https://arxiv.org/abs/1409.8286}{{\ttfamily arXiv:1409.8286}}].

\bibitem{Karagiannis:2018jdt}
D.~Karagiannis, A.~Lazanu, M.~Liguori, A.~Raccanelli, N.~Bartolo and L.~Verde,
  \emph{{Constraining primordial non-Gaussianity with bispectrum and power
  spectrum from upcoming optical and radio surveys}},
  \href{https://doi.org/10.1093/mnras/sty1029}{\emph{Mon. Not. Roy. Astron.
  Soc.} {\bfseries 478} (2018) 1341}
  [\href{https://arxiv.org/abs/1801.09280}{{\ttfamily arXiv:1801.09280}}].

\bibitem{SimonsObservatory:2018koc}
{\scshape Simons Observatory} collaboration, P.~Ade et~al., \emph{{The Simons
  Observatory: Science goals and forecasts}},
  \href{https://doi.org/10.1088/1475-7516/2019/02/056}{\emph{JCAP} {\bfseries
  02} (2019) 056} [\href{https://arxiv.org/abs/1808.07445}{{\ttfamily
  arXiv:1808.07445}}].

\bibitem{Meerburg:2019qqi}
P.D.~Meerburg et~al., \emph{{Primordial Non-Gaussianity}}, {\emph{Bull. Am.
  Astron. Soc.} {\bfseries 51} (2019) 107}
  [\href{https://arxiv.org/abs/1903.04409}{{\ttfamily arXiv:1903.04409}}].

\bibitem{Karagiannis:2019jjx}
D.~Karagiannis, A.~Slosar and M.~Liguori, \emph{{Forecasts on Primordial
  non-Gaussianity from 21 cm Intensity Mapping experiments}},
  \href{https://doi.org/10.1088/1475-7516/2020/11/052}{\emph{JCAP} {\bfseries
  11} (2020) 052} [\href{https://arxiv.org/abs/1911.03964}{{\ttfamily
  arXiv:1911.03964}}].

\bibitem{Biagetti:2022qjl}
M.~Biagetti, J.~Calles, L.~Castiblanco, A.~Cole and J.~Nore\~na, \emph{{Fisher
  forecasts for primordial non-Gaussianity from persistent homology}},
  \href{https://doi.org/10.1088/1475-7516/2022/10/002}{\emph{JCAP} {\bfseries
  10} (2022) 002} [\href{https://arxiv.org/abs/2203.08262}{{\ttfamily
  arXiv:2203.08262}}].

\bibitem{Gao:2014hea}
X.~Gao, M.~Lilley and P.~Peter, \emph{{Production of non-gaussianities through
  a positive spatial curvature bouncing phase}},
  \href{https://doi.org/10.1088/1475-7516/2014/07/010}{\emph{JCAP} {\bfseries
  07} (2014) 010} [\href{https://arxiv.org/abs/1403.7958}{{\ttfamily
  arXiv:1403.7958}}].

\bibitem{Gao:2014eaa}
X.~Gao, M.~Lilley and P.~Peter, \emph{{Non-Gaussianity excess problem in
  classical bouncing cosmologies}},
  \href{https://doi.org/10.1103/PhysRevD.91.023516}{\emph{Phys. Rev. D}
  {\bfseries 91} (2015) 023516}
  [\href{https://arxiv.org/abs/1406.4119}{{\ttfamily arXiv:1406.4119}}].

\bibitem{Collins:2011mz}
H.~Collins, \emph{{Primordial non-Gaussianities from inflation}},
  \href{https://arxiv.org/abs/1101.1308}{{\ttfamily arXiv:1101.1308}}.

\bibitem{Wang:2013zva}
Y.~Wang, \emph{{Inflation, Cosmic Perturbations and Non-Gaussianities}},
  \href{https://doi.org/10.1088/0253-6102/62/1/19}{\emph{Commun. Theor. Phys.}
  {\bfseries 62} (2014) 109} [\href{https://arxiv.org/abs/1303.1523}{{\ttfamily
  arXiv:1303.1523}}].

\bibitem{Leblond:2008gg}
L.~Leblond and S.~Shandera, \emph{{Simple Bounds from the Perturbative Regime
  of Inflation}},
  \href{https://doi.org/10.1088/1475-7516/2008/08/007}{\emph{JCAP} {\bfseries
  08} (2008) 007} [\href{https://arxiv.org/abs/0802.2290}{{\ttfamily
  arXiv:0802.2290}}].

\bibitem{DuasoPueyo:2024rsa}
C.~Duaso~Pueyo, H.~Goodhew, C.~McCulloch and E.~Pajer, \emph{{Perturbative
  unitarity bounds from momentum-space entanglement}},
  \href{https://arxiv.org/abs/2410.23709}{{\ttfamily arXiv:2410.23709}}.

\bibitem{Geshnizjani:2013lza}
G.~Geshnizjani and N.~Ahmadi, \emph{{Can non-local or higher derivative
  theories provide alternatives to inflation?}},
  \href{https://doi.org/10.1088/1475-7516/2013/11/029}{\emph{JCAP} {\bfseries
  11} (2013) 029} [\href{https://arxiv.org/abs/1309.4782}{{\ttfamily
  arXiv:1309.4782}}].

\bibitem{Bartolo:2004if}
N.~Bartolo, E.~Komatsu, S.~Matarrese and A.~Riotto, \emph{{Non-Gaussianity from
  inflation: Theory and observations}},
  \href{https://doi.org/10.1016/j.physrep.2004.08.022}{\emph{Phys. Rept.}
  {\bfseries 402} (2004) 103}
  [\href{https://arxiv.org/abs/astro-ph/0406398}{{\ttfamily
  astro-ph/0406398}}].

\bibitem{Komatsu:2001rj}
E.~Komatsu and D.N.~Spergel, \emph{{Acoustic signatures in the primary
  microwave background bispectrum}},
  \href{https://doi.org/10.1103/PhysRevD.63.063002}{\emph{Phys. Rev. D}
  {\bfseries 63} (2001) 063002}
  [\href{https://arxiv.org/abs/astro-ph/0005036}{{\ttfamily
  astro-ph/0005036}}].

\bibitem{Maldacena:2002vr}
J.M.~Maldacena, \emph{{Non-Gaussian features of primordial fluctuations in
  single field inflationary models}},
  \href{https://doi.org/10.1088/1126-6708/2003/05/013}{\emph{JHEP} {\bfseries
  05} (2003) 013} [\href{https://arxiv.org/abs/astro-ph/0210603}{{\ttfamily
  astro-ph/0210603}}].

\bibitem{Chen:2010xka}
X.~Chen, \emph{{Primordial Non-Gaussianities from Inflation Models}},
  \href{https://doi.org/10.1155/2010/638979}{\emph{Adv. Astron.} {\bfseries
  2010} (2010) 638979} [\href{https://arxiv.org/abs/1002.1416}{{\ttfamily
  arXiv:1002.1416}}].

\bibitem{Cai:2009fn}
Y.F.~Cai, W.~Xue, R.~Brandenberger and X.~Zhang, \emph{{Non-Gaussianity in a
  Matter Bounce}},
  \href{https://doi.org/10.1088/1475-7516/2009/05/011}{\emph{JCAP} {\bfseries
  05} (2009) 011} [\href{https://arxiv.org/abs/0903.0631}{{\ttfamily
  arXiv:0903.0631}}].

\bibitem{Bartolo:2021wpt}
N.~Bartolo, A.~Ganz and S.~Matarrese, \emph{{Cuscuton inflation}},
  \href{https://doi.org/10.1088/1475-7516/2022/05/008}{\emph{JCAP} {\bfseries
  05} (2022) 008} [\href{https://arxiv.org/abs/2111.06794}{{\ttfamily
  arXiv:2111.06794}}].

\bibitem{Ganz:2024ihb}
A.~Ganz, P.~Martens, S.~Mukohyama and R.~Namba, \emph{{Bispectrum from
  inflation/bouncing Universe in VCDM}},
  \href{https://arxiv.org/abs/2407.02882}{{\ttfamily arXiv:2407.02882}}.

\bibitem{Delgado:2021mxu}
P.C.M.~Delgado, R.~Durrer and N.~Pinto-Neto, \emph{{The CMB bispectrum from
  bouncing cosmologies}},
  \href{https://doi.org/10.1088/1475-7516/2021/11/024}{\emph{JCAP} {\bfseries
  11} (2021) 024} [\href{https://arxiv.org/abs/2108.06175}{{\ttfamily
  arXiv:2108.06175}}].

\bibitem{vanTent:2022vgy}
B.~van~Tent, P.C.M.~Delgado and R.~Durrer, \emph{{Constraining the Bispectrum
  from Bouncing Cosmologies with Planck}},
  \href{https://doi.org/10.1103/PhysRevLett.130.191002}{\emph{Phys. Rev. Lett.}
  {\bfseries 130} (2023) 191002}
  [\href{https://arxiv.org/abs/2212.05977}{{\ttfamily arXiv:2212.05977}}].

\bibitem{Ageeva:2024knc}
Y.~Ageeva, M.~Kotenko and P.~Petrov, \emph{{The primordial non-Gaussianities
  for non-singular Horndeski cosmologies}},
  \href{https://arxiv.org/abs/2410.10742}{{\ttfamily arXiv:2410.10742}}.

\bibitem{Ganz:2022zgs}
A.~Ganz, P.~Martens, S.~Mukohyama and R.~Namba, \emph{{Bouncing cosmology in
  VCDM}}, \href{https://doi.org/10.1088/1475-7516/2023/04/060}{\emph{JCAP}
  {\bfseries 04} (2023) 060}
  [\href{https://arxiv.org/abs/2212.13561}{{\ttfamily arXiv:2212.13561}}].

\bibitem{Khoury:2009my}
J.~Khoury and P.J.~Steinhardt, \emph{{Adiabatic Ekpyrosis: Scale-Invariant
  Curvature Perturbations from a Single Scalar Field in a Contracting
  Universe}}, \href{https://doi.org/10.1103/PhysRevLett.104.091301}{\emph{Phys.
  Rev. Lett.} {\bfseries 104} (2010) 091301}
  [\href{https://arxiv.org/abs/0910.2230}{{\ttfamily arXiv:0910.2230}}].

\bibitem{Khoury:2011ii}
J.~Khoury and P.J.~Steinhardt, \emph{{Generating Scale-Invariant Perturbations
  from Rapidly-Evolving Equation of State}},
  \href{https://doi.org/10.1103/PhysRevD.83.123502}{\emph{Phys. Rev. D}
  {\bfseries 83} (2011) 123502}
  [\href{https://arxiv.org/abs/1101.3548}{{\ttfamily arXiv:1101.3548}}].

\bibitem{Geshnizjani:2014bya}
G.~Geshnizjani and W.H.~Kinney, \emph{{Theoretical implications of detecting
  gravitational waves}},
  \href{https://doi.org/10.1088/1475-7516/2015/08/008}{\emph{JCAP} {\bfseries
  08} (2015) 008} [\href{https://arxiv.org/abs/1410.4968}{{\ttfamily
  arXiv:1410.4968}}].

\bibitem{Philcox:2025bvj}
O.H.E.~Philcox, \emph{{Searching for Inflationary Physics with the CMB
  Trispectrum: 1. Primordial Theory \& Optimal Estimators}},
  \href{https://arxiv.org/abs/2502.04434}{{\ttfamily arXiv:2502.04434}}.

\bibitem{Philcox:2025lrr}
O.H.E.~Philcox, \emph{{Searching for Inflationary Physics with the CMB
  Trispectrum: 2. Code \& Validation}},
  \href{https://arxiv.org/abs/2502.05258}{{\ttfamily arXiv:2502.05258}}.

\bibitem{Philcox:2025wts}
O.H.E.~Philcox, \emph{{Searching for Inflationary Physics with the CMB
  Trispectrum: 3. Constraints from Planck}},
  \href{https://arxiv.org/abs/2502.06931}{{\ttfamily arXiv:2502.06931}}.

\bibitem{Chen:2011zf}
X.~Chen, \emph{{Primordial Features as Evidence for Inflation}},
  \href{https://doi.org/10.1088/1475-7516/2012/01/038}{\emph{JCAP} {\bfseries
  01} (2012) 038} [\href{https://arxiv.org/abs/1104.1323}{{\ttfamily
  arXiv:1104.1323}}].

\bibitem{Chen:2011tu}
X.~Chen, \emph{{Fingerprints of Primordial Universe Paradigms as Features in
  Density Perturbations}},
  \href{https://doi.org/10.1016/j.physletb.2011.11.009}{\emph{Phys. Lett. B}
  {\bfseries 706} (2011) 111}
  [\href{https://arxiv.org/abs/1106.1635}{{\ttfamily arXiv:1106.1635}}].

\bibitem{Chen:2015lza}
X.~Chen, M.H.~Namjoo and Y.~Wang, \emph{{Quantum Primordial Standard Clocks}},
  \href{https://doi.org/10.1088/1475-7516/2016/02/013}{\emph{JCAP} {\bfseries
  02} (2016) 013} [\href{https://arxiv.org/abs/1509.03930}{{\ttfamily
  arXiv:1509.03930}}].

\bibitem{Chen:2018cgg}
X.~Chen, A.~Loeb and Z.Z.~Xianyu, \emph{{Unique Fingerprints of Alternatives to
  Inflation in the Primordial Power Spectrum}},
  \href{https://doi.org/10.1103/PhysRevLett.122.121301}{\emph{Phys. Rev. Lett.}
  {\bfseries 122} (2019) 121301}
  [\href{https://arxiv.org/abs/1809.02603}{{\ttfamily arXiv:1809.02603}}].

\bibitem{Domenech:2020qay}
G.~Dom\`enech, X.~Chen, M.~Kamionkowski and A.~Loeb, \emph{{Planck residuals
  anomaly as a fingerprint of alternative scenarios to inflation}},
  \href{https://doi.org/10.1088/1475-7516/2020/10/005}{\emph{JCAP} {\bfseries
  10} (2020) 005} [\href{https://arxiv.org/abs/2005.08998}{{\ttfamily
  arXiv:2005.08998}}].

\bibitem{Carr:2011hv}
B.J.~Carr and A.A.~Coley, \emph{{Persistence of black holes through a
  cosmological bounce}},
  \href{https://doi.org/10.1142/S0218271811020640}{\emph{Int. J. Mod. Phys. D}
  {\bfseries 20} (2011) 2733}
  [\href{https://arxiv.org/abs/1104.3796}{{\ttfamily arXiv:1104.3796}}].

\bibitem{Quintin:2016qro}
J.~Quintin and R.H.~Brandenberger, \emph{{Black hole formation in a contracting
  universe}}, \href{https://doi.org/10.1088/1475-7516/2016/11/029}{\emph{JCAP}
  {\bfseries 11} (2016) 029}
  [\href{https://arxiv.org/abs/1609.02556}{{\ttfamily arXiv:1609.02556}}].

\bibitem{Chen:2016kjx}
J.W.~Chen, J.~Liu, H.L.~Xu and Y.F.~Cai, \emph{{Tracing Primordial Black Holes
  in Nonsingular Bouncing Cosmology}},
  \href{https://doi.org/10.1016/j.physletb.2017.03.036}{\emph{Phys. Lett. B}
  {\bfseries 769} (2017) 561}
  [\href{https://arxiv.org/abs/1609.02571}{{\ttfamily arXiv:1609.02571}}].

\bibitem{Clifton:2017hvg}
T.~Clifton, B.~Carr and A.~Coley, \emph{{Persistent Black Holes in Bouncing
  Cosmologies}}, \href{https://doi.org/10.1088/1361-6382/aa6dbb}{\emph{Class.
  Quant. Grav.} {\bfseries 34} (2017) 135005}
  [\href{https://arxiv.org/abs/1701.05750}{{\ttfamily arXiv:1701.05750}}].

\bibitem{Corman:2022rqo}
M.~Corman, W.E.~East and J.L.~Ripley, \emph{{Evolution of black holes through a
  nonsingular cosmological bounce}},
  \href{https://doi.org/10.1088/1475-7516/2022/09/063}{\emph{JCAP} {\bfseries
  09} (2022) 063} [\href{https://arxiv.org/abs/2206.08466}{{\ttfamily
  arXiv:2206.08466}}].

\bibitem{Chen:2022usd}
J.W.~Chen, M.~Zhu, S.F.~Yan, Q.Q.~Wang and Y.F.~Cai, \emph{{Enhance primordial
  black hole abundance through the non-linear processes around bounce point}},
  \href{https://doi.org/10.1088/1475-7516/2023/01/015}{\emph{JCAP} {\bfseries
  01} (2023) 015} [\href{https://arxiv.org/abs/2207.14532}{{\ttfamily
  arXiv:2207.14532}}].

\bibitem{Cai:2023ptf}
Y.F.~Cai, C.~Tang, G.~Mo, S.F.~Yan, C.~Chen, X.H.~Ma et~al., \emph{{Primordial
  black hole mass functions as a probe of cosmic origin}},
  \href{https://doi.org/10.1007/s11433-023-2314-1}{\emph{Sci. China Phys. Mech.
  Astron.} {\bfseries 67} (2024) 259512}
  [\href{https://arxiv.org/abs/2301.09403}{{\ttfamily arXiv:2301.09403}}].

\bibitem{Chagoya:2016inc}
J.~Chagoya and G.~Tasinato, \emph{{A geometrical approach to degenerate
  scalar-tensor theories}},
  \href{https://doi.org/10.1007/JHEP02(2017)113}{\emph{JHEP} {\bfseries 02}
  (2017) 113} [\href{https://arxiv.org/abs/1610.07980}{{\ttfamily
  arXiv:1610.07980}}].

\bibitem{deRham:2016ged}
C.~de~Rham and H.~Motohashi, \emph{{Caustics for Spherical Waves}},
  \href{https://doi.org/10.1103/PhysRevD.95.064008}{\emph{Phys. Rev. D}
  {\bfseries 95} (2017) 064008}
  [\href{https://arxiv.org/abs/1611.05038}{{\ttfamily arXiv:1611.05038}}].

\bibitem{DeFelice:2018ewo}
A.~De~Felice, D.~Langlois, S.~Mukohyama, K.~Noui and A.~Wang,
  \emph{{Generalized instantaneous modes in higher-order scalar-tensor
  theories}}, \href{https://doi.org/10.1103/PhysRevD.98.084024}{\emph{Phys.
  Rev. D} {\bfseries 98} (2018) 084024}
  [\href{https://arxiv.org/abs/1803.06241}{{\ttfamily arXiv:1803.06241}}].

\bibitem{Pajer:2018egx}
E.~Pajer and D.~Stefanyszyn, \emph{{Symmetric Superfluids}},
  \href{https://doi.org/10.1007/JHEP06(2019)008}{\emph{JHEP} {\bfseries 06}
  (2019) 008} [\href{https://arxiv.org/abs/1812.05133}{{\ttfamily
  arXiv:1812.05133}}].

\bibitem{Grall:2019qof}
T.~Grall, S.~Jazayeri and E.~Pajer, \emph{{Symmetric Scalars}},
  \href{https://doi.org/10.1088/1475-7516/2020/05/031}{\emph{JCAP} {\bfseries
  05} (2020) 031} [\href{https://arxiv.org/abs/1909.04622}{{\ttfamily
  arXiv:1909.04622}}].

\bibitem{Faraoni:2022doe}
V.~Faraoni, A.~Giusti, S.~Jose and S.~Giardino, \emph{{Peculiar thermal states
  in the first-order thermodynamics of gravity}},
  \href{https://doi.org/10.1103/PhysRevD.106.024049}{\emph{Phys. Rev. D}
  {\bfseries 106} (2022) 024049}
  [\href{https://arxiv.org/abs/2206.02046}{{\ttfamily arXiv:2206.02046}}].

\bibitem{Miranda:2022wkz}
M.~Miranda, D.~Vernieri, S.~Capozziello and V.~Faraoni, \emph{{Fluid nature
  constrains Horndeski gravity}},
  \href{https://doi.org/10.1007/s10714-023-03128-1}{\emph{Gen. Rel. Grav.}
  {\bfseries 55} (2023) 84} [\href{https://arxiv.org/abs/2209.02727}{{\ttfamily
  arXiv:2209.02727}}].

\bibitem{Afshordi:2009tt}
N.~Afshordi, \emph{{Cuscuton and low energy limit of Horava-Lifshitz gravity}},
  \href{https://doi.org/10.1103/PhysRevD.80.081502}{\emph{Phys. Rev. D}
  {\bfseries 80} (2009) 081502}
  [\href{https://arxiv.org/abs/0907.5201}{{\ttfamily arXiv:0907.5201}}].

\bibitem{Horava:2009uw}
P.~Horava, \emph{{Quantum Gravity at a Lifshitz Point}},
  \href{https://doi.org/10.1103/PhysRevD.79.084008}{\emph{Phys. Rev. D}
  {\bfseries 79} (2009) 084008}
  [\href{https://arxiv.org/abs/0901.3775}{{\ttfamily arXiv:0901.3775}}].

\bibitem{Mukohyama:2020lsu}
S.~Mukohyama and R.~Namba, \emph{{Partial UV Completion of $P(X)$ from a Curved
  Field Space}},
  \href{https://doi.org/10.1088/1475-7516/2021/02/001}{\emph{JCAP} {\bfseries
  02} (2021) 001} [\href{https://arxiv.org/abs/2010.09184}{{\ttfamily
  arXiv:2010.09184}}].

\bibitem{Lara:2021piy}
G.~Lara, M.~Bezares and E.~Barausse, \emph{{UV completions, fixing the
  equations, and nonlinearities in k-essence}},
  \href{https://doi.org/10.1103/PhysRevD.105.064058}{\emph{Phys. Rev. D}
  {\bfseries 105} (2022) 064058}
  [\href{https://arxiv.org/abs/2112.09186}{{\ttfamily arXiv:2112.09186}}].

\bibitem{Wang:2013mea}
Y.~Wang, \emph{{MathGR: a tensor and GR computation package to keep it
  simple}},  \href{https://arxiv.org/abs/1306.1295}{{\ttfamily
  arXiv:1306.1295}}.

\end{thebibliography}\endgroup

\end{document}